\documentclass[twocolumn,appendixfloats]{aastex6}
\usepackage{xcolor}

\newcommand{\figscaleone}{\epsscale{1.15}}     
 
\newcommand{\figscalethree}{\epsscale{0.5}}

\newcommand{\htwo}{${\rm H_2}$}
\newcommand{\ph}{{\it para-}${\rm H_2}$}
\newcommand{\oh}{{\it ortho-}${\rm H_2}$}

\newcommand\msun{{\rm M_\odot}}

\newcommand\be{\begin{equation}}
\newcommand\ee{\end{equation}}

\newcommand\pn{\tilde{P}}
\newcommand\dn{\tilde{\rho}}
\newcommand\dgrad{\Delta\!\!\nabla}

\newcommand{\response}{}
\newcommand{\newresponse}{}

\shorttitle{Cosmic snow clouds}
\shortauthors{Walker \&\ Wardle}

\begin{document}

\title{\response Cosmic snow clouds:\\
self-gravitating gas spheres manifesting hydrogen condensation}

\author{Mark A. Walker}
\affil{Manly Astrophysics, 15/41-42 East Esplanade, Manly, NSW 2095, Australia}
\email{1. Mark.Walker@manlyastrophysics.org}
\author{Mark J. Wardle}
\affil{Department of Physics and Astronomy, Macquarie University, Sydney, NSW 2109, Australia}
\email{2. Mark.Wardle@mq.edu.au}

\begin{abstract}
We present hydrostatic equilibrium models of spherical, self-gravitating clouds of helium and molecular hydrogen, focusing on the cold, high-density regime where solid- or liquid-hydrogen can form. The resulting structures have masses from $0.1\,{\rm M_\odot}$ down to several $\times10^{-8}\,{\rm M_\odot}$, and span a broad range of radii: $10^{-4}\la R({\rm AU})\la10^7$. Our models are fully convective, but all have a two-zone character with the majority of the mass in a small, condensate-free core, surrounded by a colder envelope where phase equilibrium obtains. Convection in the envelope is unusual in that it is driven by a mean-molecular-weight inversion, rather than by an entropy gradient. In fact the entropy gradient is itself inverted, leading to the surprising result that envelope convection transports heat inwards. In turn that permits the outer layers to maintain steady state temperatures below the cosmic microwave background. Amongst our hydrostatic equilibria we identify thermal equilibria appropriate to the Galaxy, in which radiative cooling from \htwo\ is balanced by cosmic-ray heating. These equilibria are all thermally unstable, albeit with very long thermal timescales in some cases. The specific luminosities of all our models are very low, and they therefore describe a type of baryonic dark matter. Consequently such clouds are thermally fragile: when placed in a harsh radiation field they will be unable to cool effectively and disruption will ensue as heat input drives a secular expansion. Disrupting clouds should leave trails of gas and \htwo\ dust in their wake, which might make them easier to detect. Our models may be relevant to the cometary globules in the Helix Nebula, and the G2 cloud orbiting Sgr A*.
\end{abstract}

\keywords{ISM: clouds --- Galaxy: halo --- dark matter --- galaxies: ISM}

\section{Introduction}\label{sec:intro}
Understanding the structure of self-gravitating bodies is a fundamental aspect of many branches of astronomy: galaxies, stars, and interstellar clouds, for example. Modelling such systems is greatly simplified by assuming time-independence and spherical symmetry.  But even with these assumptions a diverse collection of structures can arise, depending on the equation of state of the fluid and the boundary conditions --- witness the variety encountered in the aforementioned disciplines. In this paper we explore a new set of time-independent, spherically-symmetric, self-gravitating equilibria, appropriate to fluids composed of helium and molecular hydrogen, in which the combination of low temperature and high density permits solid- or liquid-hydrogen to condense. 

The motivation for this study has its roots in the idea that molecular gas which is cold and dense would be very difficult to detect, and therefore large amounts of such gas could be present, yet remain undetected, in galaxies \citep{pfenniger1,pfenniger2}. Indeed Pfenniger and Combes argued that the presence of such a reservoir may help in understanding the observed properties of star-forming galaxies. In these original papers it was recognised that the molecular hydrogen component would be close to its saturated vapour pressure, and might therefore be able to condense. However, the original concept of a fractal character for the gas clouds does not lend itself to detailed structural modelling \citep[e.g.][]{pfenniger3}. Subsequent studies considered the possibility of cold, dense gas in long-lived, spherical clouds -- see, for example \citet{henriksen,gerhard,walker1998,draine1998, sciama} -- for which structural modelling is  tractable. But to date the effects of the \htwo\ phase change have been investigated only in ``one-zone'' models, where the entire body is characterised by a single, representative value of the temperature, pressure etc. \citep{wardle,fuglistaler2015,fuglistaler2016}. This paper presents models in which full radial profiles are constructed, providing the first detailed pictures of clouds manifesting \htwo\ condensation.

Because of the steep temperature dependence of the saturated vapor pressure curve (see figure 1), a basic expectation is that the outer layers of dense clouds are more favourable for condensation than the interiors. Indeed all of our models exhibit a central region which is sufficiently warm that no condensation of \htwo\ takes place there. We refer to that central region as the ``core'' of the cloud, and the outer regions, where condensation occurs, as the ``envelope''. When discussing \htwo\ condensates we will usually refer only to the solid -- i.e. hydrogen ``snowflakes'' rather than hydrogen droplets -- and we refer to our model structures as ``snow clouds''. This is not meant to be prejudicial, as liquid droplets can form under some circumstances and the physics is qualitatively similar for the two condensed phases. Rather it is a convenient brevity of expression. We will see, however, that clouds which incorporate both liquid and solid condensates occupy only a small region of the mass-radius plane compared to pure snow clouds (\S3), so a degree of emphasis on the solid form is appropriate. Moreover, our adopted description of the equation of state in phase equilibrium (\S2.6) is much more accurate for the solid-gas transition than for the liquid-gas boundary, and our rain-cloud models should be thought of as very rough sketches.

Even rough sketches can be valuable, if they are novel. The solutions we have obtained display properties that were not anticipated in earlier work, and which broaden the range of possible structures for interstellar clouds. For example, \citet{mckee} used polytropic models to highlight problems with models of self-gravitating molecular clouds that have both high densities and low temperatures --- precisely the corner of parameter space we are interested in. We address the issues in detail later in this paper (\S4.1, \S5.1, and Appendix A), but for now we note one key point of difference: \citet{mckee} assumed that the temperature of the cosmic microwave background (CMB) sets a floor on the gas temperatures within any cloud in steady state, whereas our models exhibit much lower temperatures in their outer regions. Steady-state temperatures below the CMB are permissible in our models because heat is convected from those regions to the warmer interior, whence it is radiated away. Convection of heat up a macroscopic temperature gradient is unfamiliar, even counterintuitive, and does not seem to have been dealt with previously in the literature. We give details in \S4 and Appendix A, but the main point is simply that heat flows down the entropy gradient, and entropy increases outwards in the envelopes of our model clouds, with convection being driven by a composition gradient.

The structure of this paper is as follows. We begin in \S2 by presenting the various ingredients required to construct our hydrostatic models, including a derivation of the equation of state of the fluid. In \S3 we use those ingredients to construct hydrostatic equilibria. We show one example structure in detail, and we illustrate how models populate the mass-radius plane. {\response In addition to being hydrostatic, true equilibrium structures must also be in thermal balance, so in \S4 we describe the thermal properties of the hydrostatic solutions. We consider thermal balance locally -- i.e. energy flow within each cloud -- and globally, i.e. total heating balanced by total radiative cooling. We identify a subset of the hydrostatic models which are indeed thermal equilibria; those equilibria are, however, subsequently shown to be thermally unstable.} Finally, in \S5, we consider various issues with the models and we suggest possible manifestations in the observed universe. 

\section{{\response hydrostatic model ingredients}}

\subsection{Equations of hydrostatic equilibrium}
We assume that the clouds are spherical and in hydrostatic equilibrium, so that the pressure and mass gradients are given by
\be
{{{\rm d}P}\over{{\rm d\/}r}}=-\frac{GM}{r^2}\rho,\qquad\quad{{{\rm d}M}\over{{\rm d\/}r}}=4\pi r^2\rho,
\ee
just as in the case of stars \citep[e.g.][]{kippenhahn1994}. It is convenient to work in terms of the normalised variables $\pn\equiv P/P_c$ and $\dn\equiv \rho/\rho_c$, where $P_c$ and $\rho_c$ are the central pressure and density, repectively. We are also free to choose a scaling, $r_o$, for the radial coordinate and work in terms of $z\equiv r/r_o$. Whatever radial scale is chosen we can introduce a corresponding mass scale, $M_o=4\pi r_o^3 \rho_c$, and employ the dimensionless variable $m\equiv M/M_o$. We choose the radial scale $r_o:=G M_o\rho_c / P_c$.

Instead of using $z$ as the independent variable we have found it most convenient to use $q\equiv\log\pn$. The equations of hydrostatic equilibrium then read
\be
{{{\rm d\/}z}\over{{\rm d\/}q}}=-{{z^2}\over{m\dn}}\exp(q),\quad{{{\rm d}m}\over{{\rm d\/}q}}=-{{z^4}\over{m}}\exp(q).
\ee
The main advantage of this choice is that $q$ is known at all three boundaries of the structural problem -- inner boundary ($q=0$), core-envelope boundary (see \S3.2), and outer boundary ($q=-\infty$) -- so one integrates over two predetermined ranges of the independent variable, using the appropriate equation of state in each case. By contrast, the values of $z$ and $m$, for example, are initially known only at the inner boundary. 

Use of $q$ rather than $\pn$ is motivated by our desire for an accurate structural model over a broad range of pressure. That is a significant consideration for our models because, as we will see in \S3, it is often the case that most of the volume is occupied with fluid at pressures $\pn\lll1$. A disadvantage of our coordinate choice is that the domain of numerical integration cannot extend to the true surface of the cloud at $q=-\infty$. But that is only a small disadvantage, because one can explore to arbitrarily low pressures.

\subsection{Composition}
In \S5 we suggest possible connections between our models and the observed Universe. As metals provide most of the information on the latter, there is a clear motivation to include them in our model clouds. However, adding metals greatly increases the complexity of the modelling, and the dimensionality of the parameter space in which the models are constructed. These are strong motivations to exclude metals in this initial attempt at characterizing snow clouds. For simplicity, then, we have constructed clouds of zero metallicity and with a hydrogen to helium ratio of 3:1 by mass, i.e. similar to the observed cosmic helium abundance \citep[e.g.][]{nieva2012}. (In \S5.2 we consider how the inclusion of metals might affect snow cloud models.) We further assume that all of the hydrogen is in molecular form; this is expected for long-lived clouds, as three-body reactions are efficient at converting H to \htwo\ when the gas density is high \citep{palla1983}. {\response Finally, we consider only the most abundant isotopes of hydrogen and helium, so no D, T or ${}^3$He.} Other compositions would be worth studying in future, but for this initial investigation there is plenty to explore without adding further complexity.

At the low temperatures of interest here, we need to specify the proportions of \oh\ ($J=1,3,5\dots$) and \ph\ ($J=0,2,4\dots$). On short timescales these sequences behave as if they were distinct species, because they're exclusively associated with the nuclear spin triplet and singlet states, respectively, of the \htwo\ molecule, and the rate at which nuclear spins are flipped is expected to be very low \citep{freiman2017}. Below the critical point \citep[$T_{crit}\simeq32.9\,{\rm K}$,][]{leachman2009}, where \htwo\ may condense into liquid form, the {\it ortho/para\/} ratio in thermal equilibrium is below 5\%. And below the triple point \citep[$T_{trip}\simeq13.8\,{\rm K}$,][]{leachman2009}, where solid \htwo\ may precipitate, it is more than a thousand times smaller again. We therefore proceed by neglecting the \oh\ content.

The reason that \ph\ tends to dominate at low temperatures is simply that the $J=1$ state lies roughly $170\,{\rm K}$ above the rotational ground state ($J=0$). As the energies of the excited states are proportional to $J(J+1)$, it is clear that the population of the $J=2$ level, at approximately $510\,{\rm K}$ above ground, will be very small indeed. The excited rotational levels play a key role in radiative cooling, so in \S4 we will quantify their population. But for now it suffices to make the approximation that all \htwo\ molecules are in the rotational ground state. That approximation effects two important simplifications. First it means that the \htwo\ phase boundary corresponds uniquely to that of pure \ph. Secondly, the \htwo\ can be modelled as effectively monatomic, with only translational degrees of freedom contributing to the internal energy. 

\subsection{Convection}
It is not possible to have a static model in which \htwo\ condensates are present. The reason is that the density of the solid is much higher than that of the gas, so it would precipitate out. A rough estimate of the precipitation timescale can be made for spherical particles of radius $a$, located at a radial distance $r$ from the centre of a cloud: in units of the dynamical timescale it is $\sim\sqrt{r/(\eta a)}$, where $\eta$ is the ratio of the solid density to that of the background fluid. (Dendritic snowflakes of the same mass would settle more slowly, because of their larger cross-section.) Anticipating the results we will present in \S3, we find that at the base of the envelope of the cloud the precipitation timescale for micron-sized snowflakes is $\la3\times10^6\,{\rm yr}$ for all of our models.

Precipitation creates a composition gradient in the fluid, making it more helium rich in the outer regions. While the fluid remains both saturated and static there is no limit to this process. Thus precipitation continues until buoyancy instability sets in, and the resultant mixing counters the growth of the composition gradient. We use the term ``convection'' to describe this buoyant overturn, but we caution that the character of this convection differs greatly from the more familiar case of convection in a fluid of uniform composition. We return to this issue when we consider the thermal properties of our models in \S4. Appendix A presents criteria for buoyancy instability in the presence of a composition gradient.

We note that for a fluid at the thermodynamic critical point, the density contrast between gas and liquid phases vanishes (see Appendix B). Therefore precipitation near the critical point is relatively slow, and convection can only be driven relatively weakly. The density contrast between condensed and gaseous phases increases monotonically as the temperature decreases. 

In each of our model clouds the core is too warm to support \htwo\ phase equilibrium. But it may convect, depending on the temperature gradient (equation A7) that would be needed for radiative transport of heat therein. Radiative cooling of condensate-free gas is predominantly via the narrow ${\rm S_0(0)\;(J=2\rightarrow0,\;28\,{\rm\mu m})}$ (pure-rotation) line of \htwo. Because of the high densities and low temperatures, the line is heavily optically thick (see figure 12), and heat flow takes place predominantly in the wings of the line. For similar circumstances, it was shown by \citet{clarke1997} that convection is expected if the gas is heated primarily by cosmic-rays --- as is the case in our models (see \S4.3). However, our models differ from those of \citet{clarke1997} in that (i) our clouds contain no metals, so there is no cooling by metal lines, and (ii) our models have such large central column-densities that the specific heating rate is non-uniform. Consequently convection is not guaranteed. In this initial exploration of snow cloud properties we assume, for the sake of simplicity, that the warm core is indeed convectively unstable.

\subsection{Ideal gas approximation}
It appears that helium does not alloy with solid \htwo\  \citep{leventhal1991,safa2008}, so we assume that all helium is in gas phase. For both helium and gas-phase \htwo, we use the ideal gas descriptions of pressure and entropy. This approximation is a natural one for helium at the temperatures and pressures encountered in our models, which are far from He condensation. But for \htwo\ we are specifically interested in describing its change of phase and one might expect the ideal gas model to be inadequate. Indeed it is a very poor approximation for \htwo\ near the critical point, where isotherms deviate strongly from $PV=\,{\rm constant.}$ But at lower temperatures and pressures the gaseous \htwo\ adheres closely to perfect gas behaviour right up to the point of saturation. In particular, in the vicinity of the sublimation curve ideal gas pressure and entropy are very good approximations for the gaseous components of the fluid --- see Appendix B.

For $N$ gas atoms/molecules, each of mass $\mu$, in volume $V$, the pressure of an ideal gas is
\be
P={N\over V}kT.
\ee
We are approximating the \htwo\ molecules as effectively monatomic (\S2.2) --- meaning that the internal degrees of freedom (rotation and vibration) are not excited. In this case the entropy, $S$, is given by the Sackur-Tetrode formula:
\be
{S\over{Nk}}={5\over2}+\log_e\left[{V\over N}\left({{2\pi\mu kT}\over{h^2}}\right)^{3/2}\right],
\ee
where $h$ is Planck's constant. The effect of phase equilibrium, where manifest, is then to introduce an additional constraint -- the partial pressure of \htwo\ must equal the saturated vapour pressure (\S2.5) -- and an additional freedom: the number of gas-phase molecules is not fixed.

We can gauge the accuracy of our approximations in a couple of ways. First by reference to the van~der~Waals equation of state for \htwo\ \citep{johnston2014}, which provides a simple model for non-ideality, and secondly by reference to the measured properties of \htwo\ gas near the saturation curve \citep{roder1973,leachman2009}. Those comparisons are presented in Appendix B; here we simply note that our approximations are good at temperatures below the triple point, but they worsen as the temperature is increased past that point and are very poor in the immediate vicinity of the critical point.

\begin{figure}
\figscaleone
\plotone{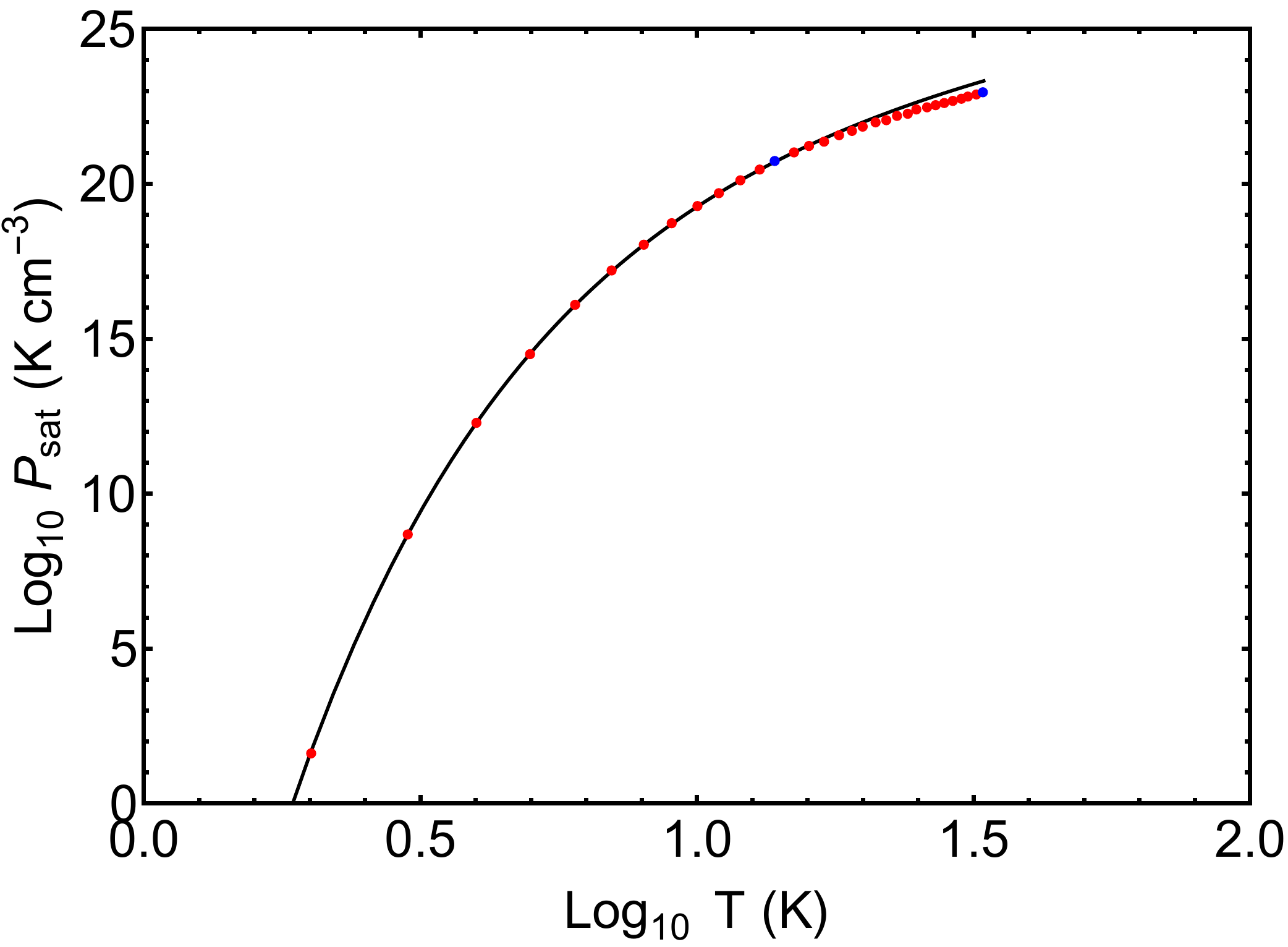}
\vskip-0.1truecm
\caption{The saturation curve of \htwo, as given by equation (5) with $b/k=91.5\,$K, is shown by the solid black line for temperatures up to the critical point, $T_{crit}\simeq32.9\,$K. Values of $P_{sat}$ taken from the literature are shown with red dots. For temperatures up to the triple point, approximately $13.8\,$K, these are the recommended values from the survey of \citet{roder1973}. Above the triple point the data are from \citet{leachman2009}. The triple point and critical point are both marked with blue dots. Equation (5) is a good approximation below the triple point, but becomes poorer as the temperature increases from there; at the critical point our model overpredicts $P_{sat}$ by a factor of 2.3.}
\medskip
\end{figure}

\subsection{Envelope phase equilibrium}
In the envelope of each cloud we assume phase equilibrium --- i.e. that the partial pressure of \htwo\ is equal to the saturated vapour pressure, $P_{sat}(T)$ at temperature $T$. The assumption of phase equilibrium is motivated by the idea that snowflakes are present in convective cells throughout the envelope, with snowflake mass growing in upwellings and shrinking in downdrafts. Furthermore snowflakes are not perfectly coupled to the gas, and to some extent must be dispersed into adjacent cells by the complicated fluid motions which convection gives rise to. In this circumstance nucleation is unlikely to be an issue and condensation/sublimation should accurately reflect the thermodynamic potentials.

In Appendix B we present laboratory data which demonstrate that, except in the vicinity of the critical point, the volume and entropy of the condensed phase are both small in comparison with those of the gas phase. In order to develop an analytic description of the equation of state, we henceforth neglect the volume and entropy of the condensed phase. The Clausius-Clapeyron equation for the phase boundary then tells us that the saturated vapour pressure is
\be
P_{sat}=kT{{(2\pi \mu kT)^{3/2}}\over{h^3}}\exp\left(-{{b}\over{kT}}\right),
\ee
where $b$ is the latent heat of sublimation, which we approximate as a constant. Equation (5) differs slightly from the saturation pressure given by E.S.~Phinney (1985, preprint), who accounted for the small heat capacity of the solid. In this paper we adopt  $b/k=91.5\,{\rm K}$, and equation (5) then describes the saturated vapour pressure of {\it para-\/}\htwo\  to within a few percent at temperatures up to the triple-point. Above the triple-point our approximation progressively worsens, overpredicting $P_{sat}$ by a factor of 2.3 at the critical-point.

Although that error of approximation is large, it should be seen in the context of the huge range in $P_{sat}$ that must be described --- as evident in figure 1. Moreover, other aspects of our microscopic description are inaccurate in the vicinity of the critical point (see \S2.4 and Appendix B). The poor performance of our $P_{sat}$ approximation in that region adds emphasis to the point, already made, that our models are only sketches in cases where the liquid condensate is present, and increasingly rough sketches as the envelope temperature approaches the critical point. 

\subsection{Equations of state}
We assume slow convective turnover -- i.e. fluid speeds that are small compared to the sound-speed -- so that hydrostatic equilibrium remains a good approximation. Slow turnover is expected if the fluid is everywhere only marginally buoyantly unstable. The timescale for radiative cooling is expected to be orders of magnitude longer than the dynamical timescale (\S4), and we therefore approximate the fluid motions as adiabatic. Marginal buoyancy instability then implies that the compressibility is given by
\be
{{{{\rm d}\log \rho}}\over{{{\rm d}\log P}}}={{{{\rm d}\log \rho}/{{\rm d}\log r}}\over{{{\rm d}\log P}/{{\rm d}\log r}}} = \left( {{\partial \log \rho}\over{\partial\log P}} \right)_{\!\!S},
\ee
everywhere in the cloud.

The total entropy of a fluid parcel can be written as the sum of two terms of the form given in equation (4): one for helium and one for hydrogen. If the number of gas-phase molecules is fixed then the differential of equation (4) is just
\be
{\rm d}S=Nk\left\{{3\over2}{\rm d}\!\log P - {5\over2}{\rm d}\!\log \rho\right\},
\ee
so that
\be
\left( {{\partial \log \rho}\over{\partial\log P}} \right)_{\!\!S} = {{3}\over{5}},
\ee
which is the familiar form for the adiabatic trajectory of an ideal gas. Equation (8) is the equation of state in the core of each cloud.

In phase equilibrium, however, the number of gas-phase molecules varies so as to maintain the partial pressure of \htwo\ at its saturated level. With that constraint, expressed in the form of equation (5), we can rewrite the Sackur-Tetrode entropy for the saturated hydrogen vapor as $S_{sat}=Nk\psi$, with
\be
\psi\equiv{5\over2}+{b\over{kT}}.
\ee
At a fixed value of the total entropy (hydrogen plus helium), the adiabatic compressibility is thus
\be
\left( {{\partial \log \rho}\over{\partial\log P}} \right)_{\!\!S} = {{2(\psi-1)^2 + 3(y+1)}\over{2\psi^2+5y}},
\ee
where $y$ is the ratio of the partial pressure of helium to that of hydrogen. Equation (10) is the equation of state in the envelope of each cloud.

As expected, equation (10) reduces to equation (8) in the limit $y\rightarrow\infty$, where there is so little hydrogen that the thermodynamics of the \htwo\ phase change become irrelevant.  On the other hand, for modest helium concentrations the phase change plays a dominant role at low temperatures, where $\psi\gg1$, and the compressibility (10) approaches unity. Thus the fluid is much softer (i.e. more compressible), under adiabatic conditions, as a result of the phase transition. The reason for this is that, under conditions of phase equilibrium, a lot of the work done during an adiabatic compression goes into liberating a small number of molecules from the condensed phase --- an energy $b\gg kT$ is required to liberate each molecule from the condensate. Consequently the fluid temperature rises only slightly on compression, and the pressure response is therefore smaller than for condensate-free gas.\footnote{\citet{fuglistaler2015} suggested that the adiabatic sound-speed of \htwo\ in phase equilibrium is zero, corresponding to infinite adiabatic compressibility. Their result was obtained by scaling the isothermal sound speed, which is zero in phase equilibrium, by the ratio of specific heats, $C_P/C_V$, which they assumed to be finite. However, at constant pressure the fluid releases/absorbs heat without any change in temperature, in response to  condensation/sublimation of the solid. Consequently $C_P$ is infinite, whereas $C_V$ is not, {\newresponse so the method of sound-speed calculation suggested by \citet{fuglistaler2015} does not yield a well-defined result.}}

In the core of the cloud, convection keeps the fluid well-mixed and the composition is expected to be uniform. By contrast, the composition of fluid in the envelope must change with radius because snowfall denudes the outer layers of hydrogen. Thus although the compressibility is everywhere equal to the locally adiabatic value -- as per equation (10) --  a different adiabat applies at each radius. We therefore need to explicitly consider the composition gradient in order to  determine the temperature and helium abundance profiles in the envelope.

\subsection{Envelope temperature and helium profile}
Ideally a structural model would be built on a microphysical description of how hydrogen snow grows and settles under the combined influence of gravity and convective fluid motions. That, however, is much more detailed than we attempt in this initial sketch of snow cloud properties. Instead we simply assume that condensate makes a negligible contribution to the fluid density at every point in the cloud. This assumption is motivated partly by simplicity, and partly by the idea that snowflakes settle out rapidly (\S2.3). For it to be a good approximation we require that the downward drift speed of the snowflakes be comparable to, or greater than, the speeds achieved by convecting fluid parcels.

It is unclear whether that condition should be expected to be met in practice, as the speed at which snowflakes settle out depends on their size -- which is unknown -- and the convection speeds are also unknown. As a fiducial, we give the settling speed for micron-sized snowflakes at the base of the envelope of a model that we will illustrate later (figure 3): it is $2\times10^{-4}$ times the sound speed in the gas.

Given the assumption of rapid precipitation, the pressure and density of the envelope fluid are fully specified by its temperature and helium content, $y$: we have $P=P_{sat}(1+y)$ and $\rho=\mu P_{sat}(1+2y)/kT$. Alternatively one can think of $T$ and $y$ as being uniquely determined by $P$ and $\rho$, and in practice that is how we proceed when constructing numerical solutions. As we step out in radius, the gas pressure declines -- in a manner quantified in the next section -- and the density of the gas declines in accord with the local adiabatic compressibility, i.e. given by equation (10). By differentiating our expressions for $P$ and $\rho$, just given, we then obtain the gradients in temperature and helium content:
\be
{{{{\rm d}\log T}}\over{{{\rm d}\log P}}}=  {{1+2y}\over{\psi+1+2y}} \left[{1\over{1+2y}}+1-{{{{\rm d}\log \rho}}\over{{{\rm d}\log P}}}\right] ,
\ee
and
\be
{{{{\rm d} y}}\over{{{\rm d}\log P}}}= {{(1+y)(1+2y)}\over{\psi+1+2y}}\left[1-\psi+\psi{{{{\rm d}\log \rho}}\over{{{\rm d}\log P}}}\right] .
\ee
Together with equations (6) and (10), these results specify conditions in the envelope of the cloud.

\subsection{Boundary conditions}
The very centre of the cloud -- where $\dn=1$, and $m=z=q=0$ -- cannot be used as the inner boundary of the numerical integration, because ${\rm d}z/{\rm d}q$ is infinite at that point. Instead we use the limiting behaviour of equations (2) at small $q$, for the equation of state (8), which is:
\be
m\rightarrow{1\over3}z^3,\quad q\rightarrow-{1\over2}z^2,\quad \dn\rightarrow1-{3\over{10}}z^2,
\ee
and we start the integration at a small, but non-zero value of $q$. In our numerical work the inner boundary was set at $q=-5\times10^{-7}$ ($z=10^{-3}$).

Using the boundary conditions just given, one can solve the equations of hydrostatic equilibrium as a set of coupled differential equations (2), together with the equation of state (8), using $q$ as the independent variable and integrating out to the core-envelope boundary, where $q=q_e$ (see \S3.2). From that point one continues the integration, but with the equation of state (10), integrating out to a predetermined, large, negative value of $q$ that in effect defines the surface of the cloud.

Our outer boundary condition neglects the ambient pressure of the external medium. An alternative procedure would be to choose a value for the external pressure -- e.g. the typical pressure of the diffuse ISM in the solar neighbourhood, if we are interested in clouds in the Galactic disk -- and use the corresponding value of $q$ as the surface boundary condition. Clearly, no single procedure yields models that are appropriate to all environments. Appendix D illustrates how our models would be truncated by an external pressure equal to that of the local, diffuse ISM. The influence of a non-zero external pressure is greatest for high-mass clouds with low central temperatures.

We obtained our solutions using the routine {\tt NDSolve\/} in the {\it Mathematica\/} software package,\footnote{www.wolfram.com} following each structure out to a minimum pressure corresponding to $q=-100$. Results are presented in \S3.

\begin{figure}
\figscaleone
\plotone{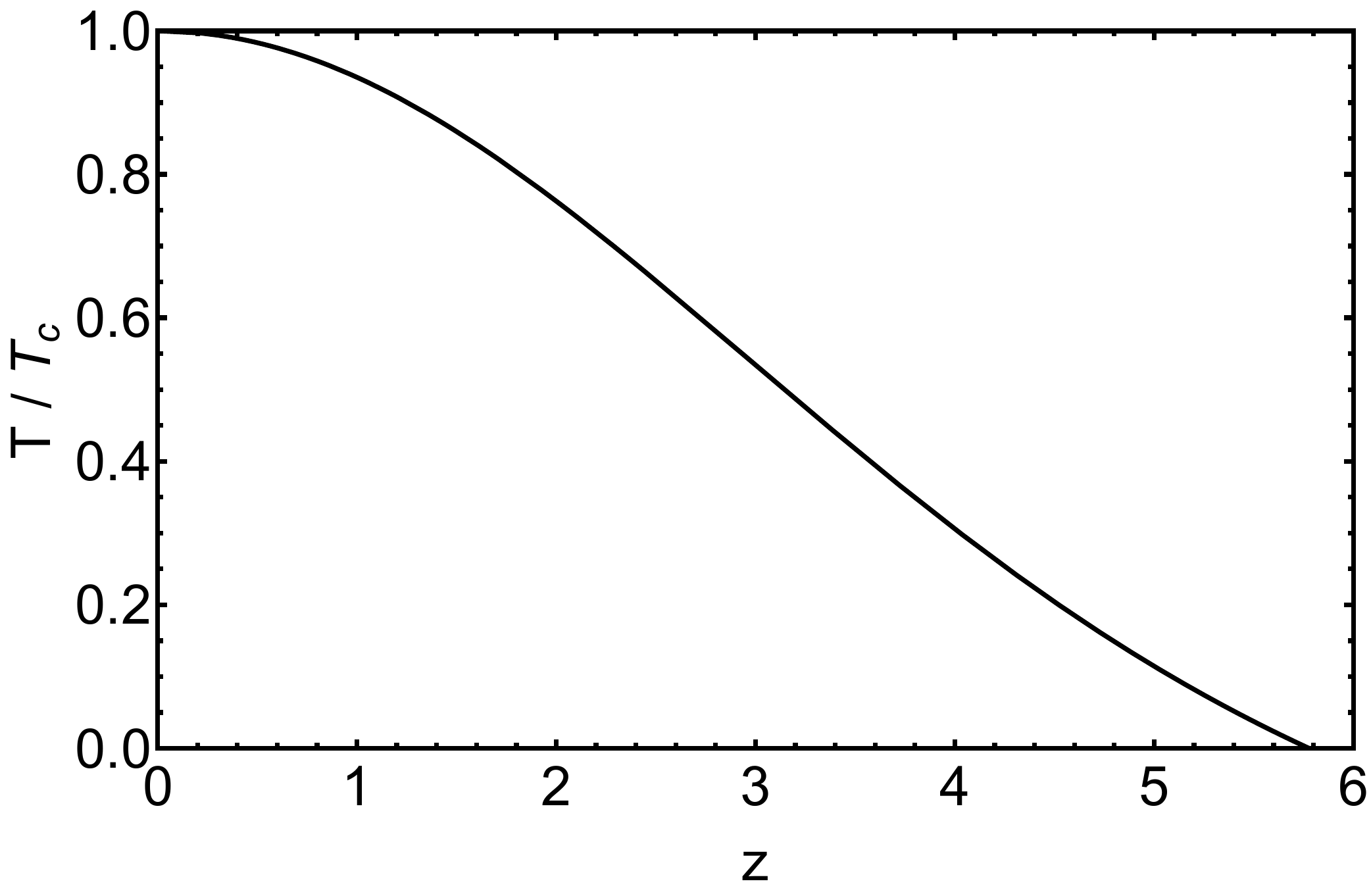}
\plotone{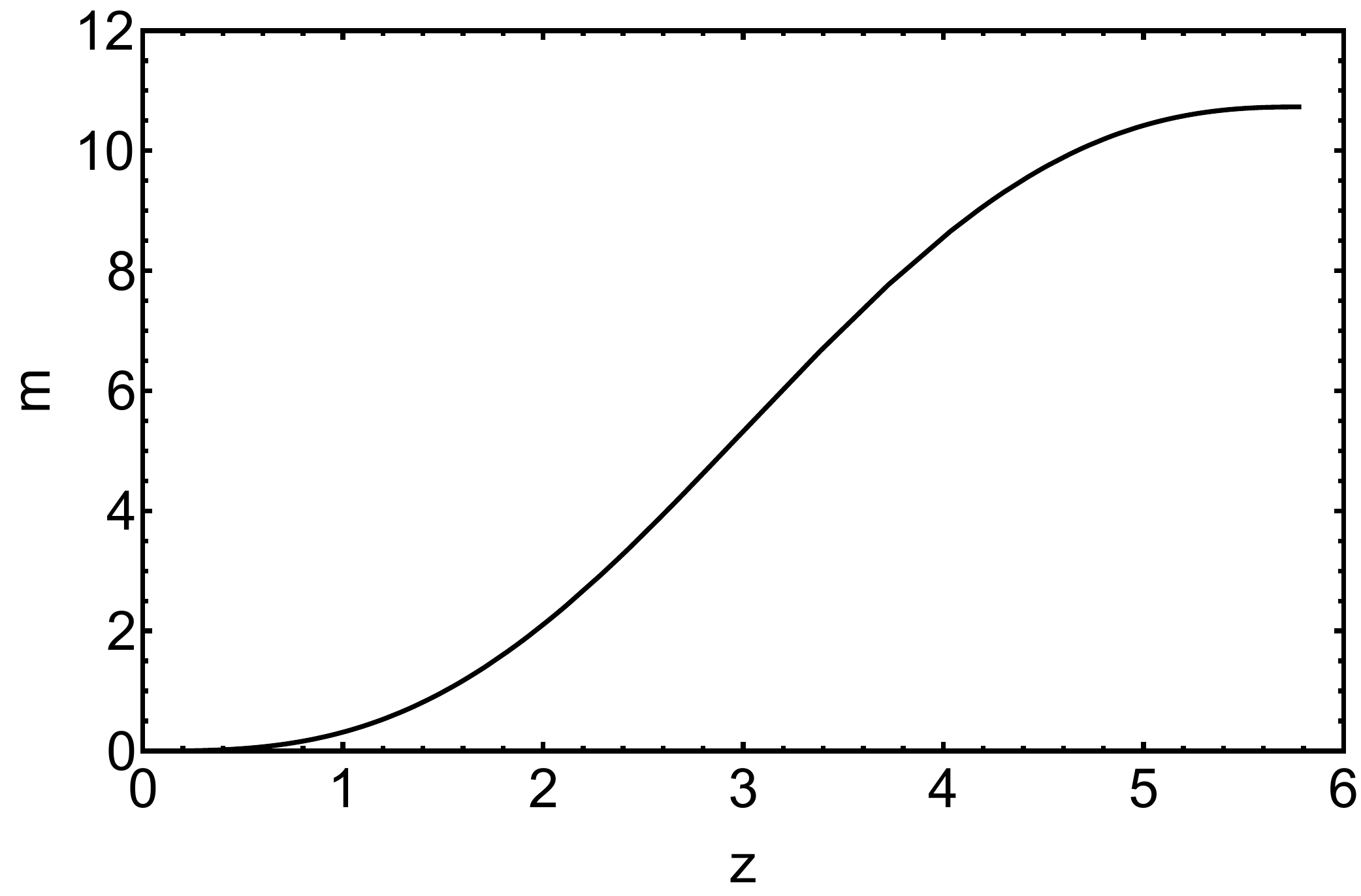}
\caption{Radial profiles for the temperature (upper panel; in units of the central temperature, $T_c$), and the scaled mass (lower panel) for an $n=3/2$ polytrope. The scalings for mass and radius are given in \S2.1. Readers familiar with polytropic models should note that the radial scaling used here makes no reference to the equation of state, so that the surface occurs at $z\simeq5.78$, a factor of $\sqrt{n+1}\simeq1.58$ larger than the value usually quoted for this polytrope. The total mass is $\simeq10.7$. With suitable choices for the central temperature, density and pressure, the profiles shown here provide the solution for the warm core of each model cloud.} 
\end{figure}
 
\section{Hydrostatic models of snow clouds} 
With the ingredients given in \S2 we can proceed to construct hydrostatic equilibria appropriate to the conditions of interest. All of our solutions exhibit two zones: a warm, dense core which does not support condensates, and a colder envelope where \htwo\ phase equilibrium obtains. Although the solution for the warm core will be familiar to most readers, it is helpful to briefly review the result before proceeding to the solution for the cloud as a whole. 

\subsection{Solution for the warm core}
In the case of a fluid which is free of condensates we use equation (8) to describe the run of density as a function of pressure in the cloud. This corresponds to the familiar case of a polytrope, with polytropic index $n=3/2$, resulting in the structure shown in figure 2. The radial profiles of density and pressure can be derived from the temperature profile simply by forming the $3/2$ and $5/2$ power, respectively, of $T/T_c$.

The profiles shown in figure 2 provide a complete structural solution for cases where the same polytropic equation of state applies throughout --- such as the case of a low-mass star that is fully convective \citep[e.g.][]{kippenhahn1994}. In this paper we are concerned with cold gas clouds, and starting from modest central temperatures it is clear that the adiabat will soon cross the \htwo\ sublimation curve (figure 1). Thereafter we need to employ the equation of state (10) rather than equation (8).

\subsection{Solutions for core-plus-envelope}
Once conditions at the centre of the cloud are fully specified (e.g. temperature, pressure, and helium abundance), figure 2 provides the unique solution out as far as the core-envelope boundary. Equation (5) gives the saturation pressure as a simple function of temperature, but there is no simple inverse-function for determining temperature from pressure. We have therefore found it convenient to specify our models using the parameter combination $\{y_c, T_c, T_e\}$, i.e. the central helium abundance, the central temperature and the temperature at the base of the envelope. Together these parameters suffice to uniquely determine conditions in the core, via
\be
P_c=\left({{T_c}\over{T_e}}\right)^{5/2} (1+y_c)P_{sat}(T_e),
\ee
for the central pressure, and
\be
\rho_c=\left({{T_c}\over{T_e}}\right)^{3/2} (1+2y_c) {{\mu }\over{kT_e}} P_{sat}(T_e)
\ee
for the central density.

Because the core obeys a polytropic equation-of-state, the value of the independent variable at the core-envelope boundary is just given by
\be
q_e={5\over2}\log_e{{T_e}\over{T_c}},
\ee
and integration over $q$ is performed separately for the ranges $0 > q \ge q_e$ and $q_e> q \ge-100$. 

As \htwo\ condensation can only be achieved below the critical temperature, the temperature at the base of the envelope must be $T_e\le T_{crit}\simeq32.9\,$K. We will see in \S4 that thermal equilibrium solutions can only be obtained if there is net radiative cooling for the envelope, taken in isolation. In turn that requires $T_e > T_{cmb}\simeq 2.73\,{\rm K}$ (the temperature of the cosmic microwave background), so acceptable values for the temperature at the base of the envelope are restricted to $T_{cmb} < T_e \le T_{crit}$.

The central temperature must be at least as large as $T_e$. However, at high central temperatures the approximations we are using break down: the \oh\ fraction becomes signicant, thus affecting the saturation pressure; and the excitation of rotations modifies the heat capacity of the gas, so equation (8) no longer represents the equation-of-state in the core. We have therefore limited our exploration to central temperatures: $T_e < T_c\le100\,{\rm K}$.

The interesting range of central helium abundance can be anticipated from the following considerations. First, the envelope is always more helium rich than the core, because $y$ monotonically increases outwards, starting from a value of $y_c$ at the base of the envelope. Secondly, the average helium abundance for the cloud must be $\langle y\rangle=1/6$ in order to yield the correct composition for the cloud as a whole (\S2.2). Thus we have $0\le y_c<1/6$.

There is, however, no way of anticipating exactly what value of $y_c$ will yield the correct average composition for a given $\{T_c, T_e\}$ pair. Thus for each combination of temperatures we must construct models for various values of $y_c$, yielding the function $\langle y\rangle(y_c)$, and then determine the particular value of central abundance which yields the correct average composition. We will see later that if $T_e\ll T_c$ then the envelope makes only a small contribution to the cloud mass and the appropriate $y_c$ is only slightly less than $1/6$.

Conversely if the ratio $T_e/T_c$ is not small then the envelope can make a large, even dominant contribution to the mass. In that case it can be impossible to construct acceptable models, because even setting $y_c=0$ yields overall too much helium relative to hydrogen. In practice, then, our models do not extend down to central temperatures as low as $T_e$. We emphasise, though, that this difficulty is found specifically for the numerical solutions we have constructed, and that those solutions rest on the assumption of negligible condensed fraction in the envelope. If one were to relax that assumption then it might be possible to construct acceptable models with smaller, or even non-existent cores --- i.e. with phase equilibrium holding throughout the cloud.

\begin{figure*}
\figscaleone
\plottwo{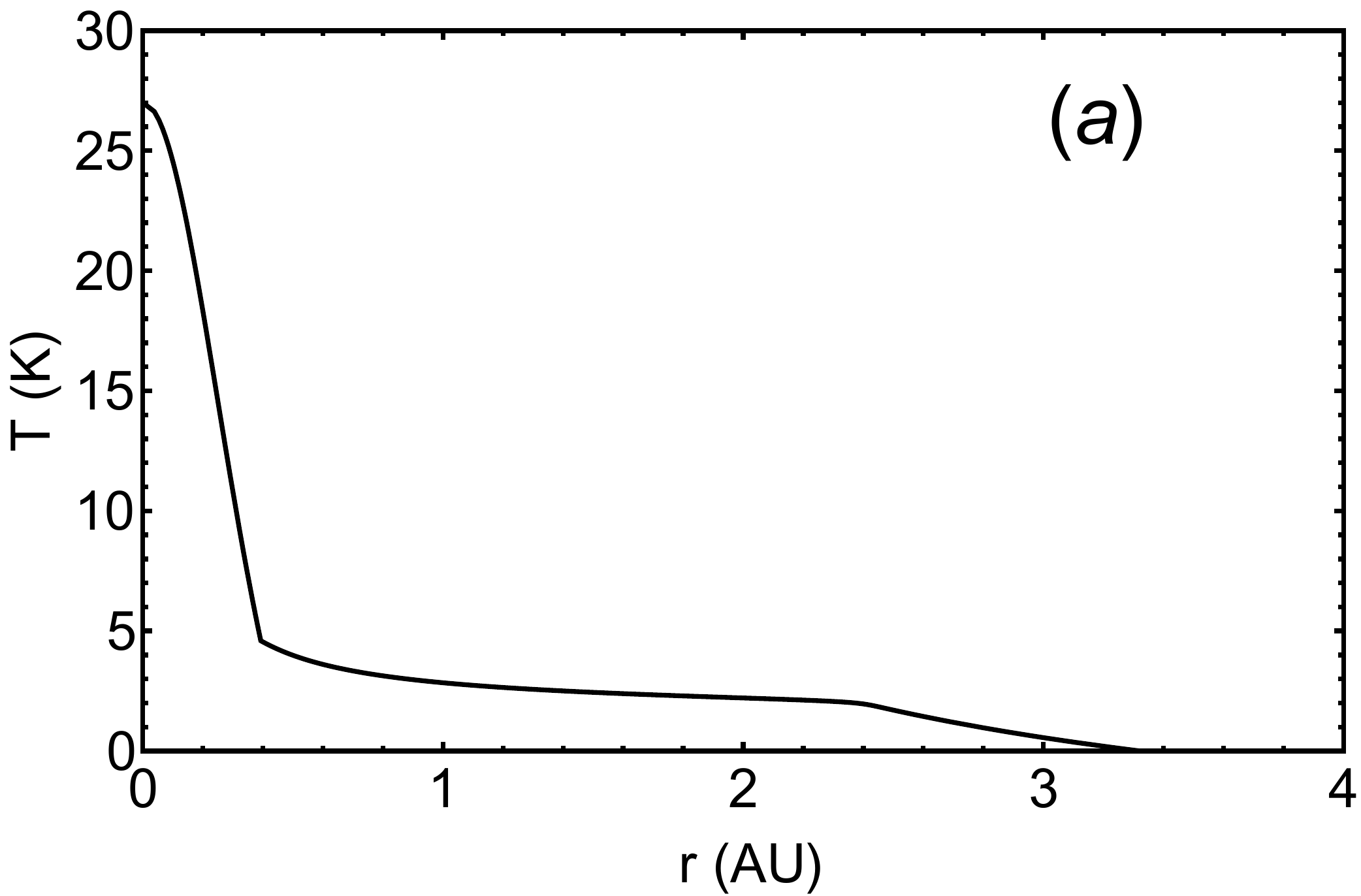}{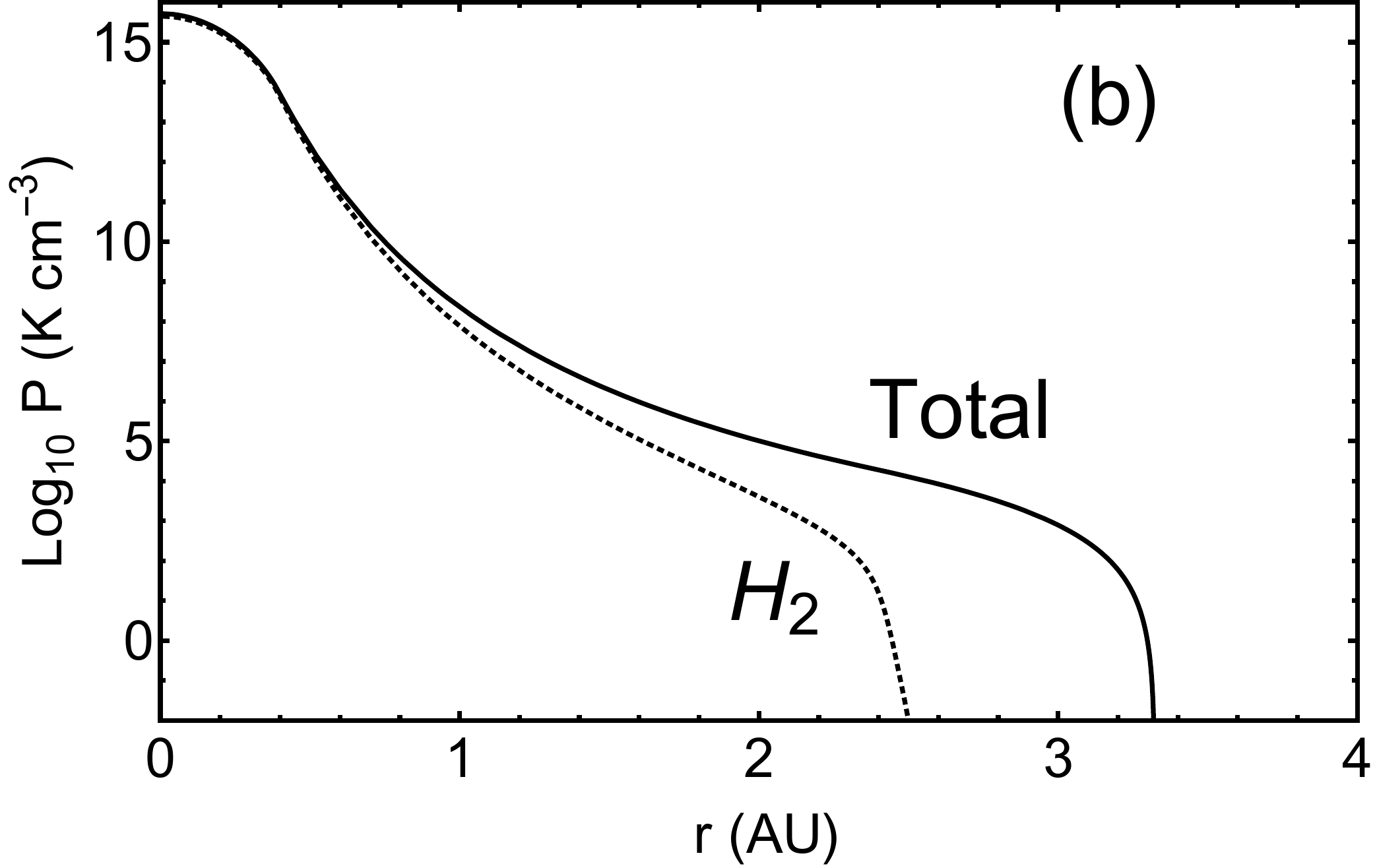}
\plottwo{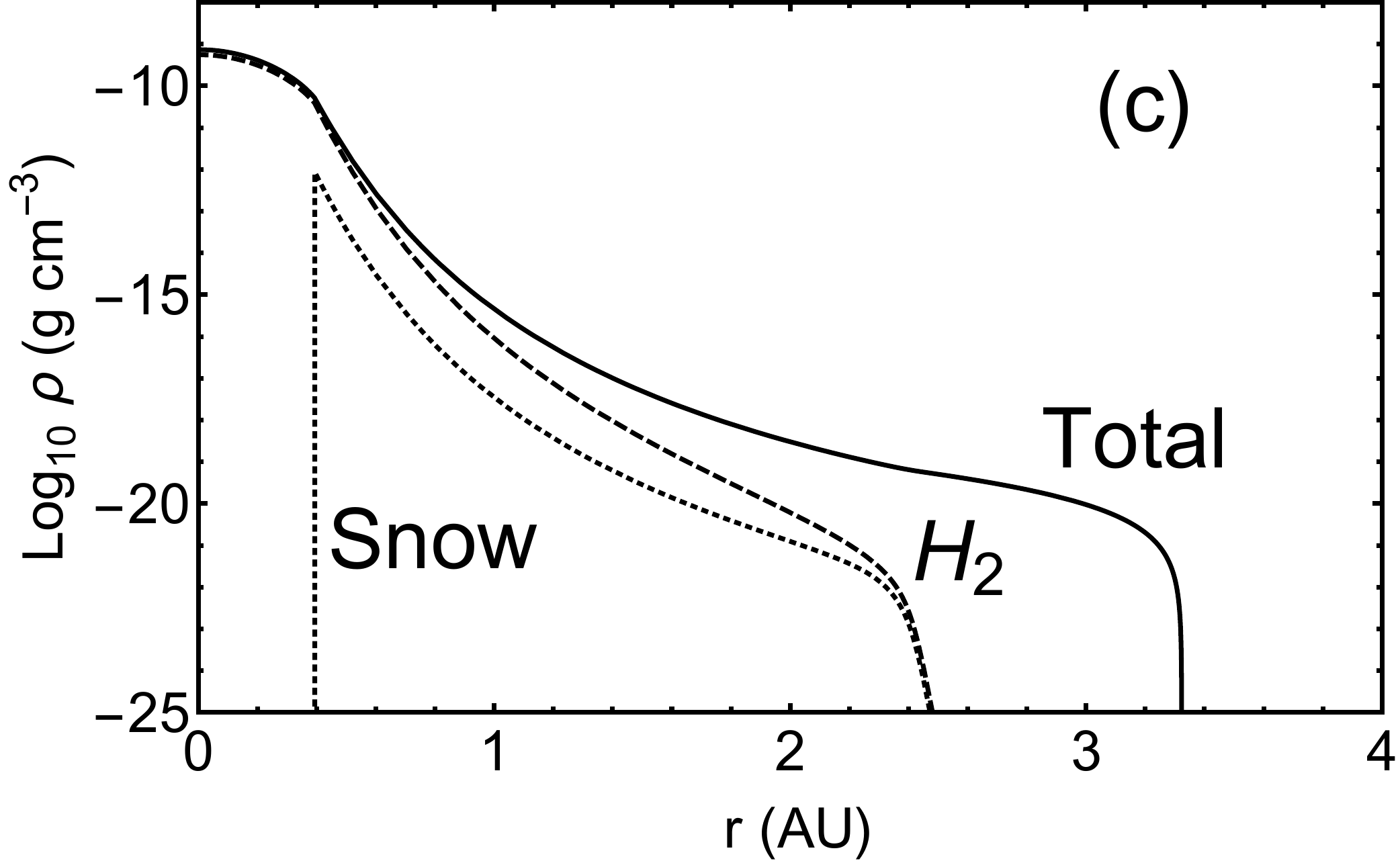}{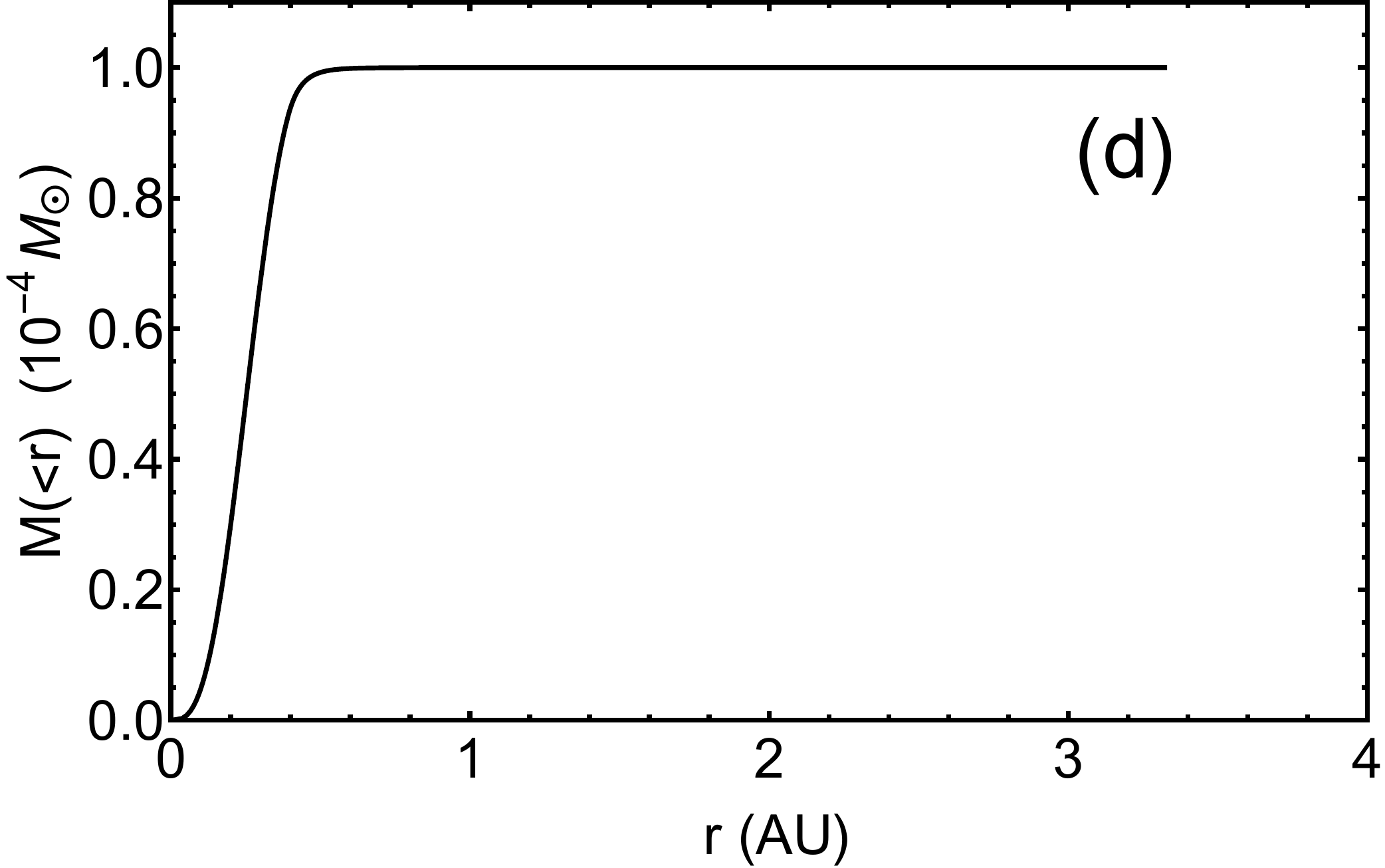}
\caption{An example of the internal structure of a model snow cloud. Here we show radial profiles for: (a) temperature;  (b) pressure; (c) density; and (d) enclosed mass, inside a cloud of mass $10^{-4}\,{\rm M_\odot}$ (25\% of which is helium), and radius approximately $3.3\,$AU.  This model was generated by the parameter combination $\{T_c=27\,{\rm K}, T_e\simeq4.5915\,{\rm K}, y_c\simeq0.16126\}$. Interior to $r\simeq0.4\,$AU the solution has the same form as that shown in figure 2. In panel (b), the solid line shows the total pressure, and the dashed line shows the partial pressure of \htwo. In panel (c), the solid line shows the total density, the dashed line shows the total density of \htwo, and the dotted line shows the density contributed by \htwo\ snowflakes.} 
\end{figure*}

\subsubsection{Amount of condensate}
Consider a single convection cell in which fluid is circulating. On the upward, expansion phase of the cycle, condensation occurs as the gas cools, and sublimation occurs on the downward, compressive phase of the cycle. The change in the number of gas-phase molecules, $N_m$,  can be determined from the adiabatic trajectory (\S2.6) to be
\be
\left( {{\partial \log N_m}\over{\partial\log P}}\!\right)_{\!\!S} =- \left( {{\partial \log y}\over{\partial\log P}} \right)_{\!\!S} = {{(1+y)(2\psi-5)}\over{2\psi^2+5y}}.
\ee
\break\noindent Starting from a condition in which no condensate is present, at the bottom of a convective cell, equation (17) tells us what fraction of the hydrogen has turned into snowflakes by the time the fluid parcel has risen one pressure scale-height, i.e. $1-\exp(\delta\log N_m)$, with
\be
\!\delta\log N_m =  \left( {{\partial \log N_m}\over{\partial\log P}}\!\right)_{\!\!S}   \delta\log P     \sim -\left( {{\partial \log N_m}\over{\partial\log P}}\!\right)_{\!\!S}\!.
\ee
\break\noindent
To calculate the total precipitate content we must average over all phases of the convective cycle, implying that the fraction of \htwo\ in the form of precipitates is $\{1-\exp(\delta\log N_m)\}/2$, with $\delta\log N_m$ as given by equations (17) and (18). The result for one particular cloud is shown in panel (c) of figure 3. There we can see that the contribution of snowflakes to the total density is everywhere very small ($\la1$\%) -- as we assumed in \S2.7 -- so our model is self-consistent in this respect.

\subsubsection{Solution for a $10^{-4}\,{\rm M_\odot}$ cloud}
The character of the structural solutions we obtain is illustrated by the example shown in figure 3, which is one possible structure for a cloud of mass $10^{-4}\,{\rm M_\odot}$. Other structures are possible for a cloud of this mass, as we will see in \S3.3. The low-density envelope, with its soft equation of state, occupies almost the entire volume of this cloud, but hardly contributes at all to the mass. Moving outwards through the envelope the helium fraction increases, as is evident from the decline of the partial pressure of \htwo\ relative to the total, and the surface of the cloud is effectively pure helium.

The kink seen in the temperature profile at $r\simeq0.4\,$AU in the top panel of figure 3 is expected: it is just the core-envelope boundary and it reflects the change in equation-of-state at that boundary. There is, however, another bend in the temperature profile at  $r\simeq2.4\,$AU which we have not yet explained. This bend is also, in effect, a change in the equation-of-state, but not a discontinuous one. It arises because the hydrogen fraction becomes so small that \htwo\ condensation no longer plays a dominant role in the thermodynamics. Consequently the equation of state rolls smoothly over into the usual adiabatic relation given in equation (8), as helium becomes increasingly dominant. The harder equation of state at the surface results in an abrupt edge to the cloud.

Although the core of our model cloud is precisely an $n=3/2$ polytrope, figure 3 is a graphic demonstration of the inadequacy of that polytropic solution for describing the cloud as a whole.

\begin{figure}
\figscaleone
\plotone{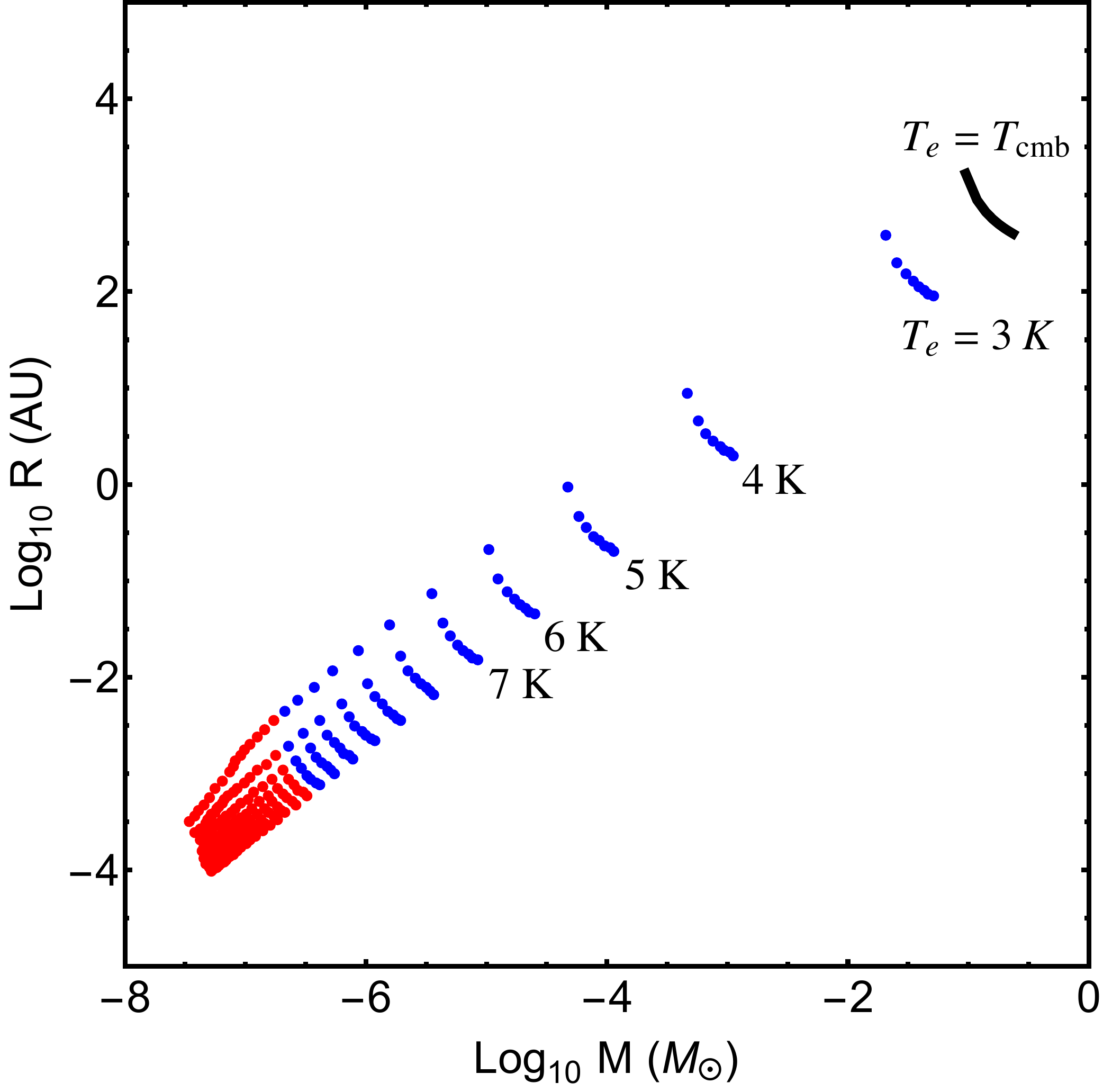}
\caption{Masses and radii for the hydrostatic equilibria described in \S3.3, having an average helium abundance $\langle y\rangle=1/6$. Each point represents a valid model on a grid of $\{T_e,T_c\}$ with $T_e({\rm K})=3,4,5,\dots31,32,33$, and $T_c({\rm K})=30,40,50,\dots90,100$. (But $T_e<T_c$.) Blue points are pure snow clouds, with $T_e<13.8\,{\rm K}$; red points correspond to $T_e\ge14\,{\rm K}$. For low envelope temperatures, the loci $T_e={\rm const.}$ are labelled. Also shown, with a solid line, is the case $T_e=T_{cmb}=2.73\,{\rm K}$. For each sequence of points with $T_e={\rm const.}$, $T_c$ increases as radius decreases. } 
\end{figure}

\subsection{Masses and radii of hydrostatic equilibria}
By constructing hydrostatic equilibrium models for each allowed parameter combination, in the manner described above, we obtain the loci of acceptable solutions in the mass-radius plane. The result is shown in figure 4 for a grid of models in which $T_e({\rm K})$ takes on integer values in the range $3\le T_e({\rm K})\le 33$, and $T_c$ varies from 100$\,$K down to 30$\,$K in steps of 10$\,$K. The lower end of the range of $T_c$ for this grid of models is only a few degrees above the limit at which we can still obtain solutions with the correct helium abundance (see \S3.3.1).

All the solutions shown in figure 4 fall in a narrow band of the mass-radius plane, with radii $R$ within a factor of a few either side of the relation $M({\rm M_\odot})=2\times10^{-4}R({\rm AU}) $. This band can be understood in the following terms. First, the range of $T_c$ covered by this grid of models is only a factor of $\simeq3$, and the $M\propto R$ relation just given reflects this narrow range in $T_c$. Secondly, although the range of $T_e$ which we explore is also modest (only a factor $\sim10$), that range corresponds to fifteen orders-of-magnitude variation in density at the core-envelope boundary, because $P_{sat}$ is a steep function of $T$ --- see equation 5, and figure 1. And the central pressure follows $P_{sat}(T_e)$ via equation (14). Consequently the large range of masses and radii seen in our models is primarily attributable to the range in $T_e$. Furthermore we can see that almost all of the spread in masses and radii is due to envelope temperatures below the triple point, i.e. the variety is almost all associated with the solid form of the condensate. Clouds which exhibit both rain and snow are exclusively found at the very low mass (and radius) end of the spectrum.

We note that in the limit $T_e/T_c\rightarrow0$ our solutions approach the usual $n=3/2$ polytropic models, as the envelope shrinks to occupy a miniscule fraction of the cloud radius. In this limit the masses and radii of our models are just those given in \S3.1, namely $R\simeq5.78 \,r_o$ and $M\simeq 10.7\,M_o$. And the scaling of mass with radius, as $T_c$ is varied at fixed $T_e$, is determined entirely by the variation of $M_o$ and $r_o$ with $T_c$. From the definitions of $M_o$ and $r_o$ given in \S2.1 it is straightforward to show that $r_o\propto T_c^{-1/4}$ and $M_o\propto T_c^{3/4}$, when $T_e$ is held constant. Thus the loci $T_e={\rm const.\/}$ of our model clouds obey $M\propto R^{-3}$, for large $T_c$, as can be verified from figure 4.

\begin{figure}
\figscaleone
\plotone{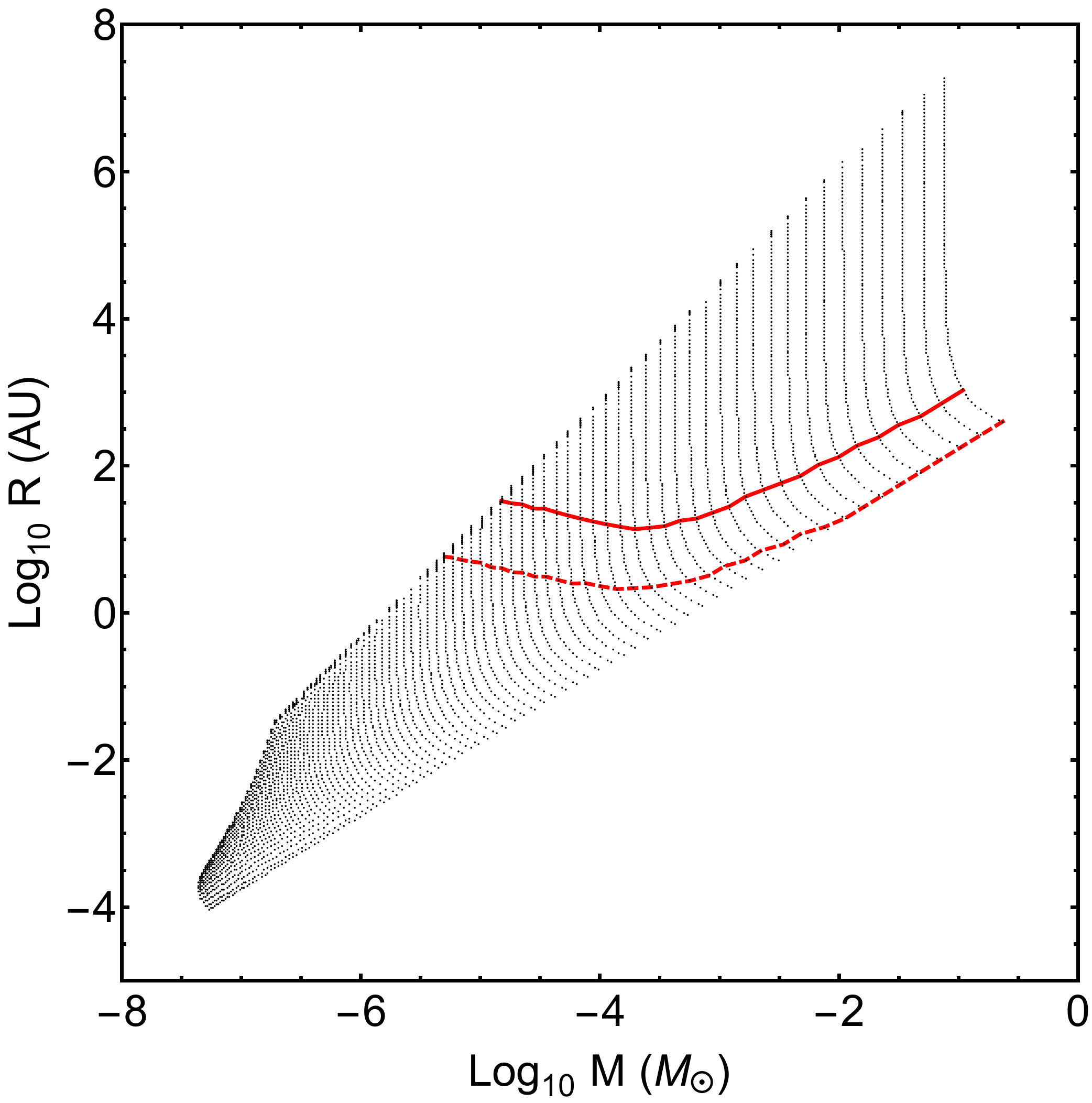}
\caption{Masses and radii for the hydrostatic equilibria described in \S3.3.1, having an average helium abundance $\langle y\rangle=1/6$. Each point represents a valid model on a grid of $\{T_e,T_c\}$ with $2.73\le T_e \le33\,{\rm K}$, and $T_c \le100$. Note that this plot extends to much larger radii than figure 4; the bigger clouds reflect valid solutions with central temperatures in the range $23\la T_c({\rm K})<30$. At any given mass, no solution with the correct $\langle y\rangle$ can be obtained for radii greater than the largest models shown here. At the dashed (solid) red line the structures would be crushed by approximately 10\% (50\%) in radius by the pressure of the diffuse ISM ($3{,}000\,{\rm K\,cm^{-3}}$).} 
\end{figure}

\subsubsection{Solutions with large radii}
The grid of models shown in figure 4 is bounded by four conditions: $33\,{\rm K}\ge T_e > T_{cmb}$, and $100\,{\rm K}\ge T_c \ge 30\,{\rm K}$. The last of these is somewhat arbitrary. Moreover, as the central temperature is lowered, the \htwo\ phase change plays an increasingly important role in the sense that the envelope becomes more extended. We have therefore explored to lower central temperatures, corresponding to larger cloud radii. 

To do so required some care because, at a fixed value of $T_e$, as $T_c$ is lowered the cloud radius becomes very sensitive to the precise value of the central temperature. To deal with this sensitivity we took the following approach. We first fixed $T_e$ and evaluated structures with large $T_c$. From those solutions we determined the numerical derivative of the cloud radius, $R$, with respect to $T_c$, and we used that derivative to estimate the value of $T_c$ that would increase $R$ by 10\%, relative to the current model. Proceeding in this way we were able to trace sequences of models out to very large radii, at fixed $T_e$. The results are presented in figure 5, where we can see additional solutions (i.e. not present in figure 4), at large radii, corresponding to $23\la T_c\,({\rm K})<30$.

Two caveats apply to the solutions with very large radii. First, the largest solutions at a given mass are very sensitive to $T_c$, with a 10\% change in radius being produced by temperature changes as small as $60\,{\rm \mu K}$ in the most extreme cases (i.e. where both $T_c$ and $T_e$ are near the lower end of their respective ranges). We therefore expect that these structures might look quite different -- indeed valid hydrostatic solutions might not even exist -- if we were to modify the slightest detail of our physical model. In other words: the largest models are unlikely to be robust to small changes in the physics, and should not be taken too seriously.  Secondly, over a large fraction of their radial extent, the largest structures exhibit pressures that are below the typical pressure of the diffuse interstellar medium. Such structures would therefore be crushed by the diffuse ISM, and those models cannot be representative of any real entity in the disk of our own Galaxy.\footnote{The largest models are more relevant to the intergalactic context, where the ambient pressure is orders of magnitude smaller.} The extent to which the ambient pressure affects the structure of the outer layers can be gauged from the profiles presented in Appendix D.

\subsection{Central helium abundances}
Readers might be curious about the central helium abundances that are required to obtain solutions with the correct mean abundance. Figure 6 shows contours of constant $y_c$ in the mass-radius plane. As expected, for very large values of $T_c/T_e$, where the core constitutes almost the whole cloud, we see that $y_c\simeq\langle y\rangle$. But the envelope is more helium-rich than the core, so every model has $y_c<\langle y\rangle=1/6$, and $y_c$ decreases as cloud radius increases at fixed cloud mass, reflecting the larger contribution of the envelope to the total cloud mass. 

\begin{figure}
\figscaleone
\plotone{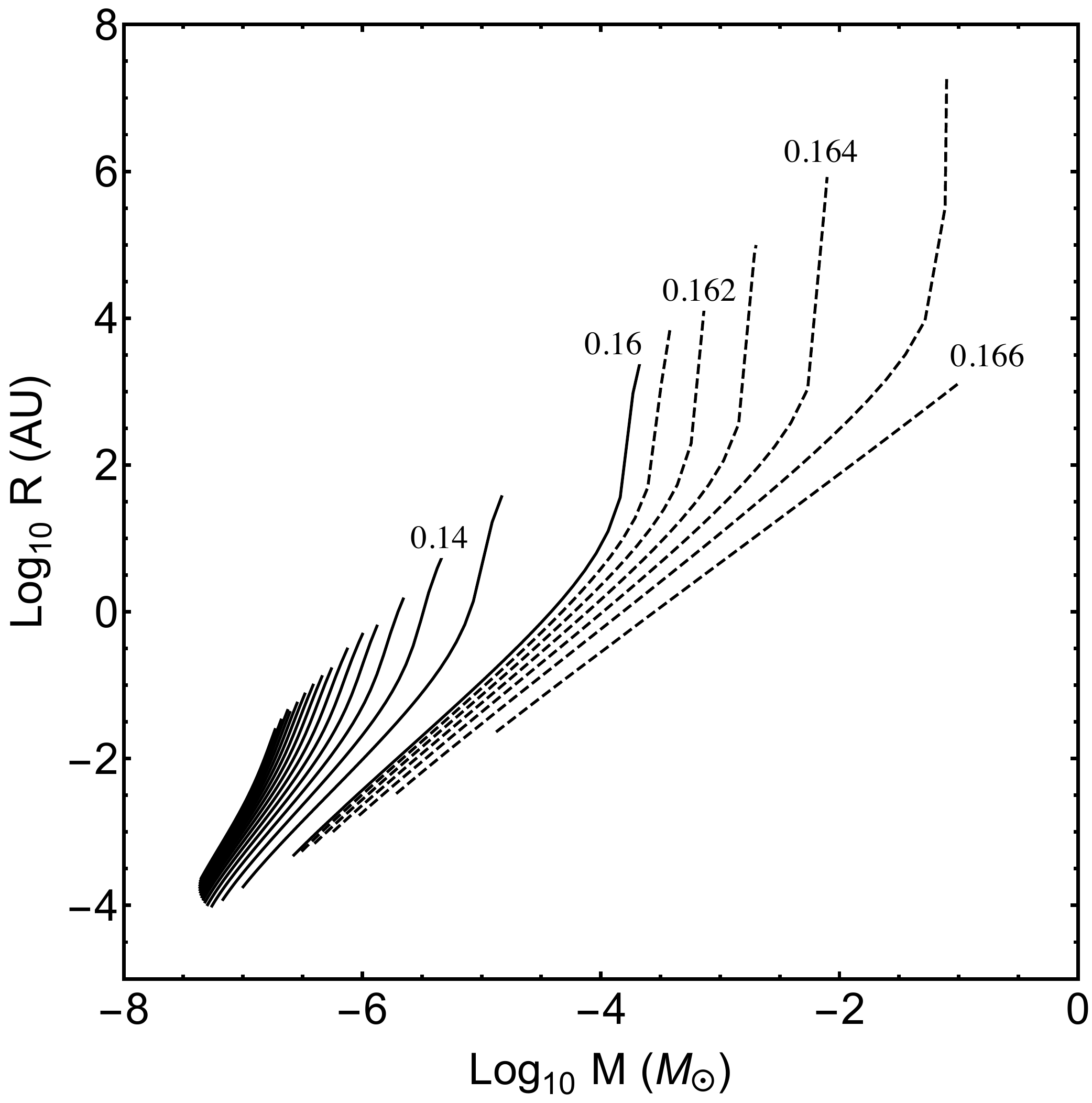}
\caption{Contours of constant central helium abundance, $y_c$, for the grid of models shown in figure 5. For $y_c\le0.16$ contours are plotted at intervals of $0.01$ in $y_c$, and are shown with solid lines. For $y_c>0.16$ contours are plotted at intervals of $0.001$, and are shown with dashed lines.} 
\end{figure}

A feature of figure 6 that is initially surprising is that $y_c$ typically does not extend down to very low values at the largest cloud radii, for any given mass. The explanation for this lies in the fact that $y_c$ influences $\langle y\rangle$ both directly, through the core helium abundance, and indirectly via the properties of the envelope. The direct influence dominates for large $T_c/T_e$. But for lower central temperatures, where the envelope becomes very large compared to the core, the indirect influence also plays an important role. What happens as the base of the envelope becomes more hydrogen rich (i.e. $y_c$ decreases) is that the thermodynamics of the \htwo\ phase change become more important in determining the adiabatic trajectory, leading to an increase in the envelope mass as $y_c$ decreases (at fixed $T_c,T_e$). This provides a countervailing trend which tends to increase $\langle y\rangle$ as $y_c$ decreases. Consequently, for low values of $T_e$ there are no solutions with $\langle y\rangle=1/6$ for $y_c\ll1/6$.

The effect just described becomes less important for large values of $T_e$. The reason is that the entropy of the saturated vapour is smaller at higher temperatures (equation 9), so that the \htwo\ phase change has a smaller influence on the adiabat (equation 10). In turn this permits valid solutions right down to $y_c=0$ for very low mass clouds.

\section{{\response Thermal properties of the hydrostatic models}}
{\response The solutions presented in \S3 are dynamical equilibria and therefore they do not evolve on the dynamical (sound-crossing) timescale. However, the equations that we solved to obtain those structures do not include an energy equation, so the hydrostatic equilibria are not necessarily thermal equilibria and are therefore only quasi-static. Specifically: if we consider a sequence of hydrostatic models of fixed mass we find that they differ slightly in binding energy -- becoming more tightly bound as the central temperature increases -- and therefore a given cloud will slowly evolve along that sequence if there is an imbalance between the rates of heating and radiative cooling. If heating exceeds cooling then the progression will be an expansion, as the excess energy goes into work done against gravity.}

{\response The primary issue that is addressed in this section is the global thermal balance of our models, and here we narrow our focus to the case of clouds located in the Galaxy, for which cosmic-rays are expected to dominate the heat input. We therefore assess the radiative output (\S4.2), and the heat input (\S4.3), and then we identify thermal equilibrium models by requiring Heating$\,=\,$Cooling (\S4.4).}

{\response Of course a time-independent model must be locally in equilibrium, as well as globally, and in \S4.1 we consider the flow of heat internal to each cloud.}

\subsection{Internal heat flow}
As mentioned in the introduction, convective heat flow within the models we have constructed is highly unusual, in that heat flows inwards throughout the envelope of each cloud. If it is not already obvious that this behavior is unusual, one need only consider that the temperature of the fluid increases inwards, and remember that the second law of thermodynamics forbids natural heat flow from a cooler body to a hotter body. The resolution of this apparent paradox is explained carefully in Appendix A; here we summarize the key points. 

The first point is that the fluid is moving, radially, as a result of a buoyancy instability, and it is these fluid motions which provide the main channel for energy transport throughout the cloud. A fluid element undergoes compression (expansion), as it moves to regions of greater (lesser) pressure, and the work done on (by) the moving fluid parcel results in a change in temperature prior to the exchange of any heat with the background fluid. So to establish the direction of heat flow -- into or out of a displaced fluid parcel -- we must determine the temperature of the fluid element {\it after\/} it has been displaced, and compare that to the temperature of the background fluid at the new location of the displaced parcel. For small radial displacements this is equivalent to comparing the temperature gradient of the background fluid with that of an adiabatic trajectory. Slow, adiabatic displacements conserve entropy, so this comparison amounts to a determination of the sign of the radial gradient of the specific entropy: heat will flow from the higher entropy regions to the lower entropy regions.

\begin{figure}
\figscaleone
\plotone{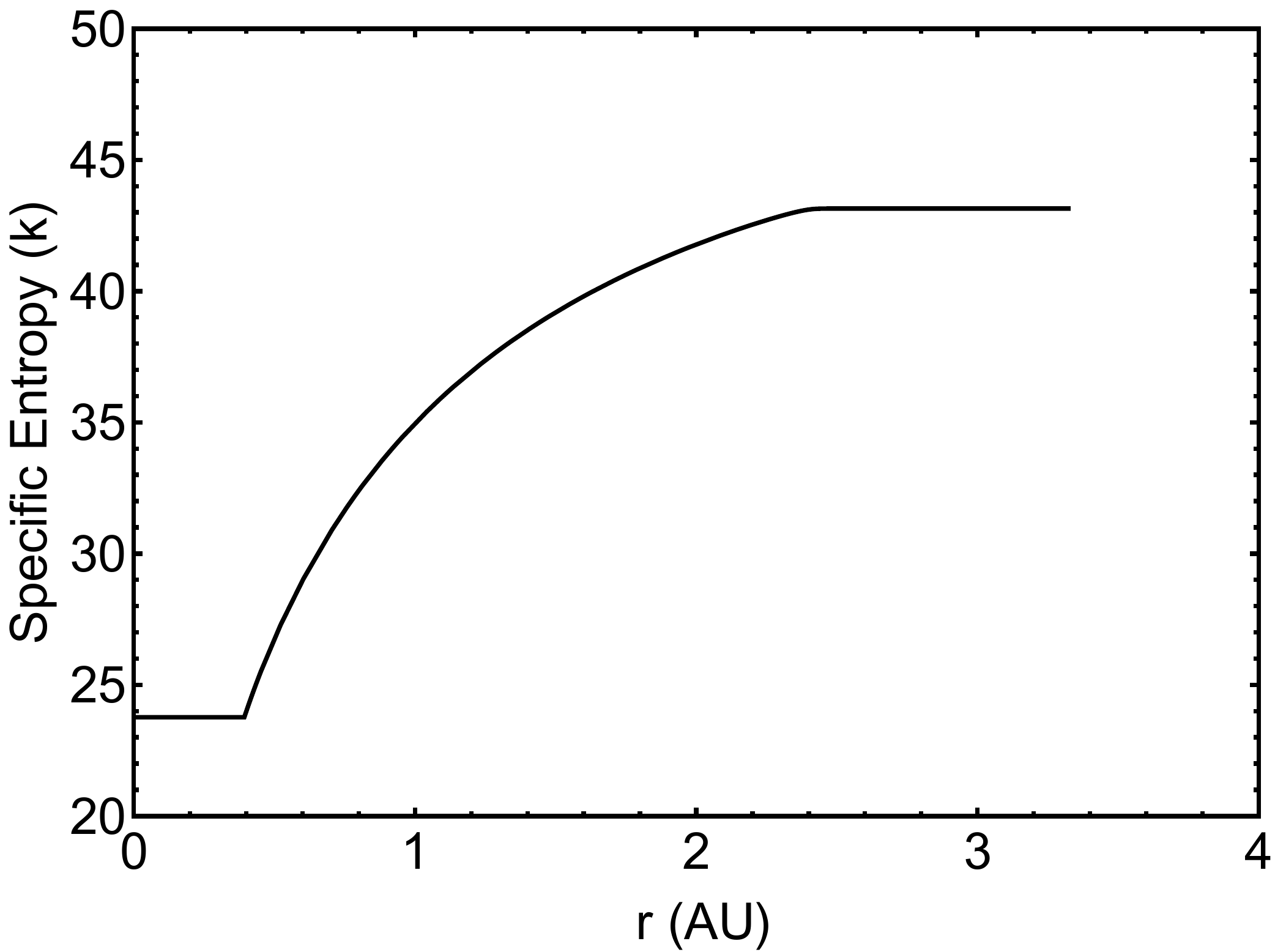}
\caption{The Sackur-Tetrode entropy-per-particle of the fluid, as a function of radius, for the same cloud model as shown in figure 3.  The outward increase in entropy, through the envelope of the cloud ($r>0.4\,{\rm AU}$), means that envelope convection transports heat inwards, up the macroscopic temperature gradient, as discussed in \S4.1} 
\end{figure}

Figure 7 shows the radial profile of the specific entropy for the same snow cloud model shown in figure 3. We see that entropy is constant in the cloud core. That is as expected, because the equation of state utilised in the core -- i.e. equation 8 -- corresponds to an isentrope. If there were no entropy gradient then convection would result in no heat flow. However, in reality the core can be only approximately isentropic: as shown in Appendix A, if a fluid has uniform composition then buoyancy instability is only present for configurations where the entropy increases with pressure. In turn that means that convective motions in the cloud core are associated with outward transport of heat.

In the envelope, however, we see that the entropy increases outwards, and consequently any fluid circulation must convect heat inwards. This remarkable result is entirely due to the changing composition of the fluid with radius, which permits buoyancy instability to exist despite the stabilising effect of the inverted entropy gradient. We remind readers that the hydrostatic equilibrium shown in figure 3 -- and indeed each of our hydrostatic models -- is, by construction, precisely neutrally buoyant in respect of adiabatic displacements at any location.

In the outermost regions of the envelope the entropy again appears to level off to a constant value. That is because the fluid is almost pure helium in that region. The entropy there is, in fact, not quite constant, but increases slightly with radius as there is a small amount of \htwo, and the \htwo\ fraction decreases outwards. Thus convection leads to inward heat flow throughout the envelope of the cloud. That conclusion is important because the outer regions of the cloud are colder than the CMB and must thus experience net radiative heating. Inward convection of heat nevertheless permits these regions to exist in steady state.

A key point to note is that heat can be convected inwards only as far as the boundary of the core, so the base of the envelope must be able to radiate away the heat (from the CMB and other sources) that is deposited throughout the cold, outer layers. In turn that means that the base of the envelope must be warmer than the CMB --- a condition which we have already imposed on our solutions in \S3.

The foregoing considerations tell us the direction of the convective heat flow -- outward in the core, and inward in the envelope -- but not its magnitude. Convection can be a very efficient means of transporting heat; however, in the limit of vanishing turnover speed the heat flux also vanishes. Thus the strength of the convection can adjust itself so as to bring about quasi-steady conditions, and we assume that it does so. {\response With this assumption it is not necessary to explicitly solve for local thermal balance.}

Radiative heat exchange between different elements within the cloud is insignificant by comparison with convection. But radiation is important because it is the only means by which the cloud as a whole can cool. We now evaluate the radiative losses.

\subsection{Radiative cooling}
We identify two sources of thermal radiation that we expect to be important in the present context: continuum emission from \htwo\ snowflakes, and $S_0(0)$ ($J=2\rightarrow0,\;28\,{\rm\mu m}$) pure rotational line emission from gas-phase \htwo. Radiative losses from those two processes are evaluated in this section.

A third source of thermal radiation which is potentially important, but which we do not include, is due to neighbouring \oh\ pairs in solid \htwo\ \citep{hardy1977,harris1977,silvera1980}. Because of the non-zero electric quadrupole moment of the \htwo\ molecule, there is a quadrupole-quadrupole interaction energy that depends on the relative orientation of the angular momenta of the two molecules and their separation. Consequently \oh\ pairs at substitutional sites in a \ph\ lattice can exist in various, discrete quantum states, and transitions between states give rise to microwave line emission. These lines are, however, very weak transitions, and furthermore the radiated power scales as the square of the \oh\ fraction, in the case of low \oh\ content. Therefore, consistent with our approximation that the clouds are made of pure \ph\ (\S2.2), we neglect the radiation from {\it ortho-}pair transitions in solid \htwo.

{\response It is worth noting that the radiation field from all processes combined is very weak indeed; consequently it has no significant influence on the temperature and density structure of the cloud.  That structure is fully determined by the hydrostatics (\S2,3), and it is thus a straightforward task to calculate the  intensity of the radiation field that arises in  each model.}

\begin{figure}
\figscaleone
\plotone{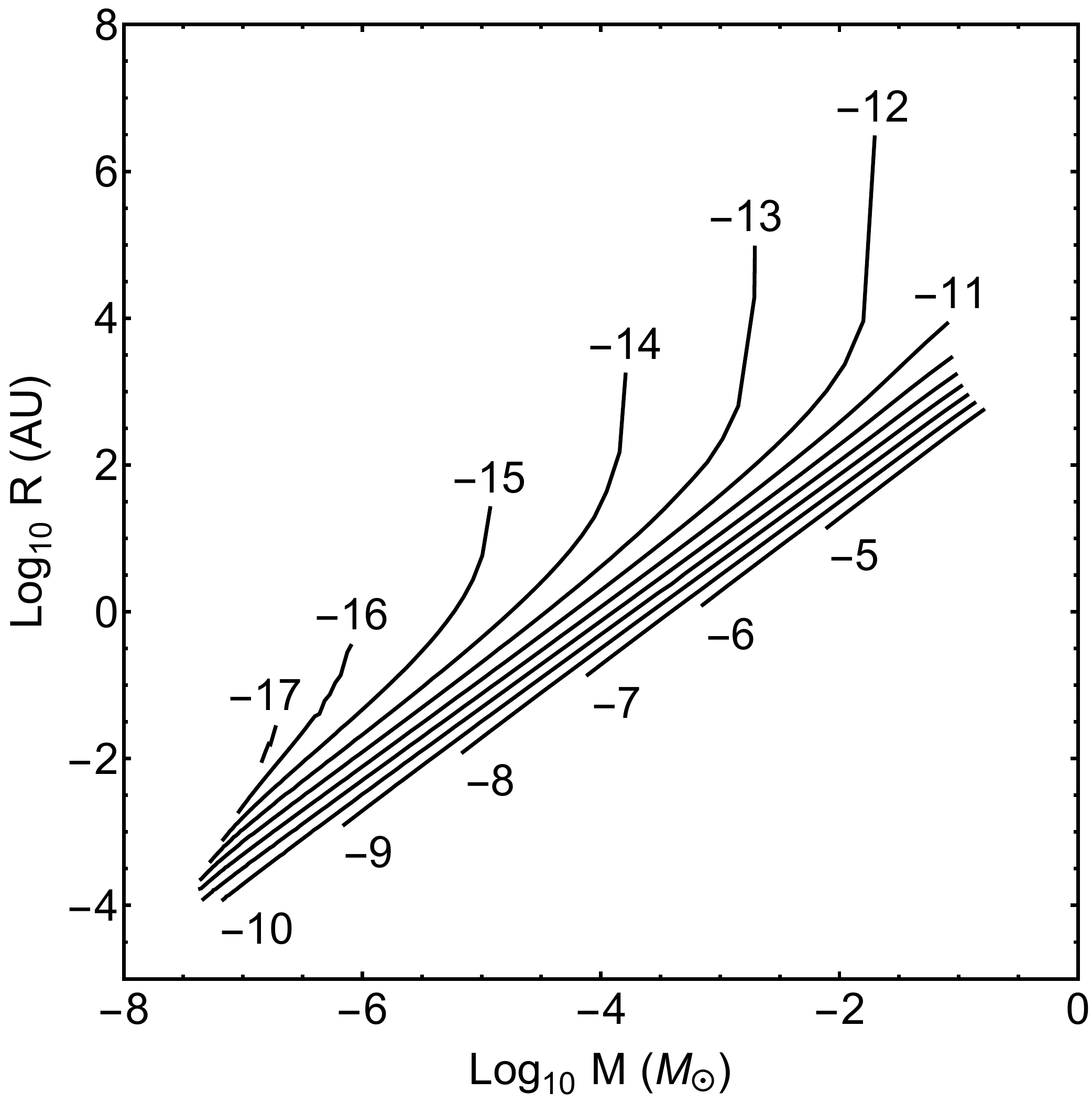}
\plotone{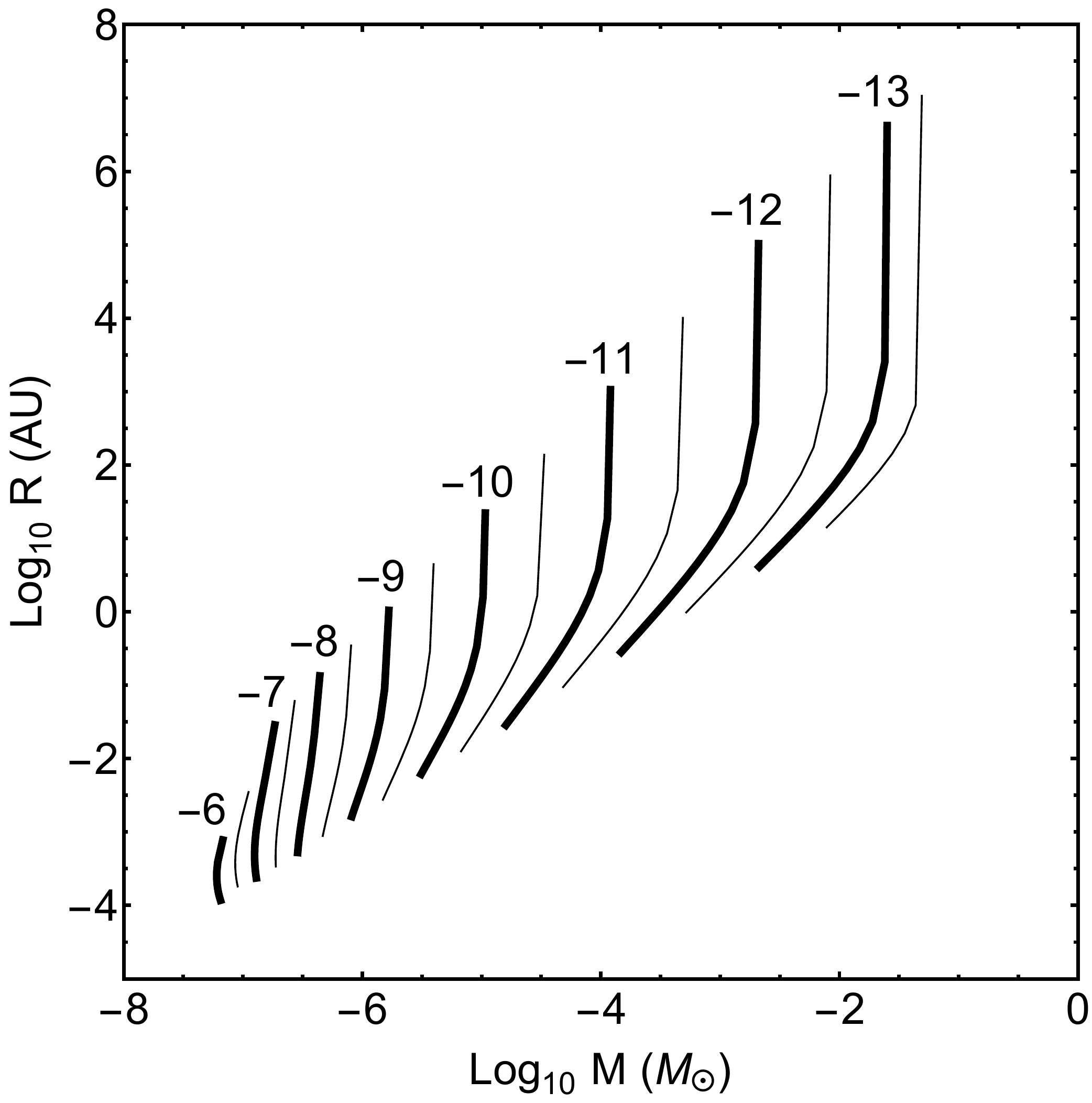}
\caption{Radiative cooling due to the $S_0(0)\;J=2\rightarrow0$, $\lambda=28\,\mu m$ pure rotational line of \htwo\ (upper panel); and the continuum emission from \htwo\ snowflakes (lower panel). In both cases the contours show $\log_{10}\Lambda$, where $\Lambda$ is the specific luminosity in units of ${\rm erg\,g^{-1}\,s^{-1}}$, net of the power that is absorbed from the CMB  (\S4.3.1).} 
\end{figure}

\ \vskip0.1cm

\subsubsection{Pure rotational transitions}
The main source of photons from pure rotational transitions of \htwo\ is the core of the cloud, where temperatures are highest. But even in the core, and at the highest temperatures we consider ($100\,$K), only the fundamental rotational transition $S_0(0)$ contributes significantly. The high density and low temperature of the gas result in a high optical depth at line centre for this transition, and an accurate formulation of the radiation transport is needed. For the hydrostatic equilibria constructed in \S3, we determine the emitted intensity along any given direction, and at any given frequency, using the usual formulae for thermal radiation transport \citep[e.g.][]{rybicki1979}.  The power radiated in the $S_0(0)$ line follows immediately by integration over frequency and angle. Results are shown in figure 8 where, as expected, we see a steep increase in radiated power as the central temperature increases.

\begin{figure}
\figscaleone
\plotone{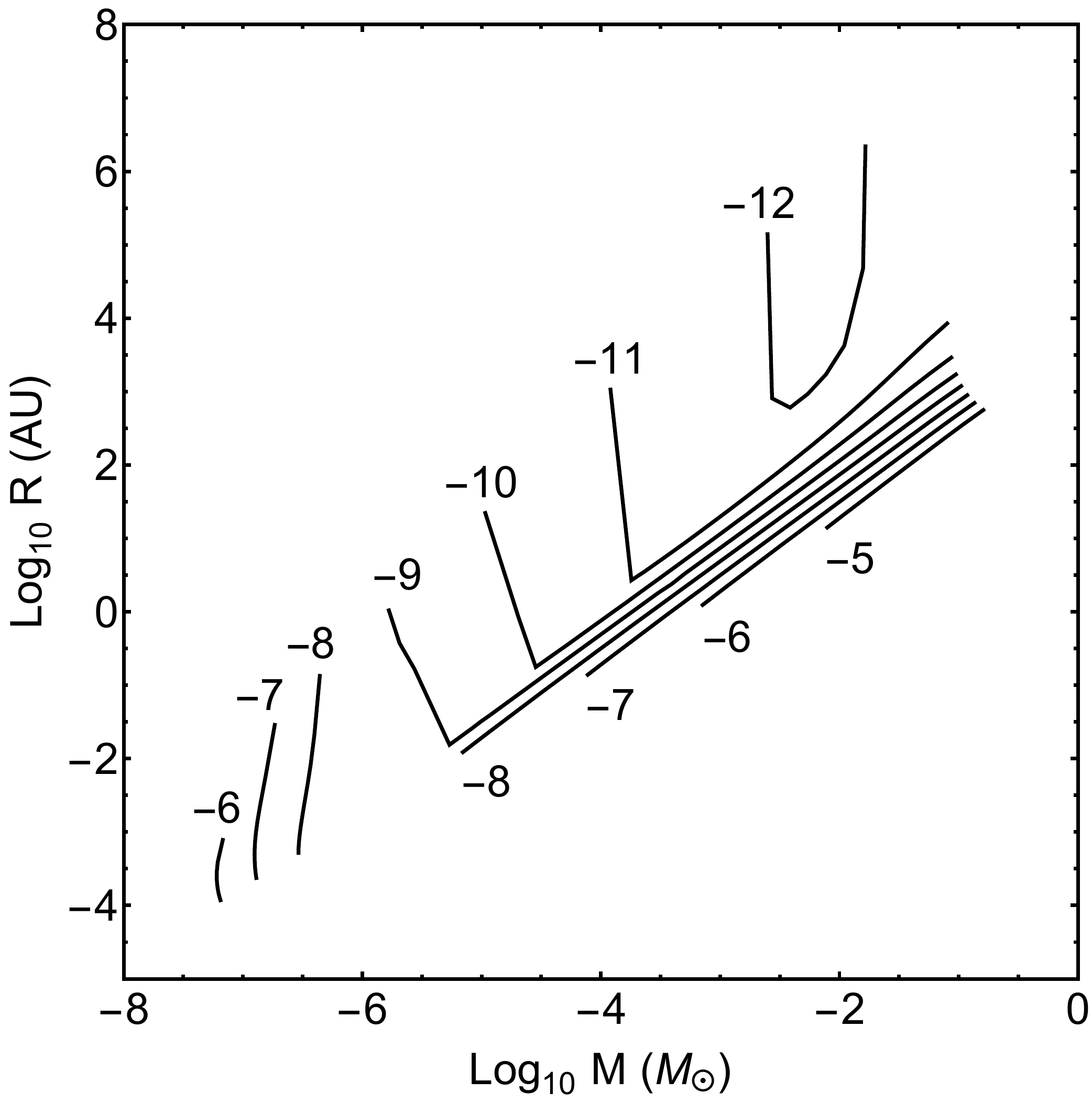}
\caption{Total radiated power, $\log_{10}\Lambda\,({\rm erg\,g^{-1}\,s^{-1}})$, for line and continuum combined. Because the binding energy per unit mass is similar for all our models, adding $7.3$ to the contour values yields a rough estimate for $\log_{10}$ of the Kelvin-Helmholtz rate in ${\rm Gyr^{-1}}$.} 
\end{figure}

\subsubsection{Snowflake continuum}
As with any dust particles, we expect thermal continuum radiation to arise from the \htwo\ snowflakes which are present in the envelope of the cloud. However, in contrast with the silicates and graphitic materials that are usually hypothesised to make up astrophysical dust \citep[e.g.][]{draine2003}, pure, solid \ph\ absorbs very weakly at low frequencies and is therefore a very poor emitter. To model the emissivity of the snowflakes we treat them as if they were small, dielectric spheroids. In this approximation the size and shape of the individual snowflakes is of no consequence, and the total radiated power depends only on the total mass in snow, its temperature, and the imaginary part of the low-frequency dielectric constant --- as described by \citet{draine1984}, whose treatment we follow. In the case of pure \ph\ at long wavelengths ($\lambda\gg28\,{\rm \mu m}$) the imaginary part of the dielectric constant  is approximately $10^{-11}/\lambda\,{\rm(cm)}$ \citep{kettwich2015}. In keeping with our approach to the hydrostatic modelling, where solid and liquid condensates are treated on a common footing, we evaluate thermal radiation from liquid \htwo\ as if it were from the same mass of solid \htwo. 

Because pure \ph\ snowflakes are only weakly absorbing, the optical depth of our model clouds to the thermal snowflake continuum is very small, and radiation transport is therefore trivial. Figure 8 shows the total power radiated by snowflakes. Unlike the contribution from rotational transitions, we see that the snowflake power decreases as the central temperature increases for all models of mass $\ga2\times10^{-7}\,\msun$.

\subsubsection{Total cooling rate}
The total cooling, which is the sum of the radiation from the two processes described above, is shown in figure 9 as the power per unit mass, $\Lambda$. As the ordinate in the figure is the mass of the cloud, it is straightforward to determine the luminosity for any model of interest. For example: the luminosity of the model shown in figure 3 can be seen to be $\sim2\times10^{29}\times3\times10^{-11}=6\times10^{18}\,{\rm erg\,s^{-1}}$. This is a very low luminosity in comparison with the Sun, for example. Indeed all of our snow cloud models have very low luminosities in comparison with main sequence stars, as can be seen by noting that (i) $\Lambda\ll {\rm L_\sun/M_\sun}\simeq2\;{\rm erg\,g^{-1}\,s^{-1}}$, and (ii) $M\ll {\rm M_\sun}$. Thus it is clear that snow clouds are intrinsically very dark -- they are a type of baryonic dark matter -- as was anticipated by \citet{pfenniger1}.

An interesting aspect of the cooling is the power per unit binding energy for the structure, which tells us the rate at which contraction would occur in the absence of any heating --- in other words the Kelvin-Helmholtz rate, ${\cal R}_{KH}$. Now the specific binding energies, ${\cal B}$,  of our model snow clouds are all quite similar, because there is little variation in their central temperatures, with $\log_{10}{\cal B}{\rm(erg\,g^{-1})}\simeq9.2\pm0.3$. Thus figure 9 is also, in effect, a contour plot of the Kelvin-Helmholtz contraction rate. Specifically, if we add $7.3$ to each of the contour values shown in figure 9 we would have an approximate contour plot for $\log_{10}{\cal R}_{KH}{\rm(Gyr^{-1})}$. Thus we see that for cloud masses $\ga10^{-6}\,\msun$ all but the most compact of our models have Kelvin-Helmholtz timescales that are comparable to or greater than the age of the Universe.

Long cooling timescales have several implications for the models we have constructed. First, thermal equilibrium may not actually be reached within the age of the Universe, even if it is possible in principle. Secondly, some pathways to thermal equilibrium may be excluded. For example: a simple collapse from large radii is strongly disfavoured for the high mass clouds shown in figure 9, because it is impossible to radiate away the gravitational binding energy within the age of the Universe. If thermal equilibria do exist, though, they could perhaps be reached by a brief period during which heat is injected into clouds that are initially more compact than those equilibria. Third, the stability of any thermal equilibria against perturbations is less of a concern if the underlying Kelvin-Helmholtz rate is very low, as instabilities should not progress very far within the age of the Universe. 

\begin{figure}
\figscaleone
\plotone{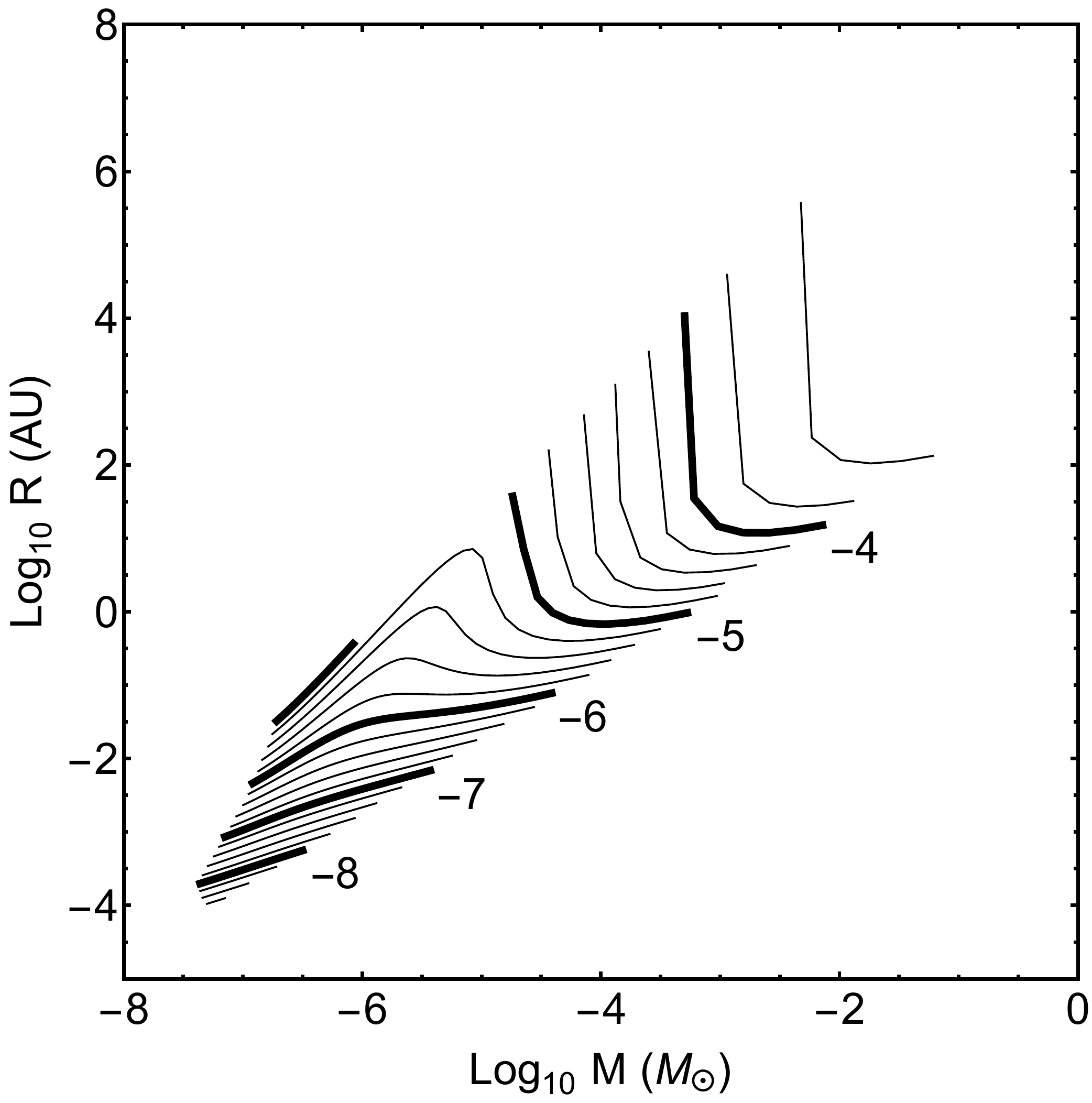}
\caption{Cosmic ray heating rate, per unit mass, $\Gamma_{cr}$, appropriate to the Galactic cosmic-ray disk, local to the Sun, for each of our hydrostatic snow cloud models. Bold contours show integer values of $\log_{10}\Gamma_{cr}\,({\rm erg\;g^{-1}\,s^{-1}})$, as marked; thin contours show intermediate values at increments of 0.2.} 
\end{figure}

\subsection{Heat sources}
Various external agents may supply heat to the clouds, with the largest contributions likely to come from: far-UV starlight, absorbed\footnote{The energy density of starlight in the optical is much higher than in the far-UV, but solid \ph\ is practically transparent in the optical \citep{kettwich2015}.} by \htwo\ snowflakes; cosmic rays; and the cosmic microwave background (CMB).  Of these three, the first is very sensitive to the location of the cloud, becoming very large at small distances from massive stars (a point that we return to in \S5.4). However, at a typical interstellar location in the solar neighbourhood, the far-UV energy density is only $\sim1$\% of that in cosmic-rays \citep[e.g.][]{draine2011}. As both species are relativistic, and both transfer their energy effectively to   molecular hydrogen, {\response this means that heat is typically supplied predominantly by cosmic-rays for clouds in the solar neighbourhood. At other locations in the Galaxy -- e.g. near star-forming regions -- heating by starlight assumes a greater importance and may become dominant. It would be interesting to explore that regime, but it is beyond the scope of this paper to do so; here we concentrate on the circumstance where heating is dominated by cosmic rays and the CMB.}

\subsubsection{Cosmic microwave background}
The photons of the Cosmic Microwave Background are a universal source of heat for snow clouds. As the CMB temperature evolves over cosmic time, so must the heat input to any cloud, and the boundary $T_e\ge T_{cmb}$ gradually relaxes.  That, however, is beyond the scope of this paper: we restrict attention to the present epoch.

Absorption of photons proceeds by the inverse of each of the two emision processes discussed in \S4.2, and it is therefore natural to include it as a negative contribution in the calculation of the radiated power. Indeed the CMB has already been accounted for in exactly that way in the results in figure 8, where the quantity plotted is the net power emitted --- i.e. emitted power minus power absorbed from the CMB.

\subsubsection{Cosmic-ray heating}
The column-density measured to the cloud centre varies from $\sim1\,{\rm g\,cm^{-2}}$ in the upper-right of figure 5, to $\sim10^8\,{\rm g\,cm^{-2}}$ in the lower-left. These columns are large enough that severe attenuation of the cosmic-ray flux is expected as one moves inward, and the specific heating rate due to cosmic-ray interactions is much less than the value appropriate to diffuse interstellar gas \citep[e.g.][]{webber,cravens}. Our calculation therefore proceeds by determining the cosmic-ray heating rate as a function of column-density, for a beam of particles, and then integrating over all incident directions. The details of this calculation are somewhat removed from the core topics of this paper, and are therefore given in Appendix C. The resulting cosmic-ray heating rate, $\Gamma_{cr}$, for clouds located in the solar neighbourhood, is shown in figure 10. 

\begin{figure}
\figscaleone
\plotone{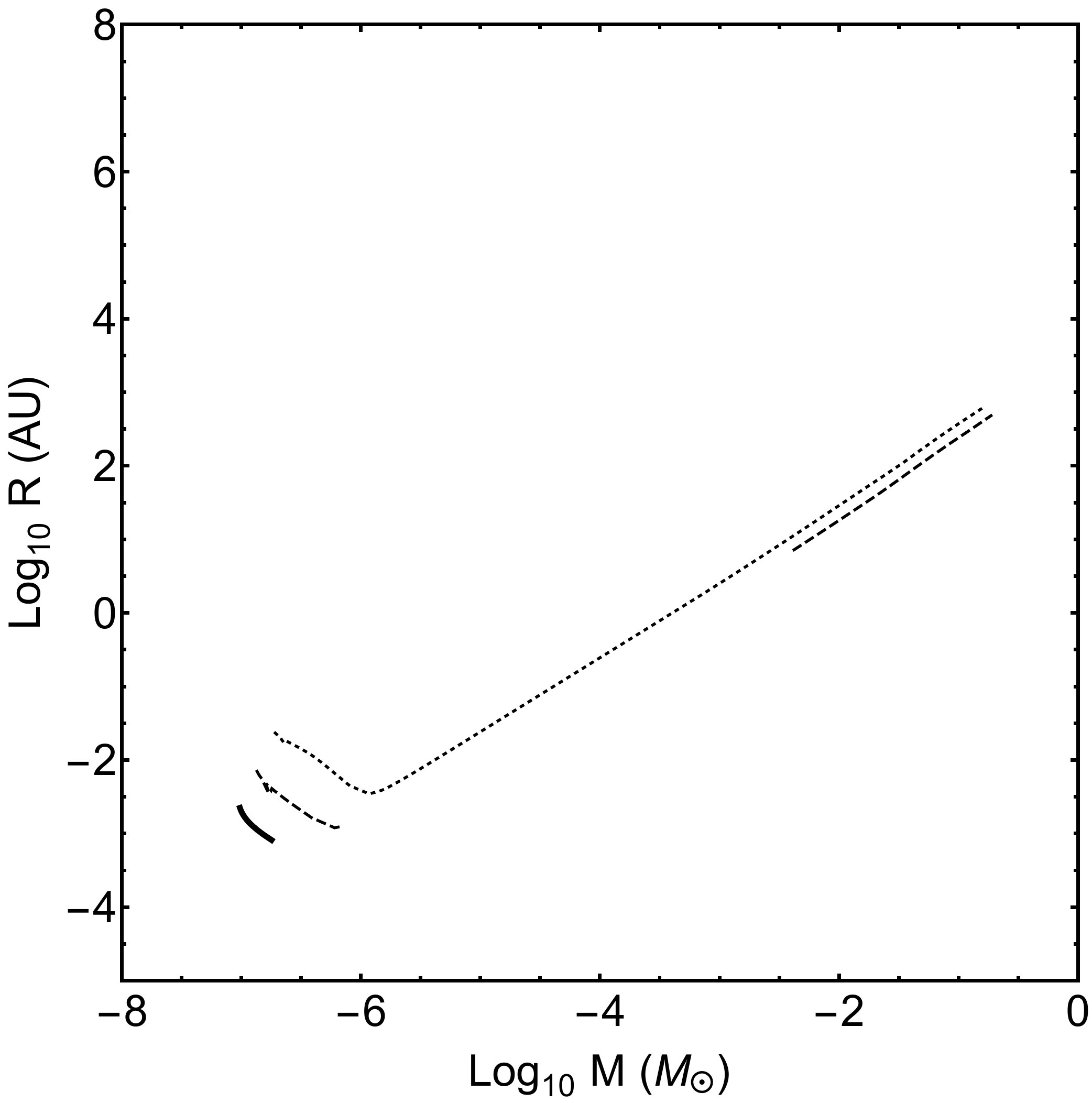}
\caption{Loci of thermal equilibria, where cosmic-ray heating balances radiative cooling, for three circumstances: (i) clouds located in the cosmic-ray disk (solid line), whose heating rate is $\Gamma_{cr}$ (as shown in figure 10); (ii) clouds heated at $0.1\times\Gamma_{cr}$ (dashed line); and, (iii) clouds heated at $0.01\times\Gamma_{cr}$ (dotted line).} 
\end{figure}

\subsection{Thermal equilibrium}
Comparing figure 10 to figure 9 we can immediately see that cosmic-ray heating, as experienced by a cloud within the cosmic-ray disk, far exceeds the cooling rate for almost all of our hydrostatic equilibria. The two rates match only for a small subset of cloud models at the low-mass end of the spectrum ($M\la2\times10^{-7}\,{\rm\msun}$), and the resulting locus of thermal equilibria is plotted as the solid line in figure 11.

In the vicinity of these equilibria, the orientation of the contours of $\Lambda$ (figure 9), and of $\Gamma_{cr}$ (figure 10), demonstrate that for models of a given mass the cooling rate is insensitive to cloud radius, whereas the heating rate is a strong function of cloud radius. And we note that the heating rate increases with increasing cloud radius (at fixed mass). It follows that these equilibria are thermally unstable: a perturbation which causes the cloud to expand (contract) slightly will lead to heating (cooling) outstripping cooling (heating), and thus the cloud will expand (contract) further.

Real clouds following orbits within our Galaxy would experience heating rates that vary with time, as they pass through different cosmic-ray environments. In particular the heating rate may achieve very low values if the cloud travels far from the cosmic-ray disk. The detailed consequences of a time-dependent heating rate are unclear, but at a simple-minded level we can consider the effect of lowering the average heating rate by rescaling $\Gamma_{cr}$. Therefore, in addition to the locus for $\Gamma=\Gamma_{cr}$, figure 11 also shows the loci of thermal equilibria corresponding to heating rates of $\Gamma=0.1\times\Gamma_{cr}$ and $\Gamma=0.01\times\Gamma_{cr}$. Within the limits of the hydrostatic equilibrium models that we have constructed (\S3), these lower heating rates both yield thermal equilibria across a wider range of cloud masses than the case $\Gamma=\Gamma_{cr}$. In particular the circumstance $\Gamma=0.01\times\Gamma_{cr}$ yields thermal equilibria across almost the entire range of masses represented in our hydrostatic models. However, as with the case $\Gamma=\Gamma_{cr}$, these equilibria are thermally unstable and as such they are not static models.

\section{Discussion}

\subsection{Reconciliation with McKee's critique}
As mentioned in the Introduction to this paper, \citet{mckee} used the known properties of polytropes to highlight some potential problems in understanding molecular clouds which are simultaneously very cold and very dense. The models constructed in this paper do employ a polytropic equation-of-state, but only in the core of the cloud (\S3.1). In the envelope, where \htwo\ phase equilibrium obtains, the equation-of-state is not described by any polytrope, and consequently the points made by \citep{mckee} do not necessarily apply.

A critical point of difference between the present paper and that of \citet{mckee} is that he assumed the microwave background temperature, $T_{cmb}$, to set a floor on the temperature throughout the cloud, whereas in our models the surface layers of every model have much lower temperatures. For this to be possible in a steady state model requires internal heat flow from the coldest regions to the warmer interior, whence it can be radiated away. Ordinarily such heat flow cannot occur, but we have shown (\S4.1 and Appendix A) that inward heat flow, up the macroscopic temperature gradient, is indeed present in our cloud models. It is a result of the composition gradient that is created by precipitation of molecular hydrogen condensates. Because our structures are not limited to surface temperatures $T>T_{cmb}$, the main thrust of McKee's (2001) critique does not apply here.

\citet{mckee} noted another potential difficulty: if the column to the cloud centre is $\Sigma\gg10^2\,{\rm g\,cm^{-2}}$ then cosmic-rays cannot penetrate to the deep interior (see Appendix C, figure C5); what, then, heats the central regions of the cloud? As most of our models have central column-densities much larger than  $10^2\,{\rm g\,cm^{-2}}$, that is a valid question for the structures we have presented. A key point here is that the fluid in our models is not static, and that opens up the possibility that the energy deposited by cosmic-rays can be transported (in chemical form) into the deep interior of the cloud, before conversion to heat.

To be more explicit about the path we are envisaging: cosmic-rays lose energy by ionising the gas, and some of the deposited energy goes into dissociating \htwo. Hydrogen atoms then circulate via convection before \htwo\ reforms, releasing $4.5\,{\rm eV}$ per molecule into the thermal pool. A rough estimate of the residence time of the hydrogen atoms suggests that dispersal of the chemical energy may well be important, as follows.

Cosmic-ray energy deposition rates for most of our models are $\Gamma_{cr}\sim10^{-6\pm2}\,{\rm erg\,g^{-1}\,s^{-1}}$, corresponding to dissociation rates $\sim10^{-20\pm2}\,{\rm H_2^{-1}\,s^{-1}}$. In steady state molecules reform from atoms at the same rate, and the dominant channel is the three-body reaction 2H+\htwo$\;\rightarrow\;$2\htwo, which has a rate coefficient of approximately $2\times10^{-31}\,{\rm cm^6\,s^{-1}}$ at a temperature of $30\,{\rm K}$ \citep{palla1983}. The density of atoms should therefore be $\sim2\times10^{5\pm1}\,{\rm cm^{-3}}$. If the molecular density is $\sim10^{12}\,{\rm cm^{-3}}$, then each hydrogen atom roams through the cloud for a time $\sim10^{13\pm1}\,{\rm s}$ before capturing another hydrogen atom. Excepting the most massive of our models, this residence time is much longer than the sound crossing time for the core of the cloud  (e.g. $\sim10^{8}\,{\rm s}$ for the case shown in figure 3). Although the timescale for convective circulation is not predicted by our model, the slow rate of \htwo\ formation suggests that hydrogen atoms may circulate throughout the core before they combine to form \htwo. As mixing occurs during convective overturn, the ratio of hydrogen atoms to molecules should not be a strong function of position. But the three-body reaction rate varies in proportion to the cube of the density, and the associated heating may thus exhibit a strong peak at the centre of the cloud.

{\response A qualitatively similar argument applies to the ionic chemistry: species that are created by cosmic-ray interactions at modest depths in the cloud ($\la 10^2\,{\rm g\,cm^{-2}}$) will subsequently be dispersed throughout the structure, with some of the chemical energy being released at the very centre. However, electron-ion recombinations typically convert much of that chemical energy into radiation, which is ineffective at heating the gas, so we expect \htwo\ re-formation to dominate.} {\newresponse Detailed models of the cosmic-ray induced chemistry in our clouds would not be easy to construct, despite their simple composition, because of the additional complexities that are introduced by \htwo\ condensation. We note in particular the following points: ionisation of the condensed \htwo\ favours the production of H$_6^+$, rather than H$_3^+$ (which dominates in gas-phase) \citep{lin2011}; clusters of \htwo\ ligands may form around any ions \citep{duley1996,bernstein2013}; electrons and ions that encounter snowflakes will tend to stick on the surface, or in the bulk of the snowflake \citep{walker2013}; and at present the relevant reaction rates in or on the condensed \htwo\ are largely unknown.}

Any concerns about heat supply to the cloud core are potentially more serious if the orbit of the cloud lies mostly away from the cosmic-ray disk of the Galaxy. And taking the argument a step further one might raise the issue of intergalactic clouds, where heat input (in any form) is very small indeed. In this context it is reassuring to note the following points: (i) very little heat is required in order to balance the very low levels of radiative cooling exhibited by our models; (ii) the Kelvin-Helmholtz contraction timescale for many of our computed structures exceeds the age of the Universe (\S4.2.3), so a lack of heat input is not necessarily a fatal problem; and, (iii) residual hydrogen in atomic form will be gradually converted to molecular form, and this source of heat alone may suffice to balance the radiative cooling --- e.g. an atomic fraction of 0.1\%, gradually converted to \htwo\ over an interval of $10^{10}\,{\rm years}$, would yield a specific luminosity of order  $10^{-8}\,{\rm erg\,g^{-1}\,s^{-1}}$. 

Finally, if we broaden the scope of this discussion to include the case of models that incorporate metals then there is even less reason to be concerned about heat supply, as discussed in the next section.

\subsection{Adding metals to the models}
Our models are constructed from \htwo\ and helium alone. But that choice was motivated only by a need for simplicity, and in future it would be appropriate to include metals. Here we consider how metals might change the character of the models.

We have concerned ourselves specifically with fluids that are so cold and dense that they manifest \htwo\ condensation, and under these conditions almost all of the metals would also be in the condensed phase. Indeed, for most of the metals that would be the case in the core of the cloud, not just the envelope, and so we expect that the metals would manifest themselves predominantly in a single, solid lump at the centre of the cloud. That lump would presumably contain $\sim1\%$ of the total mass of the cloud, and would include both refractory solids and various ices. Its gravity would have a strong influence on the pressure and density structure of the surrounding fluid. That is particularly true for the more massive clouds we have considered, as the virial temperature at the surface of a metallic core scales as $M^{2/3}$. We anticipate that the deeper gravitational potential well created by a solid metallic core might permit valid models in which the \htwo\ is in phase equilibrium throughout. In turn that would imply inward convection of heat right up to the surface of the solid core, so that heat supply to the centre would be assured even for clouds with very high column-densities. {\response We further note that the metals themselves may be a significant source of heat in the form of residual radioactivity. A rough estimate based on what is known of the terrestrial context -- where ${}^{232}$Th,  ${}^{238}$U and ${}^{40}$K decay chains dominate radioactive heating at the current epoch \citep{araki2005,bellini2013} -- indicates that the specific heating rate should be $\Gamma_{rad}\sim10^{-9}\,{\rm erg\,g^{-1}\,s^{-1}}$. Returning to figure 10 we see that, for the hydrostatic models we have constructed, this is a low rate of heat input compared to cosmic-rays in the disk of our Galaxy. However, we anticipate that models including metals would likely display higher column-densities than their metal-free counterparts, resulting in lower values for $\Gamma_{cr}$. Moreover, a radioactive core would provide heat input for the important case of intergalactic clouds, where the environmental contribution is very small.}

Thermal emission from a metallic core would contribute to the total radiative cooling of the cloud. The contribution would be approximately black-body at the core temperature, but the high density of the metallic lump means that it would have a small surface area for emission, so it is not necessarily the largest contribution to the luminosity. Metallic impurities remaining in gas phase at the core-envelope boundary may be incorporated into the \htwo\ condensate. Because pure solid \ph\ has very little absorption at low frequencies, it is possible that even a small impurity content in the solid could significantly increase the radiative efficiency of the snowflakes. {\response A further addition to the cloud's total radiation will arise in the presence of metals: line radiation from molecular species that remain in gas phase. In low temperature molecular gas, at low densities, species such as CO tend to dominate the radiated power, with strong emission resulting from high abundance, a large dipole moment and small rotational constant. However, for the high density gas that we are considering such lines have high optical depths, so strong emission is accompanied by strong absorption and metal lines diminish in importance relative to continuum emission processes.}

\begin{figure}
\figscaleone
\plotone{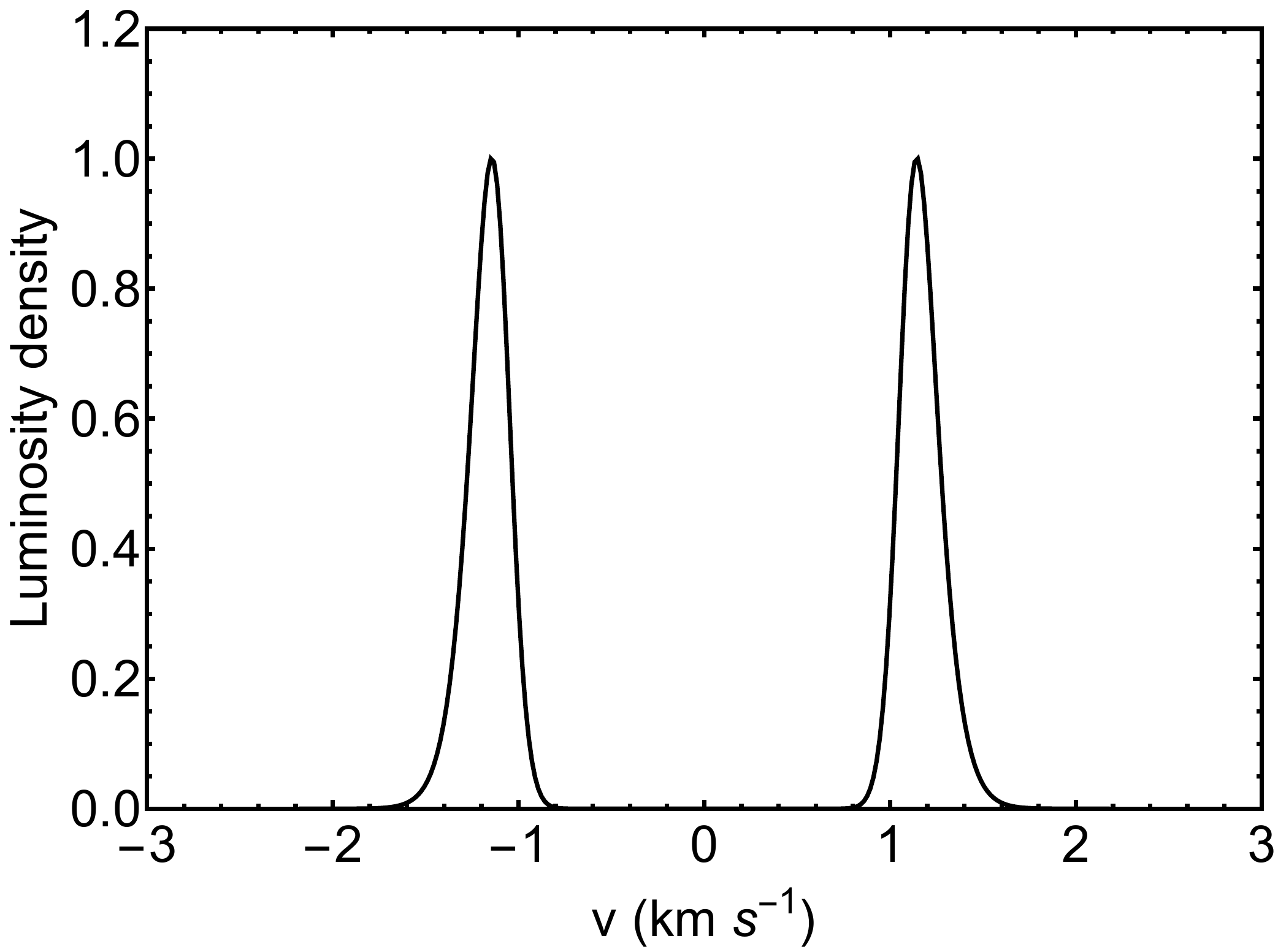}
\caption{Spectral luminosity of the $28\,{\rm \mu m}$ $S_0(0)$ \htwo\ emission line, for the same cloud model as shown in figures 3 and 11. The optical depth at line-center is very high, with emission from the core of the cloud being strongly absorbed by the overlying, cold gas in the envelope. This spectrum is appropriate to the case where convection speeds are negligible, so the line broadening is exclusively thermal.} 
\end{figure}

\subsection{Spectral line structure}
The $S_0(0)$ rotational transition of \ph\ is the principal coolant for many of our models, and it offers a possible route to discovery of \htwo\ snow clouds so its properties are of interest. Total line luminosities are given in figure 8 for all of our models, and figure 12 displays the line profile specific to the model shown in figure 3. The double-peaked structure is characteristic of all of our models, and is easy to understand. The optical depth at line centre is very high, so the intensity there reflects the very low value of the Planck function appropriate to the outer, colder regions of the cloud.  In the wings of the line, by contrast, we see into the deeper, warmer regions where the emissivity is higher.

This line profile has been calculated in the approximation that convection speeds are everywhere negligible compared to thermal speeds; but that is not necessarily true and vigorous convection could significantly alter the line profile. In that case we expect that the double-peaked structure would remain, but each of the two peaks would be broadened.

Although our models have zero metallicity, similar models which include metals will exhibit emission lines from other molecules, and those lines might be easier to detect. For example, the rotational lines of the CO molecule are excited at lower temperatures than those of \htwo, and the molecule is polar, so the CO line emission might be relatively strong. Any metal lines are likely to display qualitatively similar profiles to that shown in figure 12, for the same reasons given above. 

\subsection{Destruction of snow clouds}
Many of our models display both high central pressures, compared to the diffuse ISM, and ``hard'' surfaces (i.e. a steep increase of density with depth), as in figure 3. Consequently the diffuse ISM is not expected to have much influence on their structure --- e.g. a roughly 15\% decrease in radius can be expected for the cloud shown in figure 3, under typical conditions.\footnote{For the solutions with very large radii, on the other hand, the ambient pressure has a great deal of influence --- as noted in \S3.3.1 and illustrated in Appendix D.}  And such a cloud could tolerate a large ambient pressure jump (e.g. from a shock wave) without much effect on the core, which is where most of the mass resides. These models therefore describe well defined entities which are mechanically robust, and in that sense they are akin to stars and planets. Snow clouds are, however, quite susceptible to other destructive influences, as we now describe.

\subsubsection{Thermal disruption}
The very low luminosities of our models mean that any counterparts in the universe we inhabit would be easily overlooked. They would also be easily ``overcooked''. In other words it is easy to imagine conditions whereby a snow cloud would be heated at a much greater rate than it can cool. Indeed we have already noted that circumstance for a large fraction of our models in the case of heating by Galactic cosmic rays (\S4.4). As we have already remarked, if heating exceeds cooling the excess heat goes into work done against gravity as the cloud undergoes a secular expansion. This type of thermal imbalance is therefore disruptive if it persists, and we note that the corresponding expansion timescale may be much shorter than the Kelvin-Helmholtz contraction timescale if heating far outstrips cooling --- e.g. in the upper right of figures 9 and 10.

The other main cause of thermal disruption that we can anticipate is starlight: if a cloud happens to lie near a luminous star the ambient radiation field will be very strong. It is principally the far-UV radiation that is a concern, because much of the far-UV incident on a cloud will be absorbed. As massive stars are both hot and luminous, any snow clouds in the immediate vicinity of such a star will experience thermal disruption.

\subsubsection{Disruption by physical collisions}
Snow clouds have large column densities compared to diffuse instellar clouds; nevertheless those columns are tiny in comparison with stars. So although physical collisions are very rare for stars in most Galactic environments, they could be very important for snow clouds \citep{gerhard,walker1999}. We have already noted that our model clouds exhibit a characteristic binding energy ${\cal B}\sim10^9\,{\rm erg\,g^{-1}}$, so relative speeds $\gg2\sqrt{2{\cal B}}\sim1\,{\rm km\,s^{-1}}$ are required to unbind the material in the colliding clouds. Collisions at speeds $\la2\sqrt{2{\cal B}}$ are expected to lead to merging of the two clouds.

\subsubsection{Tidal disruption}
It is widely appreciated that massive black holes in the nuclei of galaxies can cause tidal disruption of stars which approach too close to them \citep{rees1988}. Snow clouds in galactic nuclei run the same risk. But snow clouds are not safe even if they are well removed from galactic nuclei: because the typical density inside a star is far in excess of that of a snow cloud, the latter can be tidally disrupted by the former during close encounters. The large cross-section for tidal interactions means that they are expected to be the most frequent two-body process involving a star and a snow cloud. Tidal stripping of snow cloud envelopes is expected to be more common than complete tidal disruption, as the envelope is typically extended and has lower density than the core.

\subsection{Possible observational manifestations}
Because we have been unable to find stable thermal equilibrium solutions (\S4.4), we must be circumspect in proposing connections to the observed Universe; however, some brief comments are appropriate.

\noindent 1. The neutral clouds that \citet{walker2017} inferred, from radio-wave scintillation, to be present in large numbers around main sequence stars must have low luminosities, low masses and large radii, and correspondingly low temperatures. They could therefore be interpreted as hydrogen snow clouds.

\noindent 2. Circumstellar snow clouds which survive the main-sequence phase of a low-mass star's life may become visible during post main-sequence evolution, as the UV radiation field intensifies. Thus snow clouds could explain the cometary knots that are seen in the Helix and in other planetary nebulae \citep[e.g.][]{odellhandron1996,matsuura2009}.

\noindent 3. Circumstellar snow clouds which survive the main-sequence phase of a high-mass star's life would be thermally disrupted by the UV flash from any supernova explosion, leading to a large fraction of the clouds' \htwo\ content being converted to snowflakes on the dynamical timescale. Snow clouds might therefore contribute to the rapid dust production that is observed in supernovae \citep[e.g.][]{matsuura2011}. 

\noindent 4. Near-infrared observations towards the Galactic Centre have revealed a small, dusty cloud on a highly eccentric orbit around Sgr A$^*$ \citep{gillesen2012}. The low inferred mass and large radius of this cloud place it near the upper envelope of the models shown in figure 5.

\noindent 5. X-ray absorption events seen in some active galactic nuclei give direct evidence for the existence of large numbers of dense clouds ($\sim10^{11}\,{\rm cm^{-3}}$), with sizes $\sim{\rm AU}$ \citep{maiolino2010}; these clouds could be hydrogen snow clouds.

\section{Conclusions}
We have constructed equilibrium models of cold, dense, self-gravitating gas clouds manifesting \htwo\ condensation. These structures lie in a previously unoccupied region of the mass-radius plane, having sub-stellar masses but radii which are typically very large. With hard outer edges, and high internal pressures, our models describe mechanically-robust, well-defined entities that are perhaps more akin to stars and planets than to the ISM. A key characteristic of our model clouds is their low luminosities --- they are so dim that they could be present in very large numbers yet remain undetected. They are a type of baryonic dark matter. Their thermal characteristics are surprising, with temperatures in the outer regions of each cloud ranging below that of the microwave background. That this circumstance can exist in steady state is dependent on the inward convection of heat, up the macroscopic temperature gradient --- a phenomenon which, as far as we are aware, is demonstrated here for the first time.

Amongst our hydrostatic equilibria we have identified thermal equilibria appropriate to the Galaxy, in which radiative cooling is balanced by cosmic-ray heating. These equilibria are all thermally unstable, and so we must be cautious about any possible connections between our models and observed phenomena in the real universe. However, the Kelvin-Helmholtz timescales of some of these equilibria -- at the low mass end of the spectrum -- are very long, and our solutions might therefore be fair approximations to real-world structures in a universe of age $10^{10}\,{\rm yr}$. 

In general the low luminosities of our models make them prone to thermal imbalance, and strong heating must drive secular expansion which will ultimately be disruptive. Disruption by physical collisions, and by tides, should also be commonplace if snow clouds are a significant component of the real Universe. Disrupting snow clouds should yield trails of gas and dust, and may thus be more readily detected than their undisturbed parents. It is possible that they have already been observed, in various astrophysical contexts, but not previously recognized as such. 

\acknowledgements
MAW thanks Oxford Astrophysics for hospitality, and Philipp Podsiadlowski, Bruce Draine and Ike Silvera for helpful discussions.

This paper is dedicated to the memory of Bohdan Paczy\'nski, who encouraged us and provided key insights when this work was in its early stages. His clear thinking remains an inspiration to us both.

\clearpage

\appendix

\section{Buoyancy instability and heat transport}
Suppose we have a fluid in hydrostatic equilibrium in a gravitational field. The condition for the onset of buoyancy instability is well known \citep[e.g.][]{kippenhahn1994}, and is determined by the difference between the run of density with pressure in the equilibrium fluid, and the run of density with pressure for adiabatic changes in the fluid. It is convenient to introduce the operator
\be
\dgrad\equiv{{{\rm d\/}\log\ \ }\over{{\rm d\/}\log P}}\; -\; \left( {{{\partial}\log\ \ }\over{{\partial}\log P}} \right)_{\!\!S},
\ee
which allows the instability criterion to be written as
\be
\dgrad \rho < 0 \quad \Rightarrow \quad {\rm Buoyancy\ \, Instability.\/}
\ee
This condition is the fundamental dynamical criterion. With further assumptions about the nature of the fluid it can be rewritten in other forms, as described below. 

Consider now a fluid whose composition need not be uniform. In particular let us allow for the possibility of a mean molecular mass (MMM), $\bar{\mu}$, that is a function of pressure (i.e. height within the gravitational field). We can write the density as $\rho=\rho(P,T,\bar{\mu})$, and any density interval ${\rm d}\rho$ can be expressed as a sum of the intervals  ${\rm d}\log P$,  ${\rm d}\log T$ and  ${\rm d}\log\bar{\mu}$, weighted by the corresponding partial derivatives. We can do this for the hydrostatic fluid structure, and for an adiabatic trajectory, leading to
\be
\dgrad\rho = \left( {{{\partial}\log\rho}\over{{\partial}\log\bar{\mu}}} \right)_{\!\!T,P} \dgrad\bar{\mu} \;\, +  \;\, \left( {{{\partial}\log\rho}\over{{\partial}\log T}} \right)_{\!\!P,\bar{\mu}} \dgrad T.
\ee
The criterion for buoyancy instability can thus be rewritten in terms of the right-hand-side of equation (A3). 

As a simple example we can consider the case of a gas containing a fixed number of atoms/molecules of each type, so that the adiabatic derivative of $\bar{\mu}$ is zero, and
\be
\dgrad\bar{\mu}={{{\rm d\/}\log\bar{\mu}}\over{{\rm d\/}\log P}}.
\ee
And if the gas is also ideal (i.e. $P=\rho kT/\bar{\mu}$) then the coefficients in equation (A3) are
\be
 \left( {{{\partial}\log\rho}\over{{\partial}\log\bar{\mu}}} \right)_{\!\!T,P} =1,\qquad \left( {{{\partial}\log\rho}\over{{\partial}\log T}} \right)_{\!\!P,\bar{\mu}} =-1.
\ee
We thus arrive at a form of the criterion that is specific to ideal gases with conserved particle numbers:
\be
\dgrad T> {{{\rm d\/}\log\bar{\mu}}\over{{\rm d\/}\log P}} \quad \Rightarrow \quad {\rm Buoyancy\ \, Instability.\/}
\ee
Equation (A6) is known as the Ledoux Criterion. If the fluid is uniform in composition then this becomes simply
\be
\dgrad T> 0 \quad \Rightarrow \quad {\rm Buoyancy\ \, Instability,\/}
\ee
which is known as the Schwarzschild Criterion. These results are familiar in the context of stellar structure, for example \citep[e.g.][]{kippenhahn1994}.

Henceforth we use the term ``convection'' to refer to the fluid motions which arise from a buoyancy instability.

The quantity $\dgrad T$ is just the difference between the actual temperature derivative in the hydrostatic structure and the adiabatic trajectory of the fluid which makes up that structure. If there is convection, then the sign of $\dgrad T$ tells us about the direction of convective heat flow, as follows. If $\dgrad T>0$ then a fluid parcel that has been displaced to a higher (lower) pressure location is cooler (hotter) than its surroundings, and heat will flow into (out of) the parcel. Consequently $\dgrad T>0$ results in convective heat transport from high pressure regions to low pressure regions.  And conversely, convection results in heat transport from low pressure to high pressure regions if $\dgrad T<0$. 

In a fluid of uniform composition, then, the direction of convective heat transport is always from high pressure to low pressure -- interior to exterior -- because if $\dgrad T<0$ the condition for instability given in equation (A7) will not be met, and so there would be no convection.

In a fluid of non-uniform composition, however, either direction of convective heat transport is possible, depending on the gradient of $\bar{\mu}$. Assuming that the fluid is just marginally unstable to convection, i.e. $\dgrad\rho\simeq0$, which is expected for gentle convective overturn, we have
\be
\dgrad T\simeq {{{\rm d\/}\log\bar{\mu}}\over{{\rm d\/}\log P}},
\ee
and therefore heat flows down the MMM gradient. If MMM increases inwards then convection, if it takes place, transports heat outwards. But if MMM increases outwards then convection, if it takes place, transports heat inwards. And that is true even if the structure is hotter in the interior, as is usually the case in practice. Thus we can have the paradoxical circumstance of heat being transported up a macroscopic temperature gradient.

That sounds like a violation of the Second Law of Thermodynamics, but it is not. In fact the way we determined the direction of heat flow, by considering the sign of $\dgrad T$, ensures that the Second Law is obeyed, because $\dgrad T$ reflects the temperature differences on a {\it microscopic\/} level.  The key point is that fluid elements are displaced during convection, and in the course of those displacements they change temperature as a result of adiabatic compression or expansion. If they change temperature by more than the background fluid then convection, if it occurs, will transport heat up the pressure gradient into the interior.

It is also helpful to remember that $\dgrad$ is the gradient relative to the local isentrope, thus the sign of $\dgrad T$ reflects the sign of the entropy gradient, which indeed is the quantity we expect to dictate the direction of heat flow.

For the case of an ideal gas that varies in MMM we can understand the direction of heat flow simply by considering the equation of state, as follows. In gentle convective overturn the fluid motions bring together parcels which have different MMM; but they must have essentially the same pressure and density, and therefore they have the same value of $kT/\bar{\mu}=P/\rho$. Thus for two distinct fluid elements which have been juxtaposed by convection it is always the one with the larger MMM that is hotter, and therefore heat always flows down the MMM gradient.

{\response  In studies of stellar structure one encounters examples of both positive and negative MMM gradients, as a result of nuclear burning -- either in the core or in a thin shell -- and in both cases it has been shown that convective motions can arise even in a configuration that is dynamical stable according to equation (A2). This behaviour is given different names according to the sign of the MMM gradient: if MMM increases inwards it is termed ``semi-convection'' \citep{schwarzschild1958,kato1966}, whereas if MMM increases outwards it is referred to as ``thermohaline convection'' \citep{ulrich1972,charbonnel2007}. Despite the different names there is a common aspect here: convective motions may develop in a dynamically stable configuration as a result of the exchange of heat between displaced fluid parcels and their surroundings --- a process that is excluded from the development in this Appendix by use of the adiabatic approximation for the trajectory of the displaced fluid. It would be interesting to study the onset of snow-cloud convection in the non-adiabatic case, but that is beyond the scope of this paper. Here we simply note that although thermohaline convection is slower than a dynamically driven convection, and thus yields smaller heat fluxes, the direction of the heat flow would still be inwards. The reason is that the isentrope corresponds to zero precipitation of the condensate, and one always expects some degree of precipitation to occur so entropy will increase outwards.}

\section{Thermodynamic properties of \htwo\ near saturation}
In this Appendix we demonstrate the accuracy of various facets of our adopted thermophysical description (\S2). We do so by comparing with data for \ph, taken from tables 2.2, 2.3 and 9.5 of \citet{roder1973}.

\begin{figure}
\figscalethree
\centering{\hbox{\plotone{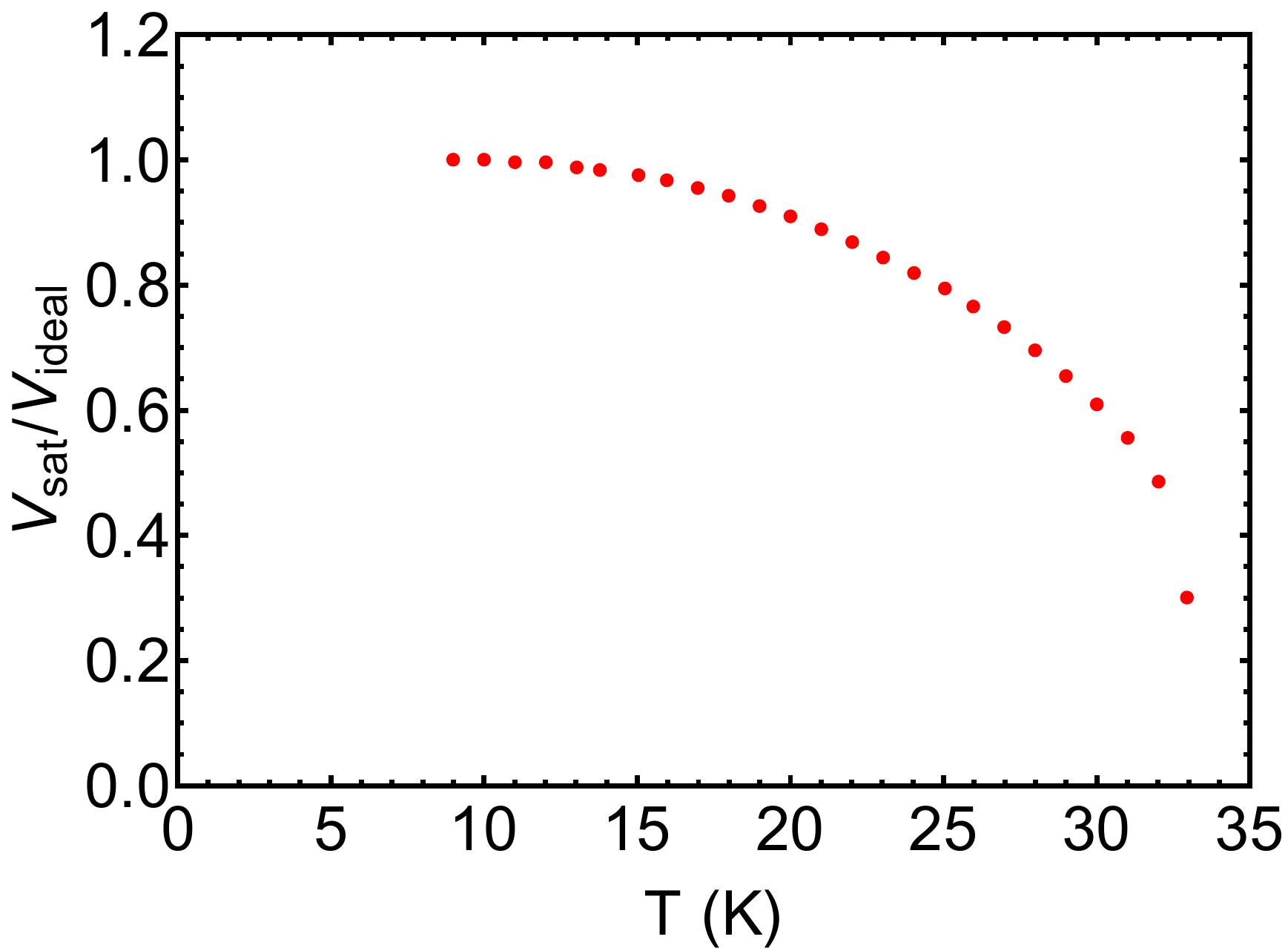}\hskip2cm\plotone{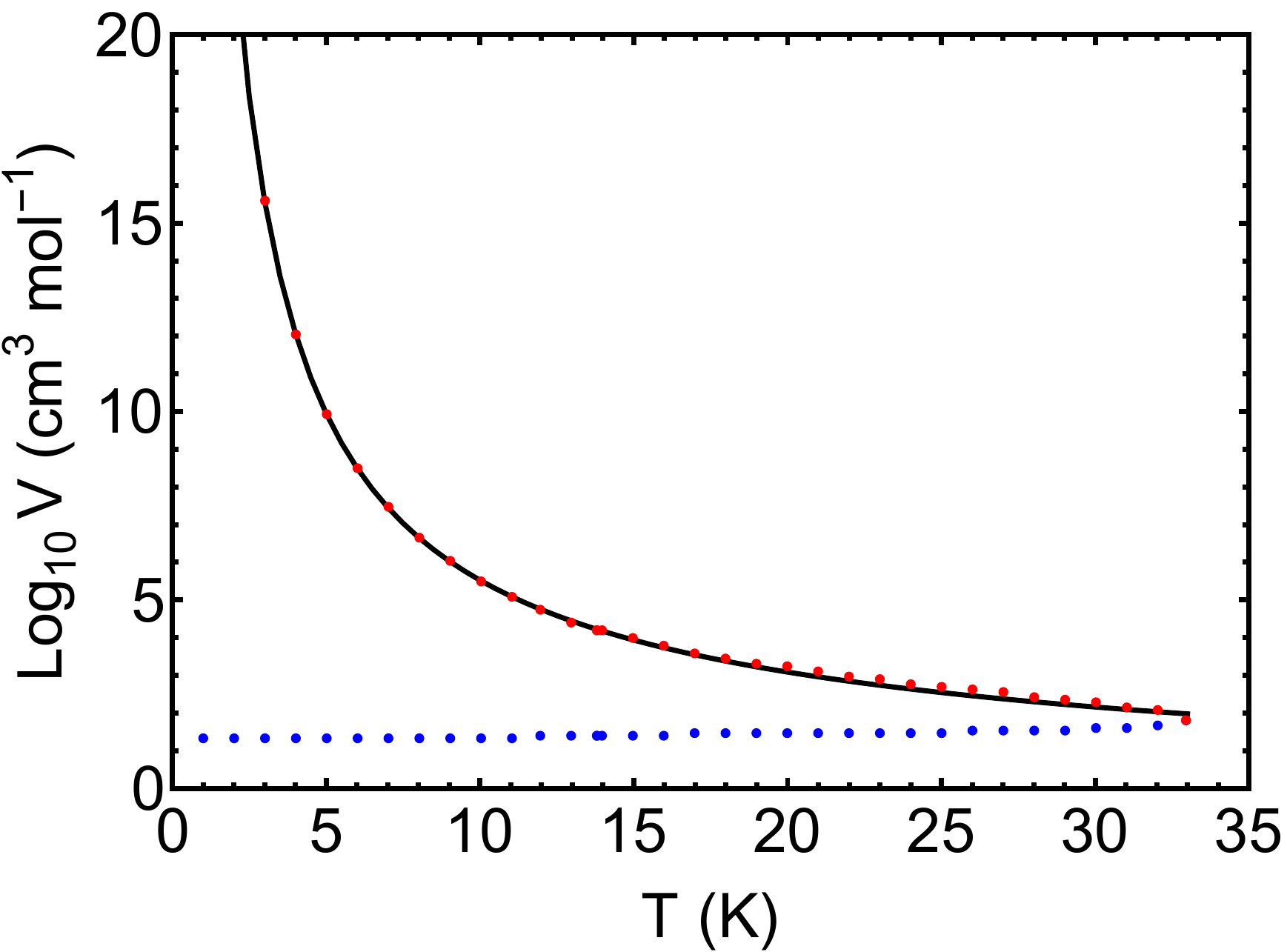}}}
\caption{Left Panel: the molar volume of the saturated vapour of \ph\ relative to that of an ideal gas of the same temperature and pressure. Right Panel: the molar volume of the saturated vapour of \ph\ (red dots), compared to our model (black line), which is that of an ideal gas at the same temperature and with the pressure as given by equation (5). Also shown is the molar volume of the condensate (blue dots). All data are as given in  tables 2.2 and 2.3 of \citet{roder1973}.} 
\end{figure}

\begin{figure}
\figscalethree
\centering{\hbox{\plotone{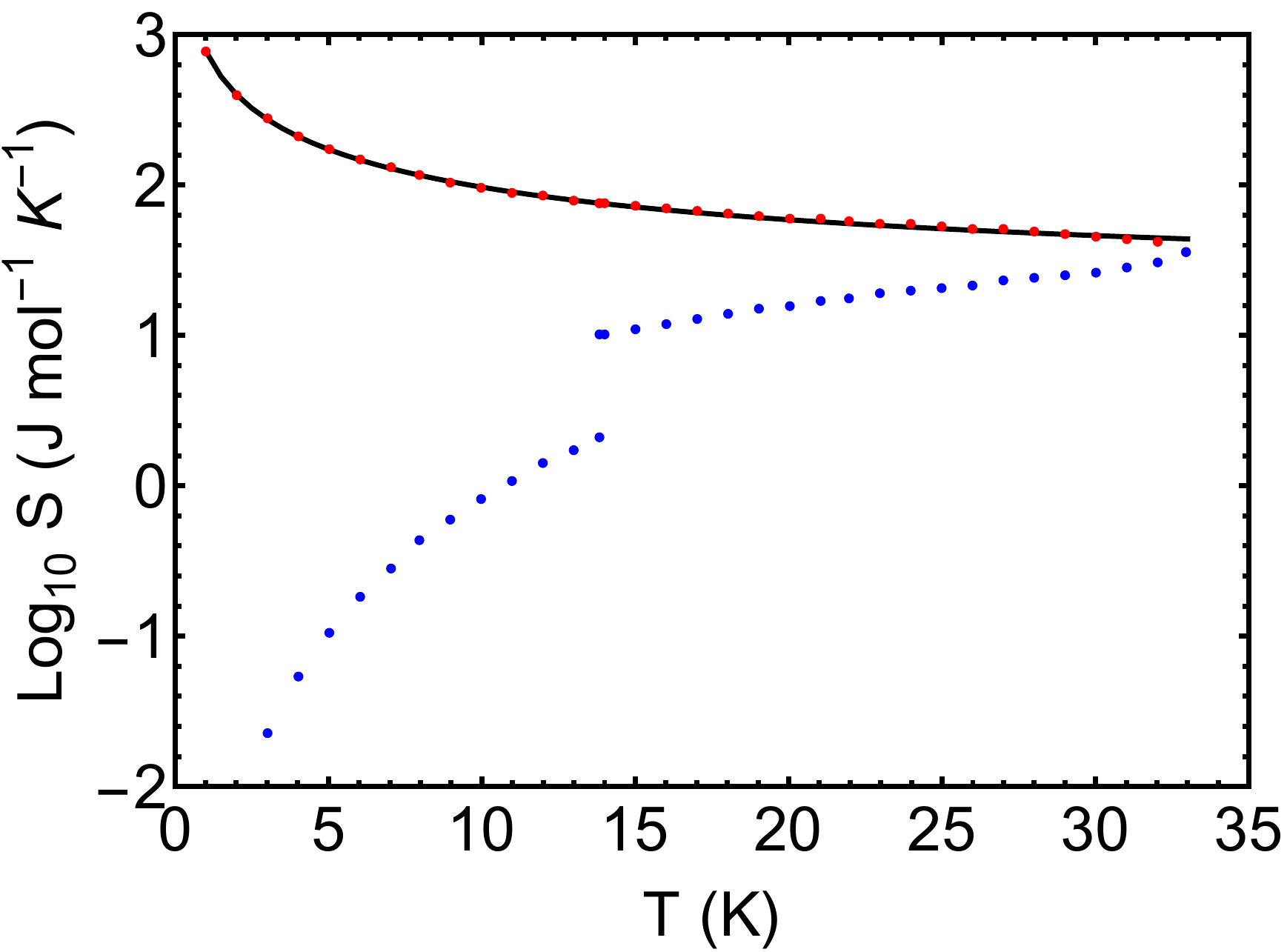}\hskip2cm\plotone{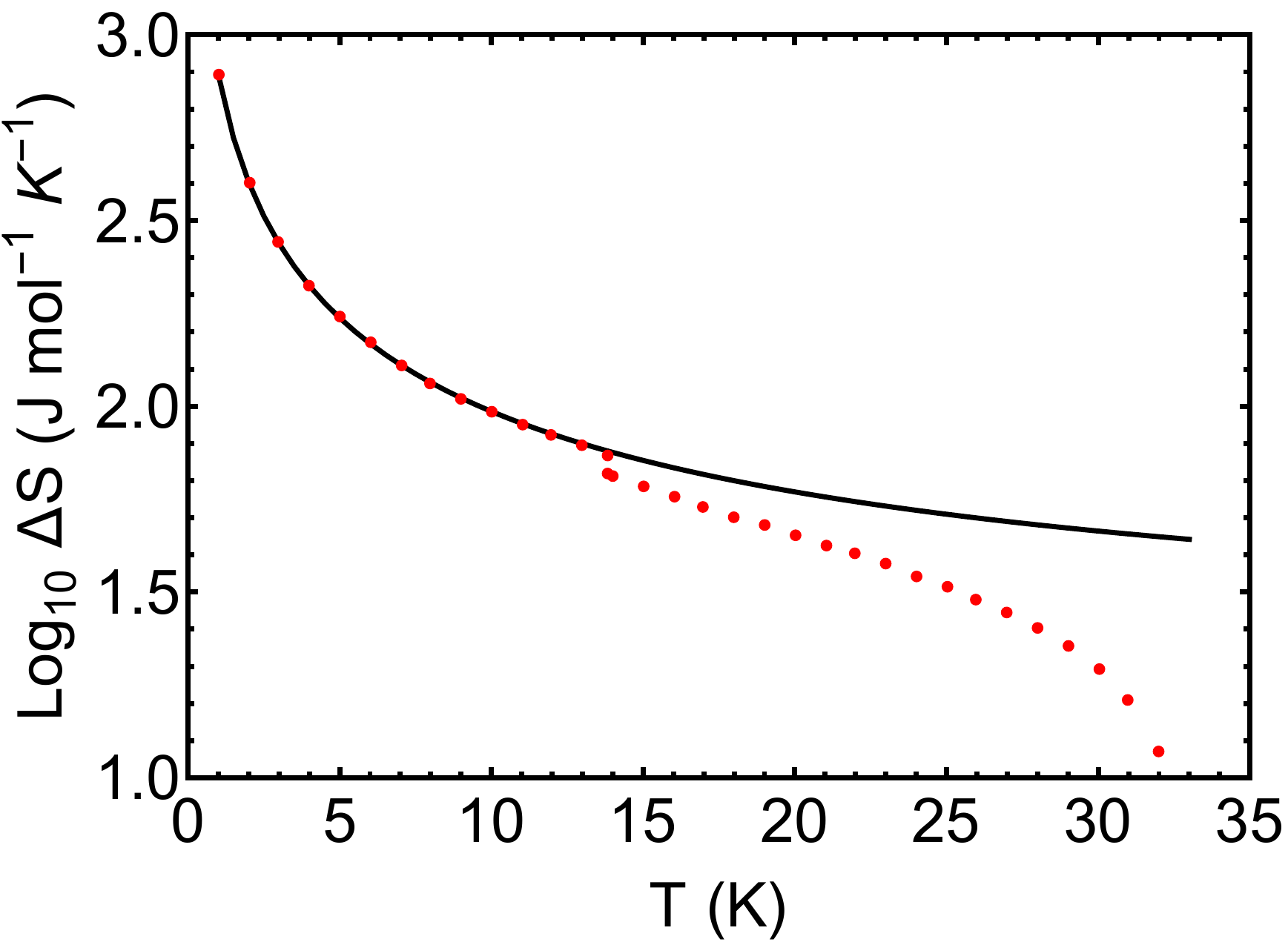}}}
\caption{Left Panel: the entropy of the saturated vapour of \ph\ (red dots), compared with our model description (equation 9; black line). Also shown is the entropy of the condensate (blue dots). The entropy discontinuity between solid and liquid \ph\ reflects the latent heat of fusion of $117\,{\rm J\,mol^{-1}}$ \citep[][tables 2.3 and 9.5]{roder1973}. Right Panel: the entropy difference between the saturated vapour and the condensate (red dots); our model for the entropy difference (black line) is the same curve shown in the left panel, because the entropy of the condensate is neglected.} 
\end{figure}

\subsection{Molar Volume}
Our adopted model of the gas pressure is the ideal gas law, $P=NkT/V$, for both He and \htwo, but with the partial pressure of \htwo\ limited to the saturation pressure, $P_{sat}(T)$. The left-hand panel of figure B1 shows departures from ideality of the saturated vapour, as gauged by its volume relative to that of an ideal gas of the same temperature and pressure. From that graph we can see that departures from ideality are very small right up to the triple point ($T_{trip}\simeq13.8\,{\rm K}$), and are still below 10\%\ at $T=20\,{\rm K}$, but for temperatures close to the critical temperature ($T_{crit}\simeq32.9\,{\rm K}$), an ideal gas is a poor model of the saturated vapor. The right-hand panel of figure B1 shows the molar volume of both the saturated gas (red dots) and the condensate (blue dots), as a function of temperature. Also shown is our model for the molar volume of the saturated gas, as given by the ideal gas law in combination with equation (5) for the saturation pressure. We neglect the molar volume of the condensate.

To determine how the ideal gas law performs for pressures below the saturation pressure, we have used the van~der~Waals equation of state as a guide \citep{johnston2014}. The van~der~Waals equation of state predicts departures from ideal gas behavior which scale approximately as $P/P_{sat}$ for a given isotherm.  Thus, even for $T\sim T_{crit}$, the ideal gas law should be accurate to a few percent for pressures that are $\la P_{sat}/10$. We conclude that the ideal gas law is an adequate representation of gaseous \ph\ except in the imediate vicinity of the critical point.

\subsection{Molar Entropy}
Figure B2 shows the entropy of gaseous \ph\ under saturated conditions, as a function of temperature, together with our adopted description as given in equation (9). Except for temperatures very close to the critical temperature, the entropy of saturated, gaseous \ph\ is well described by equation (9). 

Also shown in figure B2 is the entropy of the condensate, which in our model is neglected entirely. That is clearly a good approximation at all temperatures below the triple point. The latent heat of fusion of \ph\ introduces a discontinuity in the entropy of the condensate at the triple point. Above that point our neglect of the condensate entropy becomes progressively worse until, at the critical point, the entropy of the liquid is identical with that of the saturated vapor.

\section{Evaluation of the cosmic-ray heating rate}
One can evaluate the cosmic-ray power input to a parcel of gas, of mass $\Delta M$, from $\Gamma_{cr}\,\Delta M$, where $\Gamma_{cr}$ is the specific heating rate. For diffuse interstellar gas one usually assumes that $\Gamma_{cr}$ is approximately constant --- i.e. independent of column-density, in the limit of low column-density. Its value has previously been estimated as  $\Gamma_{cr}\simeq3\times10^{-4}\;{\rm erg\,g^{-1}\,s^{-1}}$ \citep{cravens,webber}, with most of the heating coming from cosmic-ray protons. The clouds discussed in this paper are much denser than the diffuse ISM, and the centre-to-surface column densities are so high that the interior cosmic-ray spectrum is substantially attenuated relative to the interstellar spectrum. In this circumstance we must evaluate $\Gamma_{cr}$ as a function of depth in the cloud, accounting for the change in spectrum with depth. In this Appendix we present details of that calculation.

We wish to evaluate the heating rate for a large number of different structures, so we employ a simple calculation based on the continuous slowing down approximation, and we assume that the energy of all secondary particles -- bremsstrahlung photons, pions, electrons etc -- is absorbed on-the-spot. Although this approach is motivated mainly by a need for simplicity, it is a sensible approximation to make because: (i) charged secondaries are stopped by Coulomb interactions with the gas; (ii) far-UV (and shorter wavelength) photons are absorbed by electronic transitions in the He atoms and \htwo\ molecules; (iii) near-IR (and shorter wavelength) photons are likely to be absorbed by \htwo\ snowflakes; and, (iv) excited \htwo\ rovibrational states may de-excite more rapidly by collisions than by quadrupole radiation emission.

As the starting point of our calculation we adopt the stopping power (${\rm d}{\cal E}/{\rm d\Sigma}$) in hydrogen and helium of electrons and protons given by the ESTAR and PSTAR models.\footnote{\tt http://www.nist.gov/pml/data/star/} Those tables cover particle energies $10^{-3}-10^4\,{\rm MeV}$. For higher energy cosmic-rays we calculated ionisation losses using the Bethe-Bloch formula, with the effective ionisation energies for hydrogen (19.2$\;$eV), and helium (41.8$\;$eV), quoted in the ESTAR/PSTAR database. And in the case of cosmic-ray electrons we added bremsstrahlung losses, which are important at high energies, being a factor ${\cal E}/{\cal E}_{crit}$ larger than the ionisation losses \citep{eidelman2004}. We determined ${\cal E}_{crit}$ appropriate to hydrogen and helium by matching to the ESTAR tables at $10^4\,{\rm MeV}$.

The PSTAR tables do not include ``pionization'' losses --- due to pion production when cosmic-ray protons interact with nucleons. Pionization becomes important at energies of $10^2\,{\rm MeV}$ and above. A convenient approximation to these losses, valid for  ${\cal E}\ga10^{3}\,{\rm MeV}$, is provided by equation (34) of \citet{krakau}:
\be
-\left({{{\rm d}{\cal E}}\over{{\rm d\Sigma}}}\right)_{\!\pi}\simeq4.4\times10^{-3} \beta^{-1} {{{\cal E}^{1.28}}\over{(2\times10^5 + {\cal E})^{0.2}}} \qquad {\rm MeV\;g^{-1}\;cm^2},
\ee
with $\beta$ being the proton speed, in units of $c$, and ${\cal E}$ the kinetic energy in ${\rm MeV}$. This approximation was presented by \citet{krakau} for cosmic-ray protons interacting with target protons in the interstellar gas; we adopt it also for target neutrons, so that equation (C9) is used here for the pionization losses of both hydrogen and helium.

\begin{figure}
\figscalethree
\plotone{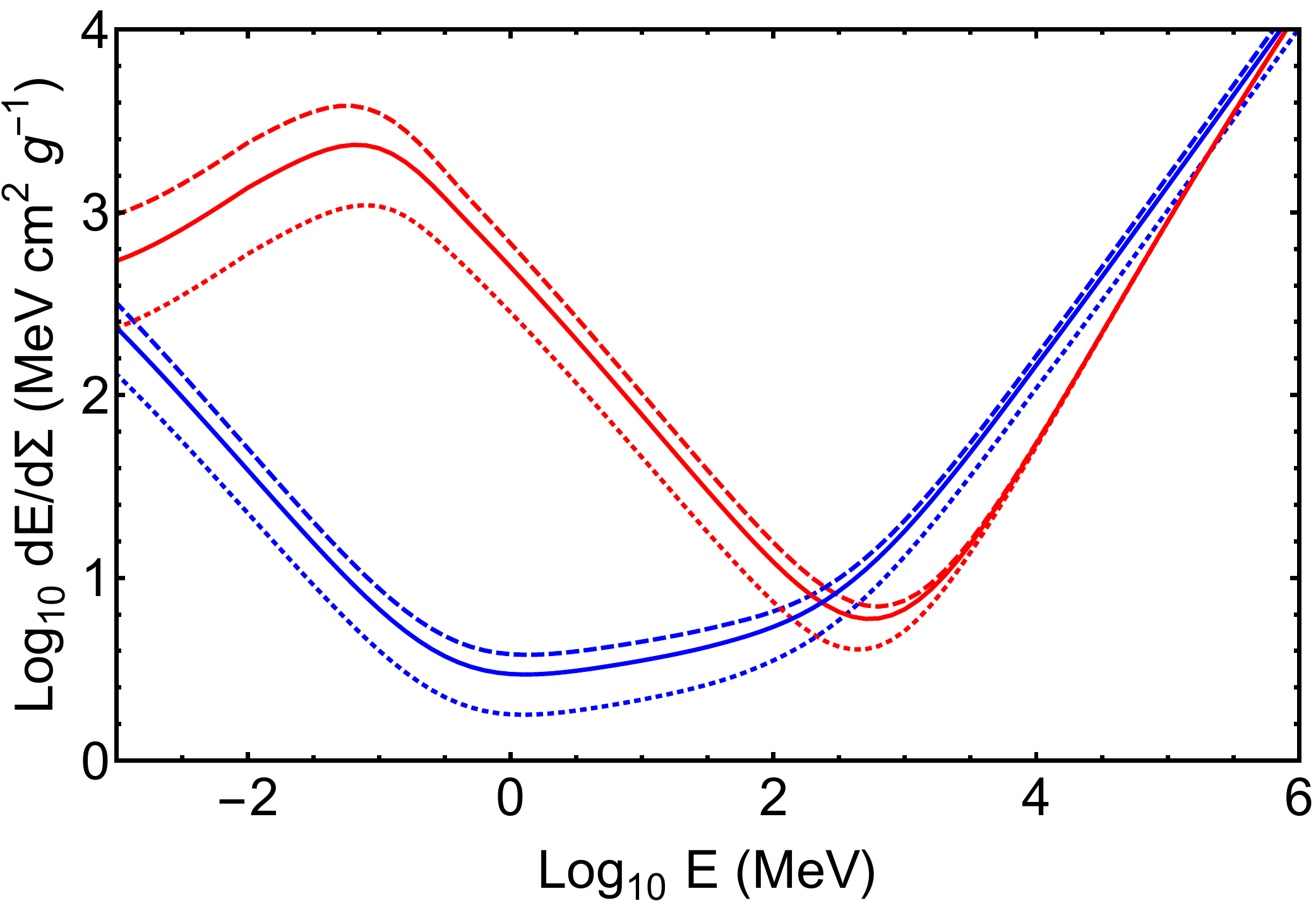}
\caption{Stopping power (${\rm d}{\cal E}/{\rm d\Sigma}$) for protons (red) and electrons (blue), of pure hydrogen (dashed), pure helium (dotted), and a hydrogen-helium mixture that is 25\% He by mass (solid curves). } 
\end{figure}

In this way we extended the stopping power calculations up to $10^6\,{\rm MeV}$ for both electrons and protons. The combined inverse stopping-power (i.e. ${\rm d\Sigma/d}{\cal E}$) of the hydrogen-helium mixture was then determined by adding the inverse stopping-powers of the hydrogen and the helium, weighted according to their 0.75, 0.25 average mass-fractions, respectively. This procedure yields the results shown in figure C3.

As detailed in \S3 of this paper, each cloud has a compositional gradient, being helium rich in the outer regions. Using the mean cloud composition to arrive at a single stopping power, as a function of energy, for each particle species, is thus an approximation. It is a convenient approximation to make because the heating rate due to cosmic-rays coming from a particular direction is then only a function of depth (expressed as column-density, $\Sigma$), rather than being a function of both the hydrogen column and the helium column.  A more precise treatment of the energetic particle energy losses does not seem warranted in this initial sketch of cloud properties, given the other uncertainties involved in the calculation.

The stopping power of the fluid tells us the differential heating rate (${\rm MeV\,g^{-1}\,s^{-1}\,sr^{-1}}$) due to a particle beam of intensity $I_f$ (${\rm cm^{-2}\,s^{-1}\,sr^{-1}\,MeV^{-1}}$):
\be
{{{\rm d\Gamma_{cr}}}\over{{\rm d\Omega}}}=\int_0^\infty\! {\rm d}{\cal E}_f \;{{{\rm d}{\cal E}}\over{{\rm d\Sigma}}}\bigg|_{{\cal E}_f} I_f({\cal E}_f),
\ee 
where $I_f$ is the beam intensity at the point where we wish to calculate the heating. If this lies at depth $\Sigma$, measured along the direction of incidence of the beam, then we can relate the particle kinetic energy at the site of interest, ${\cal E}_f$, to the initial particle kinetic energy, ${\cal E}_i$, via
\be
\Sigma=\int_{{\cal E}_f}^{{\cal E}_i}{\rm d}{\cal E}\,{{\rm d\Sigma}\over{{\rm d}{\cal E}}}.
\ee
And particles in the beam are conserved, so $I_f\Delta {\cal E}_f=I_i \Delta {\cal E}_i$, where  $I_i$ is the particle spectrum at zero column -- i.e. the interstellar particle spectrum -- whence
\be
{{{\rm d\Gamma_{cr}}}\over{{\rm d\Omega}}}=\int_0^\infty\! {\rm d}{\cal E}_f \;{{{\rm d}{\cal E}}\over{{\rm d\Sigma}}}\bigg|_{{\cal E}_f} I_i({\cal E}_i)\left({{\partial{\cal E}_i}\over{\partial{\cal E}_f}}\right)_{\!\Sigma}.
\ee 
In practice the integration in equation C12 is taken over the finite domain $10^{-3}\,{\rm MeV}\le{\cal E}_f\le10^{6}\,{\rm MeV}$, for which we have a good description of the particle energy losses (figure C3).

\begin{figure}
\figscalethree
\centering
\plotone{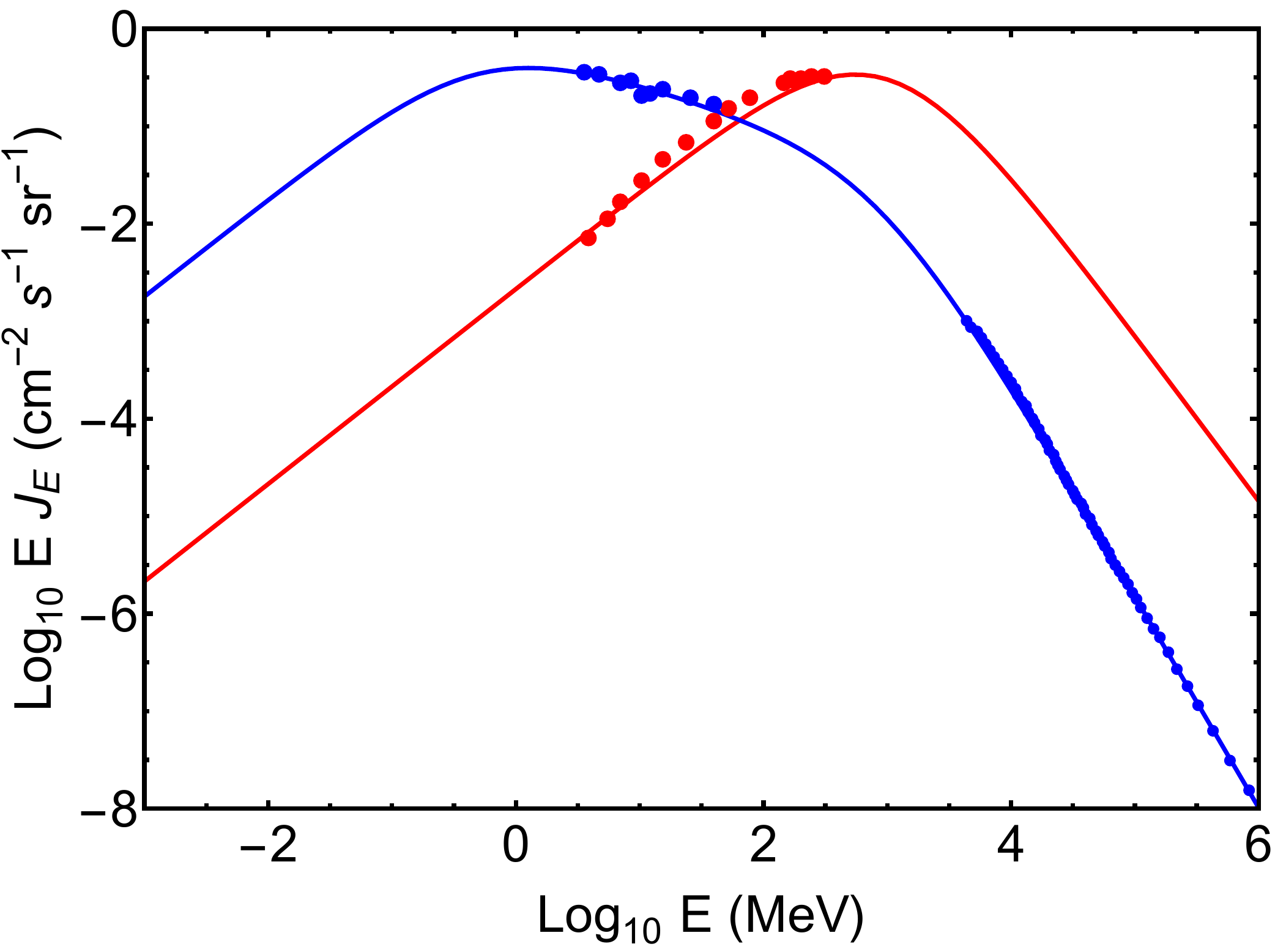}
\caption{Our adopted cosmic-ray spectra for protons (red line; equation C13) and  electrons (blue line; equation C14). Also shown are the interstellar spectra measured by {\it Voyager 1\/} \citep[large circles;][]{cummings2016}, along with the {\it AMS\/} electron spectra at high energies \citep[small circles;][]{aguilar2014}.} 
\end{figure}

Low energy electron and proton spectra are strongly modulated by the solar wind, making it difficult to determine the interstellar cosmic-ray spectra. The {\it Voyager 1\/} spacecraft crossed the heliopause in August 2012, and subsequently made direct measurements of the interstellar proton spectrum in the energy range from 3 to $350\,{\rm MeV}$, and the interstellar electron spectrum in the range 3 to $70\,{\rm MeV}$ \citep{cummings2016}. The {\it Voyager 1\/} data are shown in figure C4, along with our adopted model spectra. For protons we use the spectrum given by \citet{beringer2012}, which lies close to the {\it Voyager 1\/} data:
\be
I_p({\cal E})={1.8\times10^{-3}}\left({{1000}\over{m_pc^2+{\cal E}}}\right)^{2.7}\qquad{\rm cm^{-2}\,s^{-1}\,sr^{-1}\,MeV^{-1}},
\ee
where ${\cal E}$ is the kinetic energy of the particle, in MeV, and $m_pc^2$ is the proton rest energy. For electrons we have used an analogous form (i.e. with the proton rest mass replaced by the electron rest mass), but with an additional spectral break at $850\,{\rm MeV}$:
\be
I_e({\cal E})= 0.70\left({{1}\over{m_ec^2+{\cal E}}}\right)^{1.4}\left(1+{{\cal E}\over{850}}\right)^{-1.77}\qquad{\rm cm^{-2}\,s^{-1}\,sr^{-1}\,MeV^{-1}}.
\ee
As can be seen in figure C4, this function approximately reproduces both the {\it Voyager 1\/} data at low energies \citep{cummings2016}, and the {\it AMS\/} data at high energies \citep{aguilar2014}. The associated energy-densities are: $0.66\,{\rm eV\,cm^{-3}}$, for protons, and $0.024\,{\rm eV\,cm^{-3}}$ for electrons.

Using the cosmic-ray spectra given in equations (C13) and (C14) leads to differential heating rates, ${{{\rm d\Gamma_{cr}}}/{{\rm d\Omega}}}$, as a function of depth, $\Sigma$, shown in the left panel of figure C5. By integrating over column-density we then obtain the total absorbed beam intensity, $I_{abs}(\Sigma)$, as shown in the right panel of figure C5. At low column-densities the differential heating rate is only a weak function of column-density, and the absorbed intensity is roughly linear in $\Sigma$. At large columns essentially all the cosmic-ray particles are stopped, and the absorbed intensity saturates at a value equal to the total cosmic-ray intensity: $I_{abs}\rightarrow\int{\rm d}{\cal E}\,{\cal E}\{I_p({\cal E})+I_e({\cal E})\}\simeq2.63\times10^{-3}\,{\rm erg\,cm^{-2}\,s^{-1}\,sr^{-1}}$.  

Given the absorbed intensity, as a function of column-density, the spherical symmetry of our models renders it straightforward to compute the total heating rate. Consider a point on the surface of the cloud. Particles incident at polar angle $\theta$ encounter a column-density $\Sigma(\theta)$, and the total cosmic-ray power input, ${\cal P}_{cr}$,  is just $4\pi R^2$ times the total absorbed flux, whence
\be
{\cal P}_{cr}=(2\pi R)^2 \int_0^{\pi/2}\!\!\!{\rm d\theta}\,\sin2\theta\,I_{abs}(\Sigma(\theta)).
\ee

\begin{figure}
\figscalethree
\centering{\hbox{\plotone{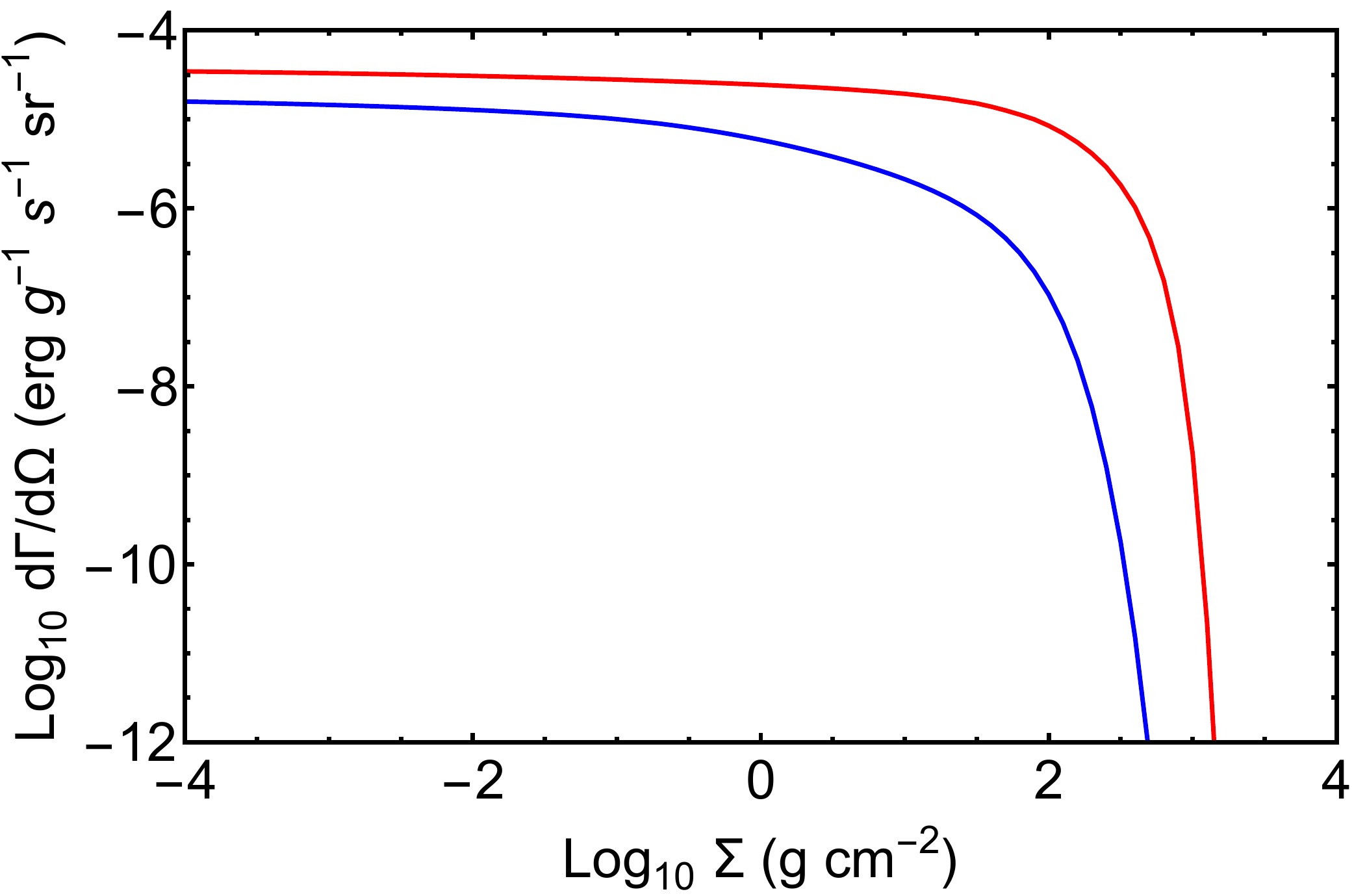}\hskip2cm\plotone{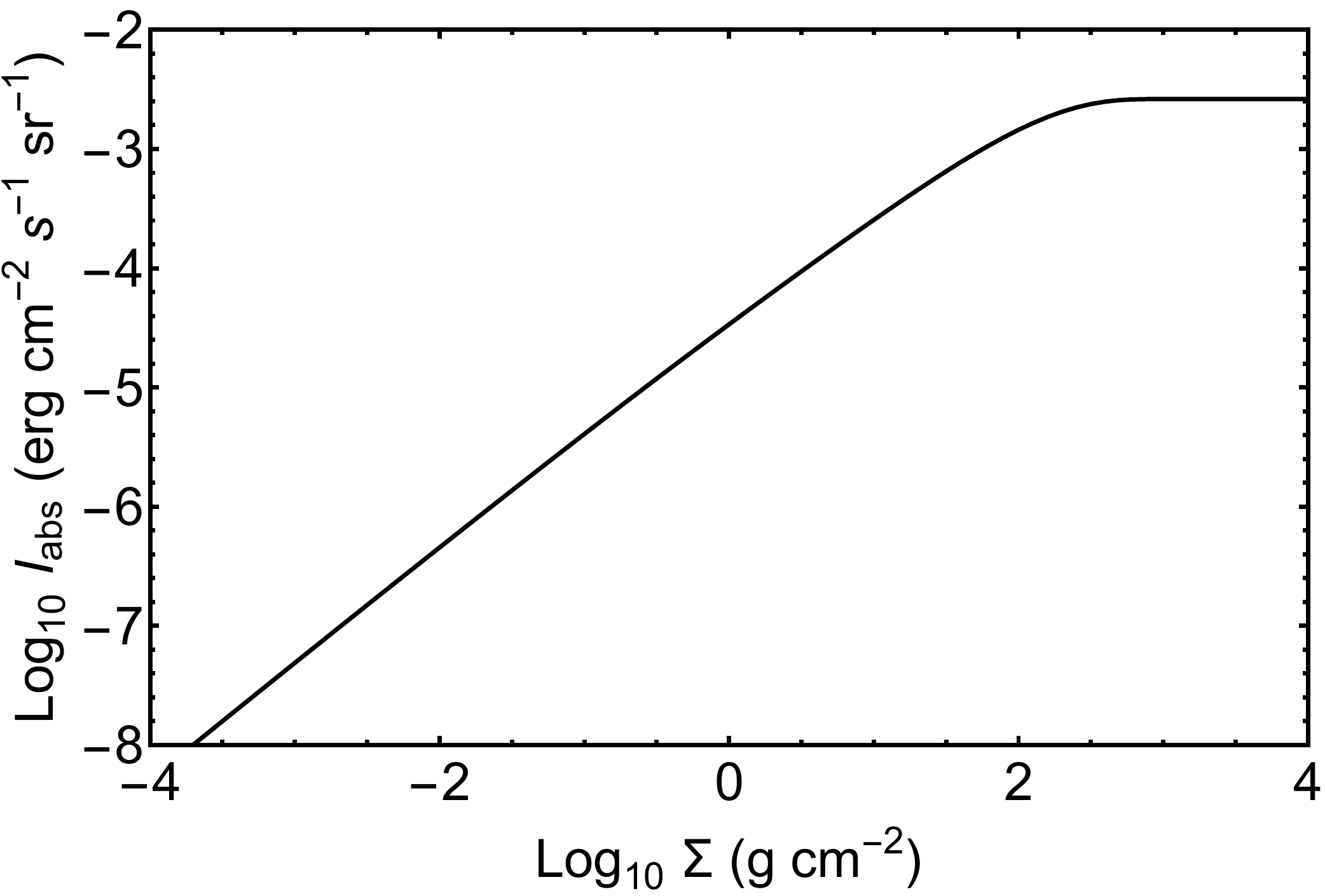}}}
\caption{Left panel: the heating rate per unit solid angle, as a function of column-density, for cosmic-ray electrons (blue curve) and protons (red curve), calculated for the spectra given in equations C13 and C14, and the stopping power of the H$_2$-He mixture as shown in figure 15. Right panel: the absorbed intensity, $I_{abs}=\int{\rm d\Sigma\;d\Gamma_{cr}/d\Omega}$, for cosmic-ray protons and electrons combined.} 
\end{figure}

\vfill\eject
\section{Pressure Profiles}
All of our cloud models comprise a core that is an $n=3/2$ polytrope, and an envelope where \htwo\ is in phase equilibrium. For small values of $T_e/T_c$, the resulting structure hardly differs from a pure $n=3/2$ polytrope; but as $T_e/T_c$ increases the envelope becomes increasingly prominent. Figure D6 shows this progression in the form of pressure profiles, for a sample of 30 snow clouds spanning most of the range in masses and radii that is displayed in figure 5. We note that where the pressure is small compared to that of the ambient medium the computed structure is not relevant in practice, as the outer layers of the cloud would be crushed by its surroundings. 
\begin{figure*}
\figscaleone
\plottwo{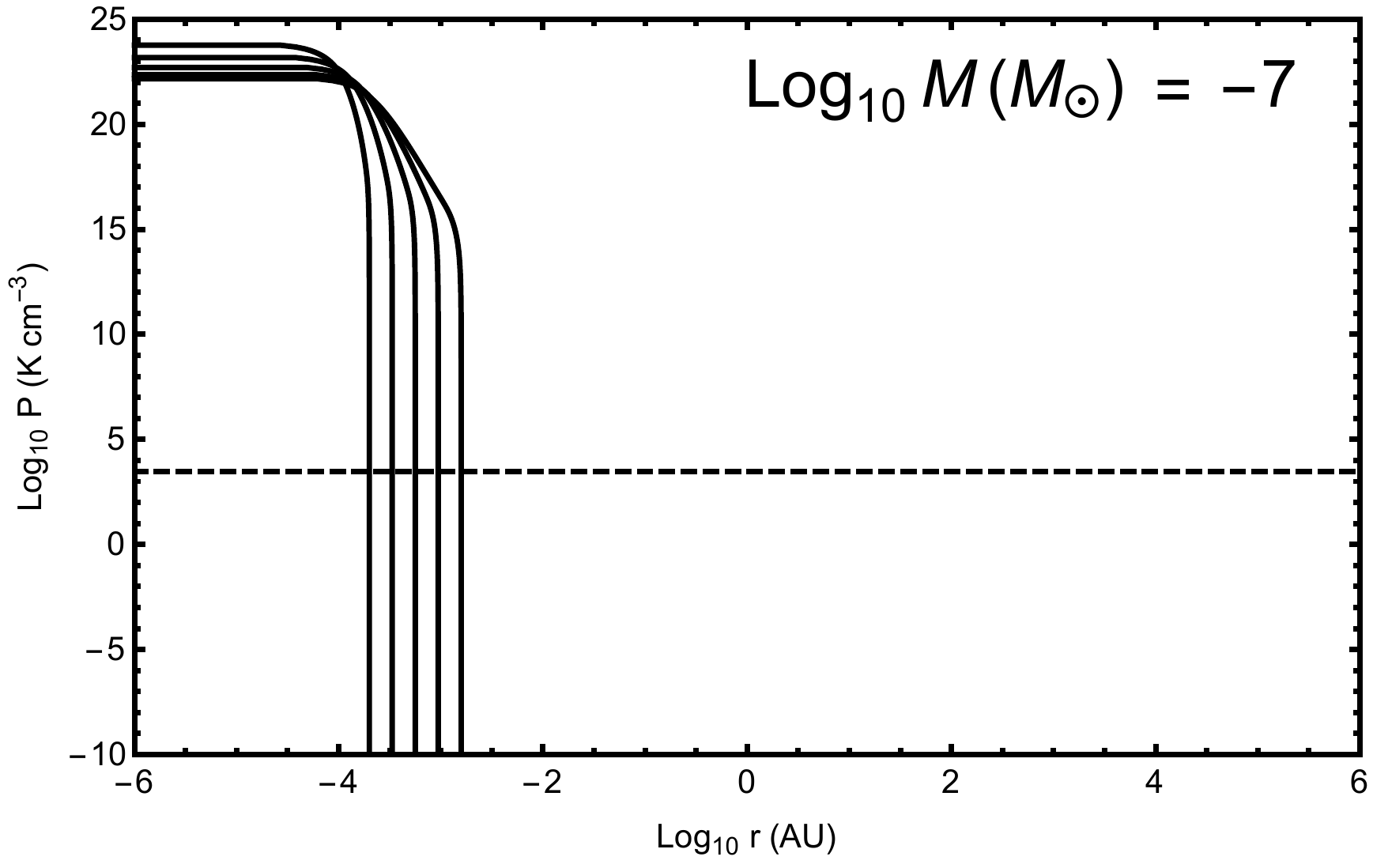}{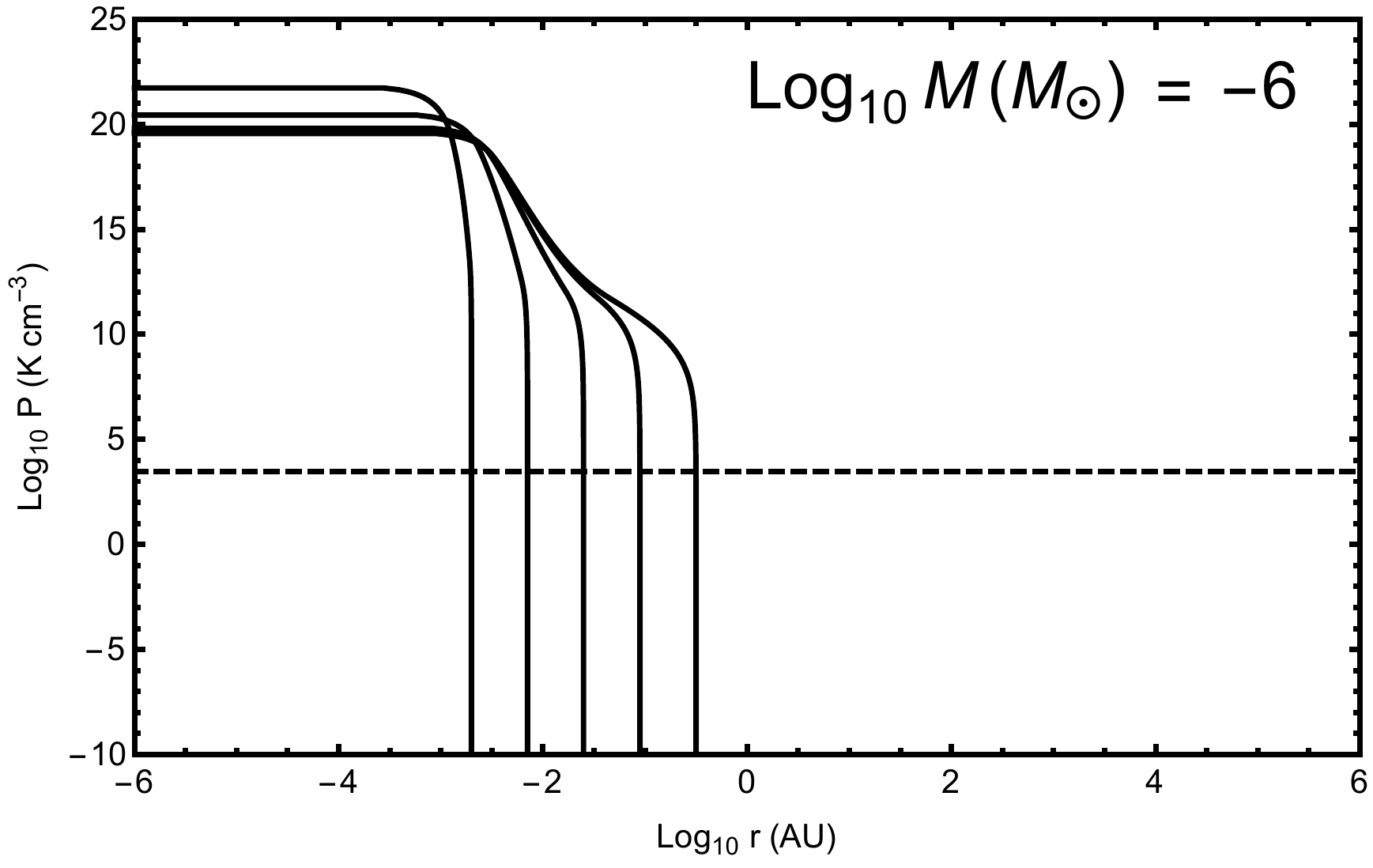}
\plottwo{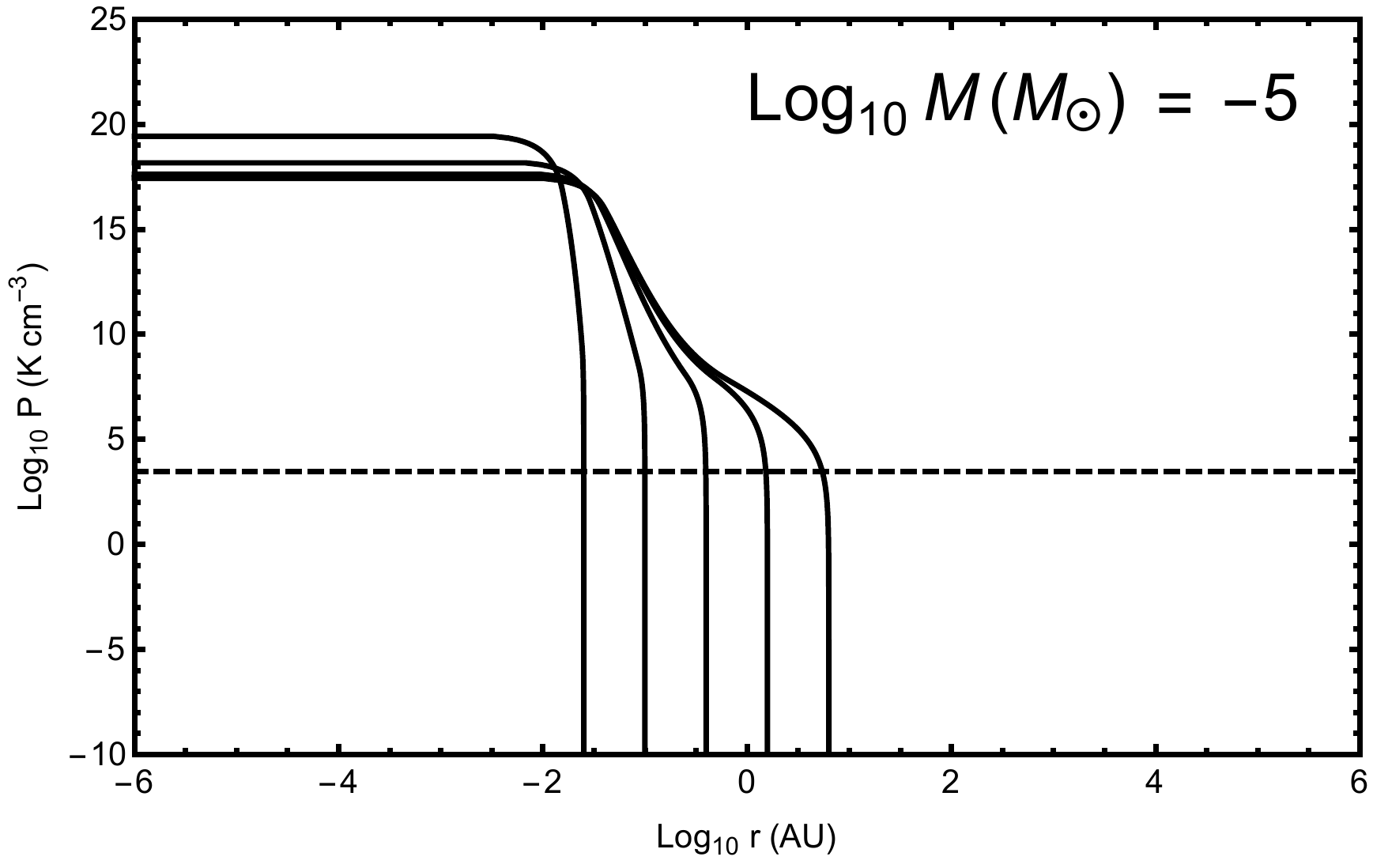}{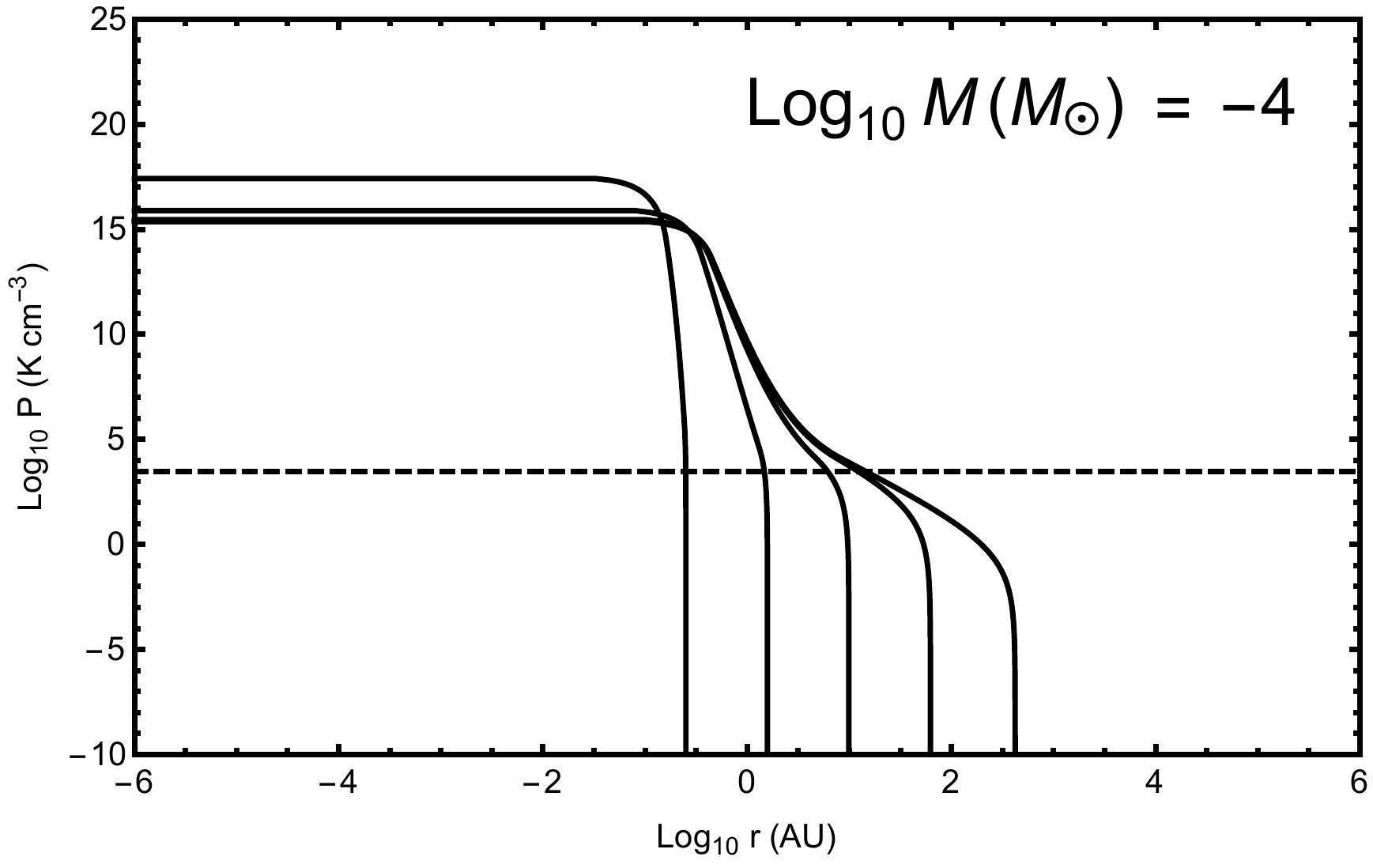}
\plottwo{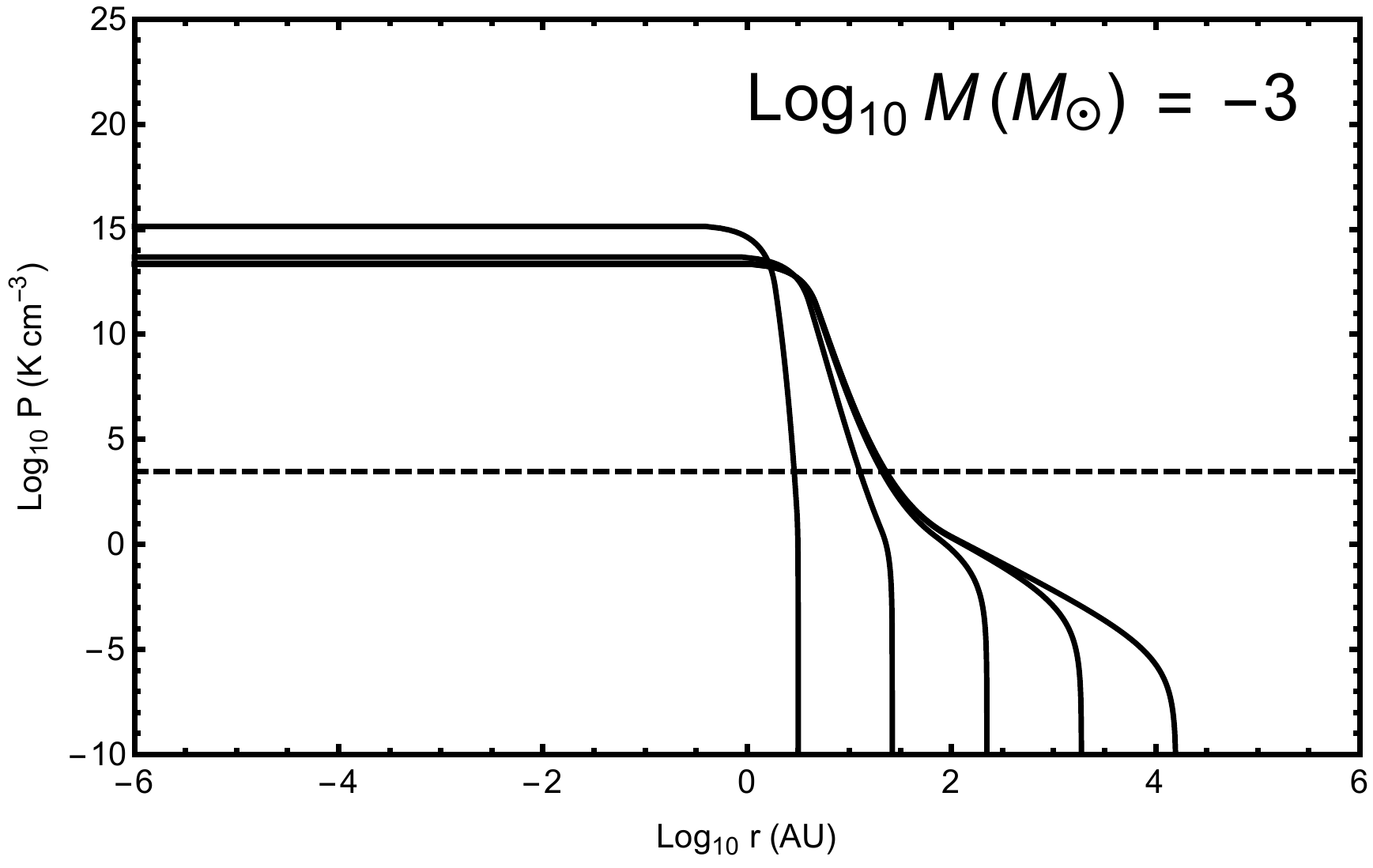}{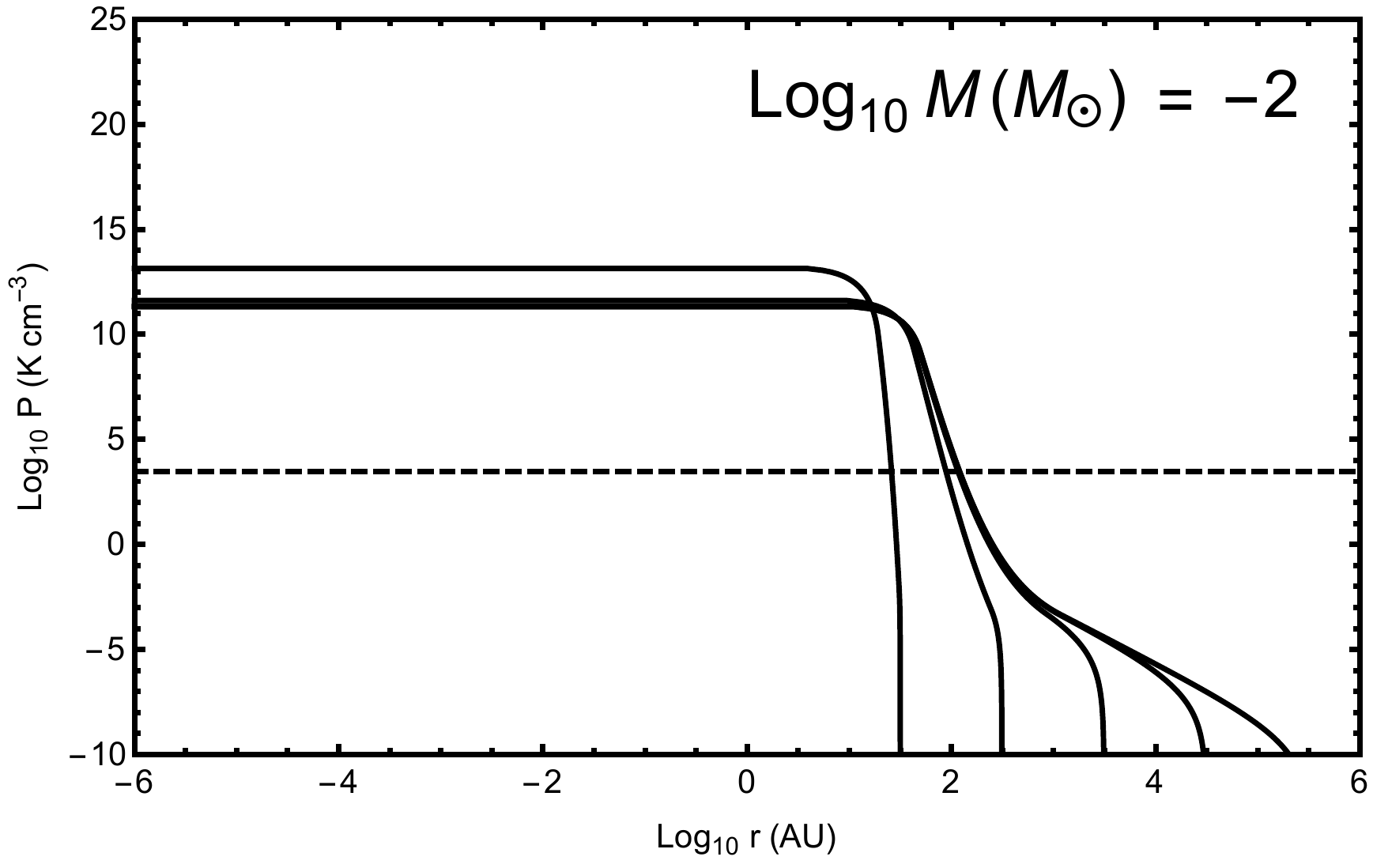}
\caption{Examples of pressure profiles for model snow clouds of various masses. For each mass, five profiles are plotted, chosen to span most of the range of cloud radii exhibited by the models shown in figure 5. The dashed line represents the pressure of the diffuse interstellar medium of our Galaxy \citep[$\sim3{,}000\,{\rm K\,cm^{-3}}$,][]{jenkinstripp2011}.} 
\end{figure*}

\end{document}